\documentclass[10,journal,compsoc]{IEEEtran}
\usepackage{microtype}                 
\PassOptionsToPackage{warn}{textcomp}  
\usepackage{textcomp}                  
\usepackage{mathptmx}                  
\usepackage{times}                     
\usepackage{tabu}                      
\usepackage{booktabs}                  

\usepackage{multirow}
\usepackage{amsmath}
\usepackage{amsthm}
\usepackage{hyperref}
\usepackage{float}

\usepackage[pdftex]{graphicx}
\graphicspath{{../pdf/}{../jpeg/}}
\DeclareGraphicsExtensions{.pdf,.jpeg,.png, .jpg, .eps}

\ifCLASSOPTIONcompsoc
  \usepackage[nocompress]{cite}
\else
  \usepackage{cite}
\fi
\ifCLASSINFOpdf
\else
\fi
\begin{document}

%
\title{Interactive Focus+Context Rendering for \\ Hexahedral Mesh Inspection}
%
%
%
%

\author{Christoph~Neuhauser,~Junpeng~Wang,~and~R\"udiger~Westermann
\IEEEcompsocitemizethanks{\IEEEcompsocthanksitem All authors are with the Computer Graphics \& Visualization Group, Technische Universit{\"a}t M{\"u}nchen, Garching, Germany.\protect\\
E-mail: \{christoph.neuhauser, junpeng.wang, westermann\}@tum.de.}}

\IEEEtitleabstractindextext{%
\begin{abstract}
The visual inspection of a hexahedral mesh  with respect to element quality is difficult due to clutter and occlusions that are produced when rendering all element faces or their edges simultaneously. Current approaches overcome this problem by using focus on specific elements that are then rendered opaque, and carving away all elements occluding their view. In this work, we make use of advanced GPU shader functionality to generate a focus+context rendering that highlights the elements in a selected region and simultaneously conveys the global mesh structure in the surrounding. To achieve this, we propose a gradual transition from edge-based focus rendering to volumetric context rendering, by combining fragment shader-based edge and face rendering with per-pixel fragment lists. A fragment shader smoothly transitions between wireframe and face-based rendering, including focus-dependent rendering style and depth-dependent edge thickness and halos, and per-pixel fragment lists are used to blend fragments in correct visibility order. To maintain the global mesh structure in the context regions, we propose a new method to construct a sheet-based level-of-detail hierarchy and smoothly blend it with volumetric information. The user guides the exploration process by moving a lens-like hotspot. Since all operations are performed on the GPU, interactive frame rates are achieved even for large meshes.
\end{abstract}

}

\maketitle

\IEEEdisplaynontitleabstractindextext

%
\IEEEpeerreviewmaketitle

\begin{figure*}[h]
  \centering
  \includegraphics[width=0.33\linewidth]{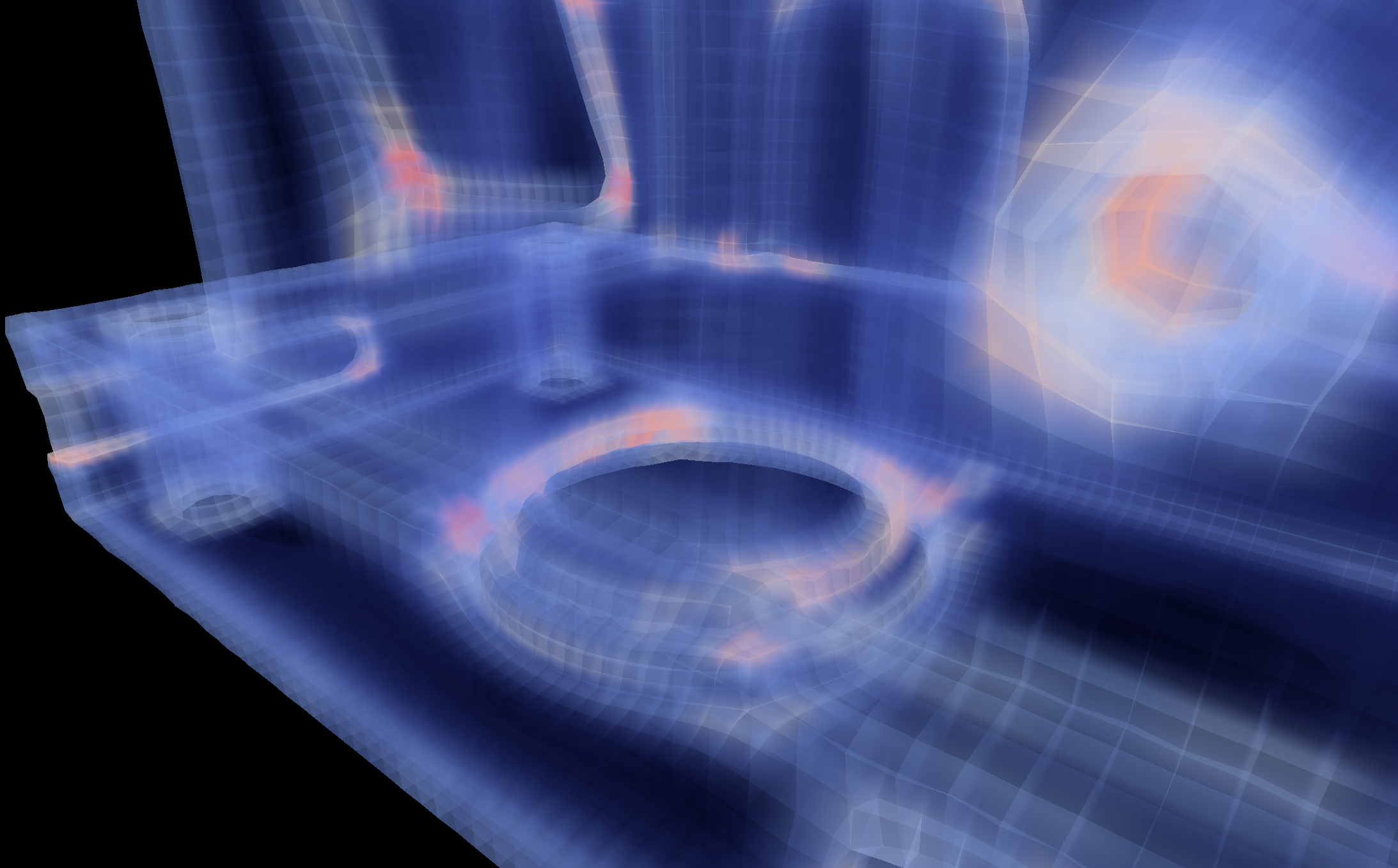}  
  \includegraphics[width=0.33\linewidth]{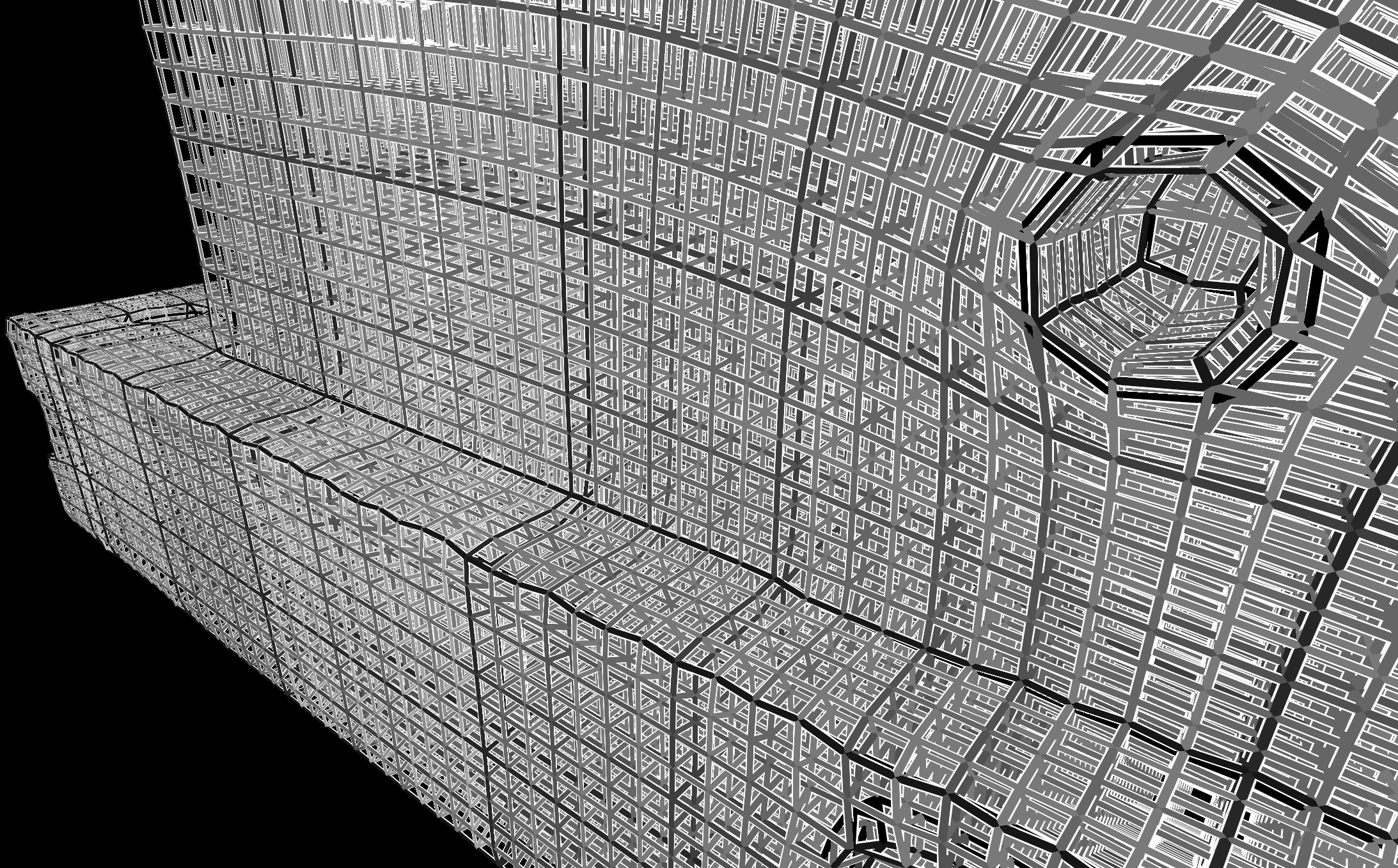} 
  \includegraphics[width=0.33\linewidth]{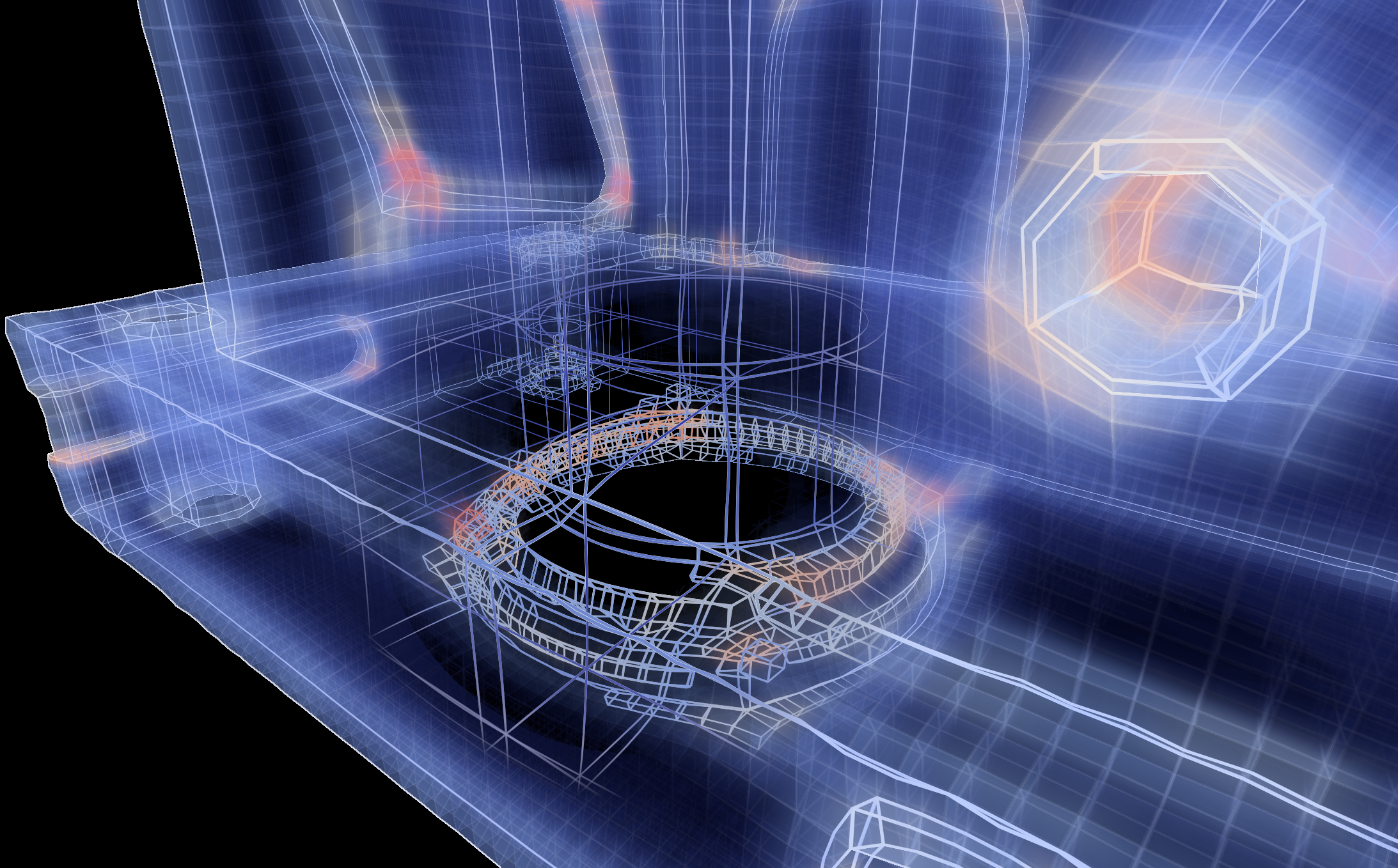} 
  \caption{From left to right: Contextual visualization using face-based volume rendering, edge visualization using fragment-based rendering, our proposed focus+context (F+C) rendering using smooth blending between edge and volumetric rendering including contextual lines. Model courtesy of~\cite{HexMeshSGP2011}.}
  \label{fig:teaser}
\end{figure*}

\section{Introduction}
Hexahedral elements have widespread use in numerical simulation methods using finite elements and finite volumes. 
Therefore, hexahedral mesh generation (hex-meshing) has become a topic of intense research. On the other hand, for all but simple volumetric bodies it is impossible to construct a distortion-free hexahedral mesh, i.e., where the elements are rectilinear cubes (cuboids), that accurately represents the boundary of the body or aligns with specific material features in the interior of the body. Thus, it is one of the grand challenges in hexahedral meshing to construct meshes with as less distortions as possible.

Many different hex-meshing techniques have been proposed over the last years, and for a thorough overview let us refer to \cite{Gao:2017:ACM, Huang2014:LCO, livesu2019loopy}. 
The mesh quality is majorly determined by the scale of deformation of its elements, which can be deduced from the Jacobian matrix that is related to the transformation of a cuboid into the deformed cell. 
A thorough review of the Jacobian ratio metric and alternative deformation measures for assessing the quality of hex-mesh designs is provided by \cite{Gao:2017:EHQ}. 
Interpreting these measures as scalar per-cell or per-vertex attributes yields a volumetric saliency map that indicates important mesh regions.
Depending on the used meshing technique and deformation measure, significantly different saliency maps are obtained and need to be inspected to assess the resulting mesh quality. A visual inspection is difficult, however, since often high deformations occur in the interior of the volumetric body, for instance, along notches or at degenerate points when aligning elements along major stress directions. Yet it is especially such structures which are important, since they reveal the specific differences between different meshing techniques and hint to problematic regions in the body.

\begin{figure}[h]
 \centering
\includegraphics[trim=0cm 0cm 0cm 0cm, clip=true, width=0.99\linewidth]{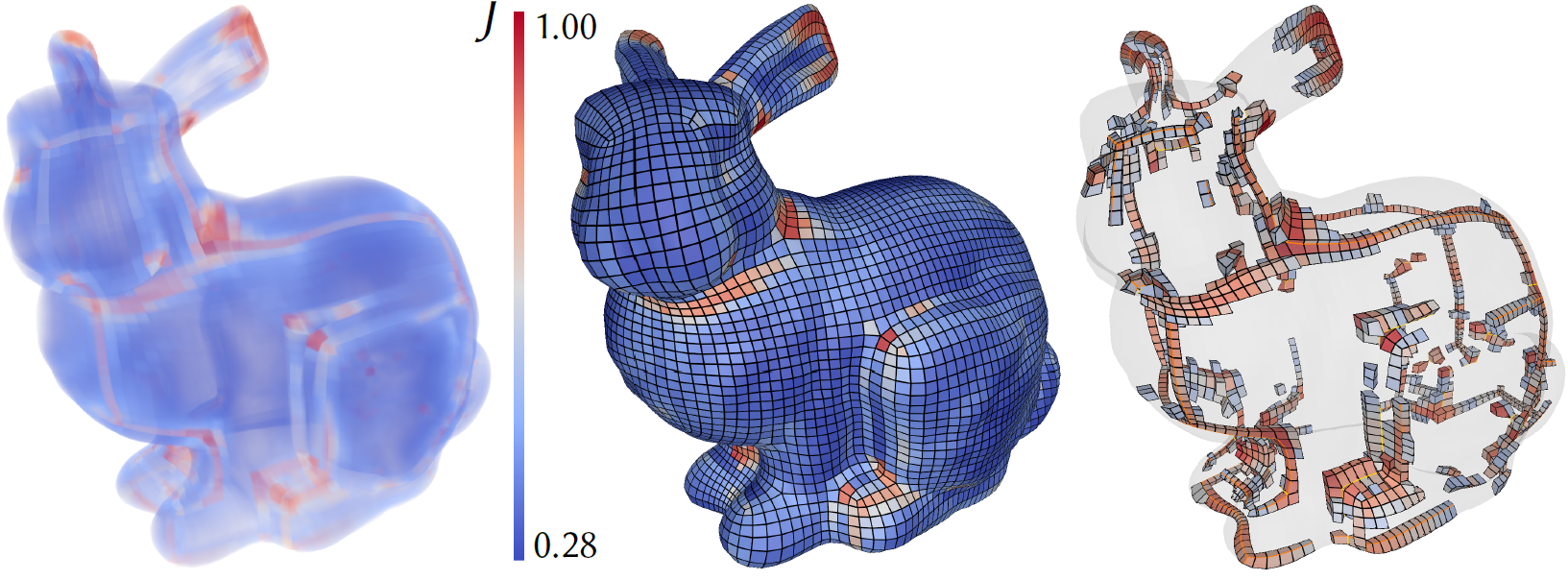}
 \caption{Conventional hex-mesh visualizations. $J$ is the Jacobian ratio of each cell. From left to right: Direct volume rendering, the boundary surface, filtering cells with low deformation. Model courtesy of \cite{Huang2014:LCO}.}
 \label{fig:convMeshVis}
\end{figure}

In principle, direct volume rendering techniques for unstructured volumetric meshes can be used to render the 3D saliency map (\autoref{fig:convMeshVis}~(left)). By using color and opacity, this can effectively locate regions containing highly deformed cells, yet the edge structure is entirely lost. Drawing all edges, on the other hand, results in extreme clutter and occlusions (\autoref{fig:teaser}~(middle)), and prohibits an intuitive understanding of the underlying mesh structure. Thus, common visualization tools for hexahedral meshes often restrict the analysis to the boundary faces of the 3D mesh (\autoref{fig:convMeshVis}~(middle)), and provide options to clip subsets of elements so that interior faces appear (\autoref{fig:convMeshVis}~(right)). 

The most recent approach by Bracci et al.~\cite{Bracci:2019} improves on this by providing additional means to peel away layers of elements from outside to inside, filter cells below a certain distortion value, and reveal hidden irregular edges via transparency rendering. This enables an interactive user-guided visual exploration of hex-meshes. On the other hand, since there is no guidance at first hand to the important regions, the user needs to either actively search through the mesh or filter out a large number of cells to reveal those with high distortion. Furthermore, peeling and filtering can make it difficult to maintain the global mesh structure and spatial relationships between mesh regions with different properties. The latter problem has been addressed by Xu and Chen \cite{Xu:2018:TVCG} via the computation of a global topological mesh structure, which is decomposed into a set of contiguous sub-structures to support a part-based topological complexity analysis. 




\subsection{Contribution}
In this work, we extend on current visualization techniques for hex-meshes by a combination of face-based volume rendering with fragment-based edge rendering. Our goal is to effectively communicate the mesh structure by embedding a carefully designed detail view into a surrounding context that conveys important positional cues. 
In the global context view the saliency map is rendered as a semi-transparent volumetric field in combination with few contextual edges, so that important regions and the spatial mesh structure can quickly be recognized. The detailed focus view is selected via a user-defined screen space lens with depth focus, in which edge-based rendering is used. 

To obtain a smooth transition from edge-based focus rendering to volumetric context rendering, we introduce a GPU renderer for hex-meshes that solely renders cell faces and performs all operations that change the mesh appearance in a fragment shader. The shader smoothly blends between different rendering options depending on where a fragment is located w.r.t.\ the focus region and whether a fragment is an edge or interior face fragment.
Furthermore, we incorporate an edge-based level-of-detail (LoD) structure into the renderer, to adapt the density of rendered mesh edges depending on cell distortion and distance to the focus center. The coarser LoDs further serve as shape cues in the context region. 
Since all rendering options are performed solely on the fragment level, a smooth blending from sharp details to a more fuzzy appearance with embedded characteristic edges as shape cues can be performed efficiently.
Our proposed visualization technique builds upon the following specific contributions:
\begin{itemize}
 \item A single-pass GPU renderer with fragment shader-based edge and volume rendering including transparency.
 \item A LoD line hierarchy that is extracted from a hex-mesh using a topological subdivision scheme based on mesh sheet elements.
 \item A rendering technique that smoothly blends between focus and context, by continuously adapting edge density as well as edge and face appearance.
\end{itemize}  

We demonstrate our approach for a number of hex-meshes, including a detailed quality and performance analysis that justifies its feasibility even for meshes comprised of millions of elements. The code is published on \url{https://github.com/chrismile/HexVolumeRenderer}.

\section{Related Work}
\textbf{Hexahedral Mesh Visualization} Visualization techniques for hexahedral grids can be divided into surface-based and direct volume rendering techniques. Surface-based techniques render hexahedral elements as opaque cuboids, including wireframe rendering and face coloring to emphasize certain element properties. For a thorough overview of the different rendering styles that are used in modelling applications, let us refer to the recent work by~\cite{Bracci:2019}. They also introduced novel line-based visualization options to maintain the overall model structure and emphasize singular edges in a hex-mesh. For computing the deformation of cells, our implementation uses the code from~\cite{Bracci:2019}, which implements various measures for cell deformation supported by the Verdict library~\cite{VerdictLibrary}. A summary and discussion of different quality metrics for hex-meshes is given by~\cite{Gao:2017:EHQ}. The code from~\cite{Gao:2017:ACM} is used for loading and processing hexahedral meshes. 

Recently, \cite{Xu:2018:TVCG} proposed to visualize the mesh structure of hexahedral meshes by using a subset of the most important base-complex sheets and dual chords, and show their interrelation using adjacency matrices. We take inspiration from their approach utilizing base-complex mesh sheets to reduce the structural complexity of a mesh (cf. Section~\ref{sec:lod}). Our approach uses hexahedral sheets \cite{Woodbury:2011,Borden2002HexahedralSE} instead of base-complex sheets, and merges sheets for creating a LoD structure instead of directly visualizing a subset of them. The use of hexahedral sheets for hex-mesh construction, simplification and reparameterization is thoroughly discussed in the work by \cite{wang2017sheet}.  

Direct volume rendering of hexahedral meshes has a long tradition in volume visualization, and many of the concepts that are used by more recent works are discussed in the survey by~\cite{silva2005survey}. Our GPU-based approach shares similarities with cell projection techniques w.r.t. how the cells are rendered and their visibility order is established. 
Cell projection techniques exploit the GPU to efficiently render triangles and perform linear interpolation of per-vertex attributes for each rendered fragment. Cuboids are first decomposed into tetrahedra, and then rasterized and blended using the GPU~\cite{weiler2003hardware, callahan2005hardware, marroquim2006gpu,georgii2006generic}. \cite{callahan2005hardware} utilized the GPU for visibility sorting of rendered fragments, which is conceptually similar to the approach we employ for visibility sorting using per-pixel fragment lists~\cite{Yang2010}, a GPU realization of the A-buffer~\cite{carpenter1984buffer} to store the unordered set of fragments falling into each pixel. These fragments are then sorted explicitly based on the stored depth information. Recently, SparseLeap~\cite{Beyer2017:SparseLeap} has been introduced as a pyramidal occupancy map to generate geometric structures representing non-empty regions, which makes use of per-pixel fragment lists to determine occupied space and accelerate volume ray-casting. 

\textbf{Focus and Context} Focus+context (F+C) visualization techniques aim at smoothly combining different aspects of the data into one single visual representation. While the contextual visualization provides cues to understand the overall shape and spatial arrangement of the model or scene, the focus emphasizes specific aspects of the data. In F+C data visualization, especially lens-based approaches have a long tradition ~\cite{Tominsks2017}. Distortion lenses have been used in volume rendering applications to magnify structures in focus \cite{lamar2001magnification, ikits2004focus, wang2005magic}.  \cite{viola2005importance} discuss how two obtain different sparsity levels depending on the importance of structures, and propose importance-based rendering styles. \cite{Traore2019} use an object-space lens in combination with a fish-eye view to distort structures and push them out of focus when occluding features of interest. \cite{Krueger:2006:ClearView} combine renderings of an exterior and interior isosurface using a screen-space lens. \cite{Dick:2009:StressTensorFields} demonstrate the application of a screen-space lens for F+C stress visualization, by letting the thickness and number of stress lines being controlled by the lens. We make use of a circular screen space lens to let the user select a cylindrical focus region in object space and smoothly blend into a volumetic context view with increasing distance to the lens center. 

Related to our proposed edge-based F+C rendering is the use of transparency and adaptive primitive density for streamline rendering. When too many lines are shown simultaneously, occlusions and visual clutter are quickly introduced. While we address this by smoothly blending into a volumetric context and using few representative edge sequences from coarser LoDs, others have proposed importance- and similarity-based criteria in screen-space to select the rendered lines dynamically on a frame-to-frame basis. Screen-space approaches determine for each new view the subset of lines to be rendered so that occlusions are reduced and more important lines are favored over less important ones~\cite{marchesin2010view, lee2011view, ma2013coherent}. The amount of occlusion is determined by the “overdraw”, i.e., the number of projected line points~\cite{marchesin2010view, ma2013coherent} or the maximum projected entropy~\cite{lee2011view} per pixel. \cite{gunther2013opacity} decrease the opacity of non-important foreground lines using per-frame opacity optimization. A summary and evaluation of different GPU transparency rendering techniques for large line sets is given by~\cite{Kern:2020:TVCG}. 
\cite{kanzler2016line} build a line hierarchy to continuously decrease the density of less important lines.

\section{Method Overview}

\begin{figure}[h]
 \centering
 \includegraphics[width=0.99\linewidth]{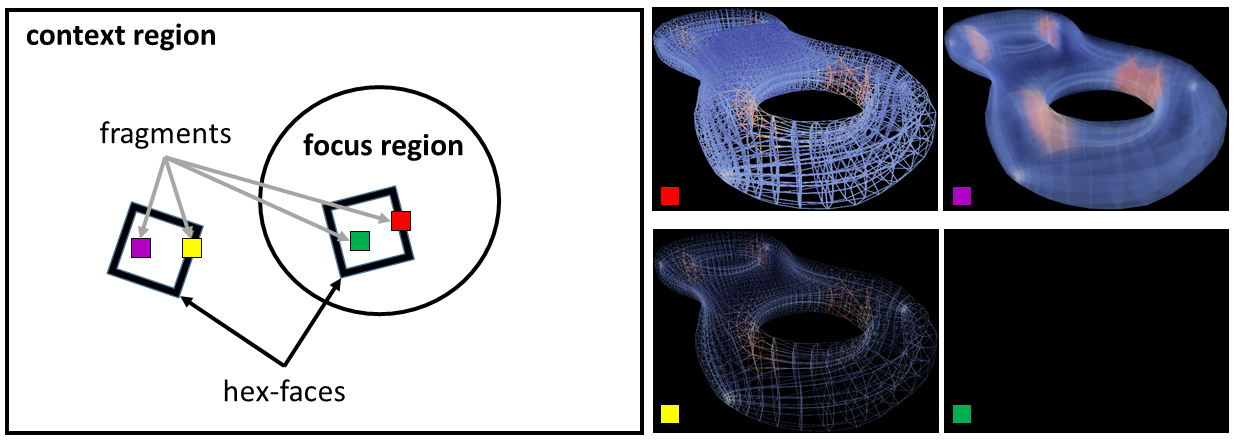}
 \caption{Left: Classification of fragments depending on whether they are in focus or context, and whether they are close to a face edge or not. Right: Depending on the classification, the fragments take on different appearances. For each fragment, the image shows how the rendering looks like if only fragments of this type are rendered.}
 \label{fig:fragclass}
\end{figure}

Our method renders all hex-faces as a quadrilateral formed by two triangles. A fragment shader determines the appearance of each fragment depending on whether it lies in the focus or context region, and further depending on whether it is an “edge fragment”, i.e., lying closer to a face edge than a given edge thickness, or a “face fragment”, i.e., lying too far away from any of the face edges. This classifies each fragment into 4 different types that determine how it is shaded (see~\autoref{fig:fragclass}). While in the context region a more volumetric appearance with subtle edge accentuation is used, in the focus region only the edges are clearly emphasized. Edge and face colors and opacities are made dependent on the importance measure, i.e., the strength of cell deformation, so that also in the context region important cells are emphasized. Since the importance measure is cell-based, every vertex gets assigned the maximum importance value of all cells sharing this vertex. The triangle rasterizer then brings the interpolated importance values to the fragments. In addition, the maximum importance value of all cells sharing an edge are made available in the fragment shader for that edge. 

The renderer changes the appearance smoothly from edge-based to volumetric with increasing distance to the focus center in screen space, as described in Section~\ref{sec:fcc}.
Therefore, the opacity and width of edges in focus is smoothly decreased towards the focus border, and the color is blended towards the face colors used for rendering the context region. 
For each pixel, all fragments falling into that pixel are stored in a per-pixel fragment list on the GPU, and they are sorted w.r.t. increasing distance to the camera. This enables opacity-based blending, i.e., $\alpha$-blending, of fragments in the correct visibility order. For sorting, we use a GPU-friendly implementation of priority queues \cite{Kern:2020:TVCG}. 

\begin{figure}[h]
 \centering
 \includegraphics[width=0.99\linewidth]{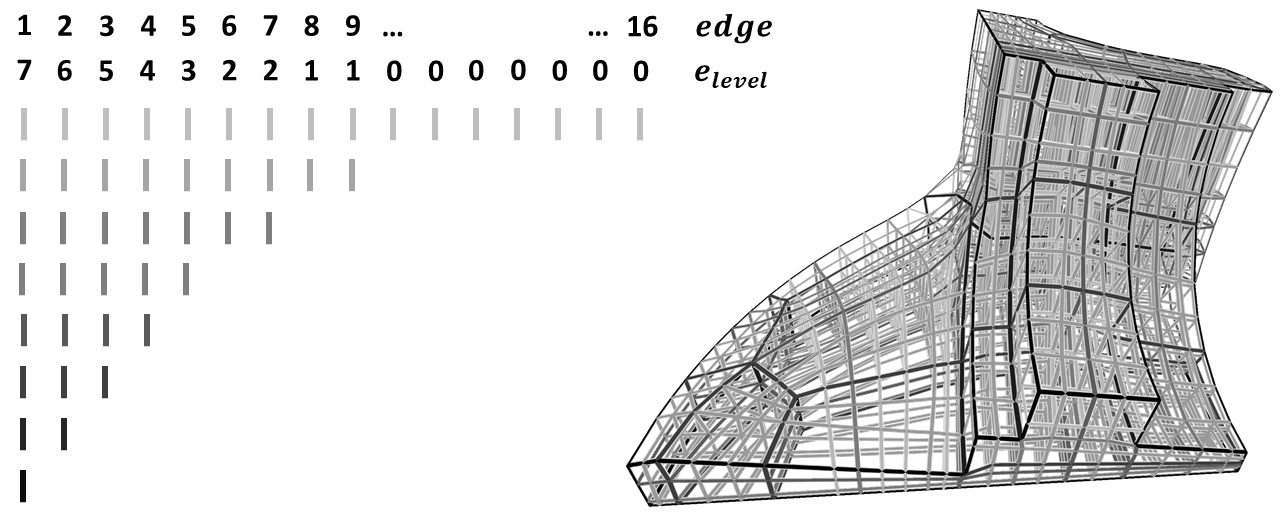}
 \caption{Left: Edges are continually thinned out from level to level in the LoD hierarchy. Single edges get assigned the level at which they are last contained. Right: The LoD edge structure for a given hex-mesh. Greyscales from bright to dark encode LoD levels from fine to coarse. Model fandisk courtesy of~\cite{DualSheetMeshing2019}.}
 \label{fig:lod}
\end{figure}

The described rendering approach has two limitations: Firstly, in the focus region there can be many non-important edges that occlude important ones. Secondly, in the context region the basic mesh structure gets lost due to increasing volumetric appearance. To address these limitations, we construct a LoD line structure (Section~\ref{sec:lod}), in which mesh edges are continually removed at coarser hierarchy levels. \autoref{fig:lod} illustrates how the LoD structure is used, by assigning to every edge the maximum level at which this edge is still present in the LoD structure. In the focus region, instead of removing edges with an importance value below a selected threshold, these edges are rendered if they are also present at some coarse LoD. We call these edges \emph{contextual edges}. In the context region, only contextual edges are rendered to provide an overview of the shape of the hex-mesh. 


\section{Focus+Context} \label{sec:fcc}
In the following, we describe how focus and context rendering is performed, and in particular how a smooth transition between both is achieved. A detailed discussion of the reference GPU implementation is given in Section~\ref{sec:implementation_details}. The user defines the focus region by positioning a circular lens with a \emph{center} and controlled \emph{radius} in screen space. The \emph{focus} is 1 at the lens center and goes smoothly down to 0 towards its boundary.

Regardless of whether a fragment is finally shaded to appear as part of an edge or a face, hex-faces are rasterized with two triangles, and per-vertex attributes like the cell importance are barycentrically interpolated. For every fragment, a fragment shader determines whether it should appear as an edge or a face. This is performed by first computing a fragment's screen space coordinate and its distance $dist$ to the focus center (normalized to range from 0 at the focus center to 1 at the focus boundary), and evaluating the \emph{focus} as $1 - smoothstep(0.7, 1, dist)$.

Then, a fragment's edge opacity $\alpha_e$, which determines whether the fragment belongs to a face ($\alpha_e=0$) or an edge ($\alpha_e > 0$), is determined as follows:
\begin{equation} 
\begin{gathered}
w = (1 + 0.3 \cdot focus) \cdot w_{base} \\
e_{focus} = (d_{edge} \leq w \land (e_{level} \geq lod \lor e_{attr} \ge \delta)) \; ? \; 1 : 0 \\
e_{context} = (d_{edge} \leq w \land e_{level} \geq lod) \; ? \; 1 : 0 \\
\alpha_e = lerp(e_{context}, e_{focus}, focus)
\end{gathered}
\label{equ:edges-primary}
\end{equation}
Here, $w_{base}$ is a minimum edge width, $d_{edge}$ is the fragment's shortest distance to any of the face edges, $e_{level}$ is the LoD level of the edge (\autoref{fig:lod}), and $e_{attr}$ is the edge importance. 
First, the edge width is decreased with increasing distance to the focus center. 
Then, via $e_{context}$ and $e_{focus}$, respectively, it is determined whether the fragment belongs to an edge that should be rendered when lying in the focus or context region. In focus, an important edge is always rendered, i.e., $e_{attr}$ is greater than a selected importance threshold $\delta$. An unimportant edge is rendered only if it's at a user-selected coarse LoD level $lod$. In context, every edge with $e_{level} \geq lod$ is rendered. The final distance-based linear interpolation between $e_{context}$ and $e_{focus}$ transitions smoothly from focus to contextual edges. 

\begin{figure}[h]
 \centering
 \includegraphics[width=0.48\linewidth]{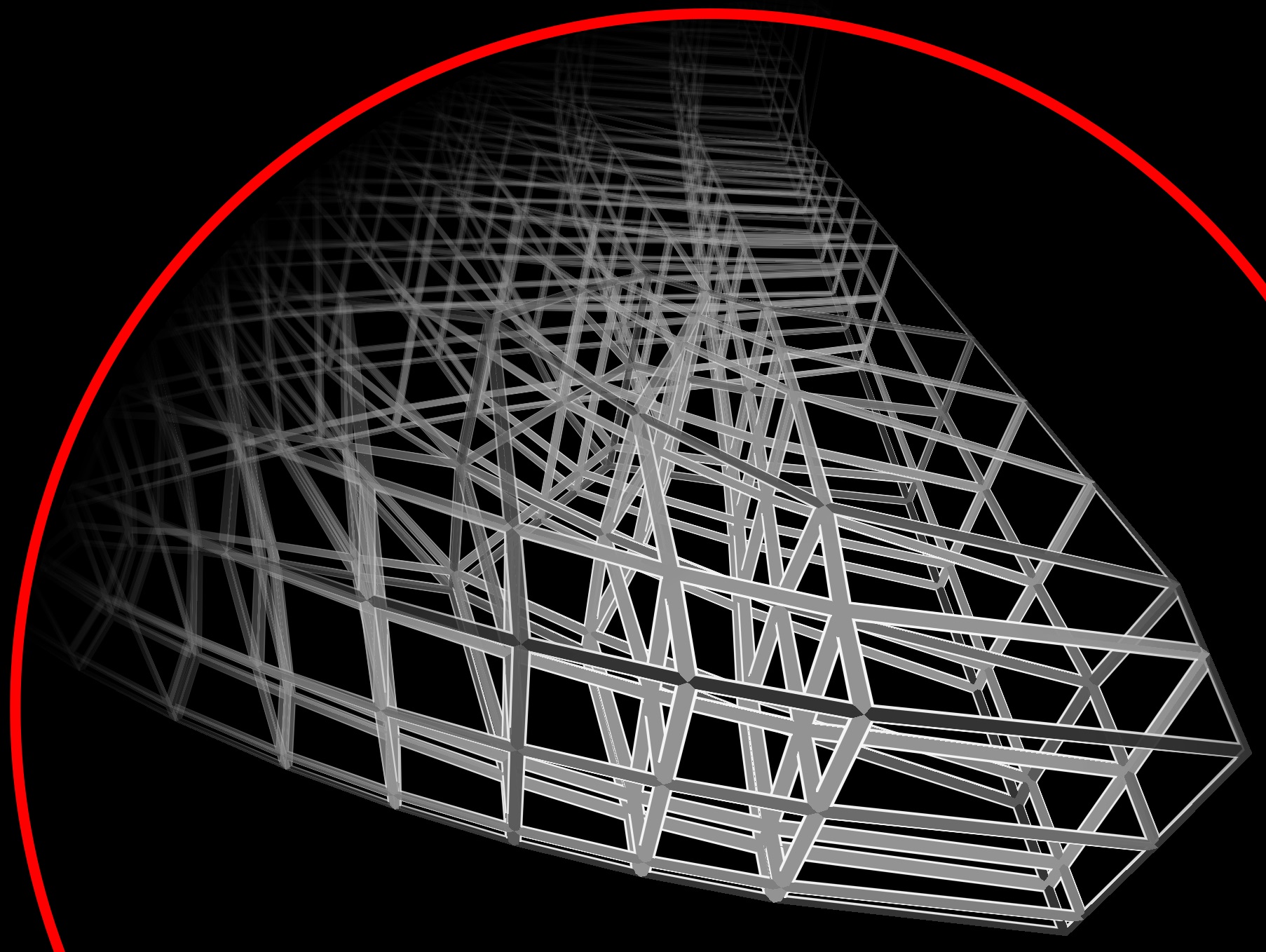}
 \includegraphics[width=0.48\linewidth]{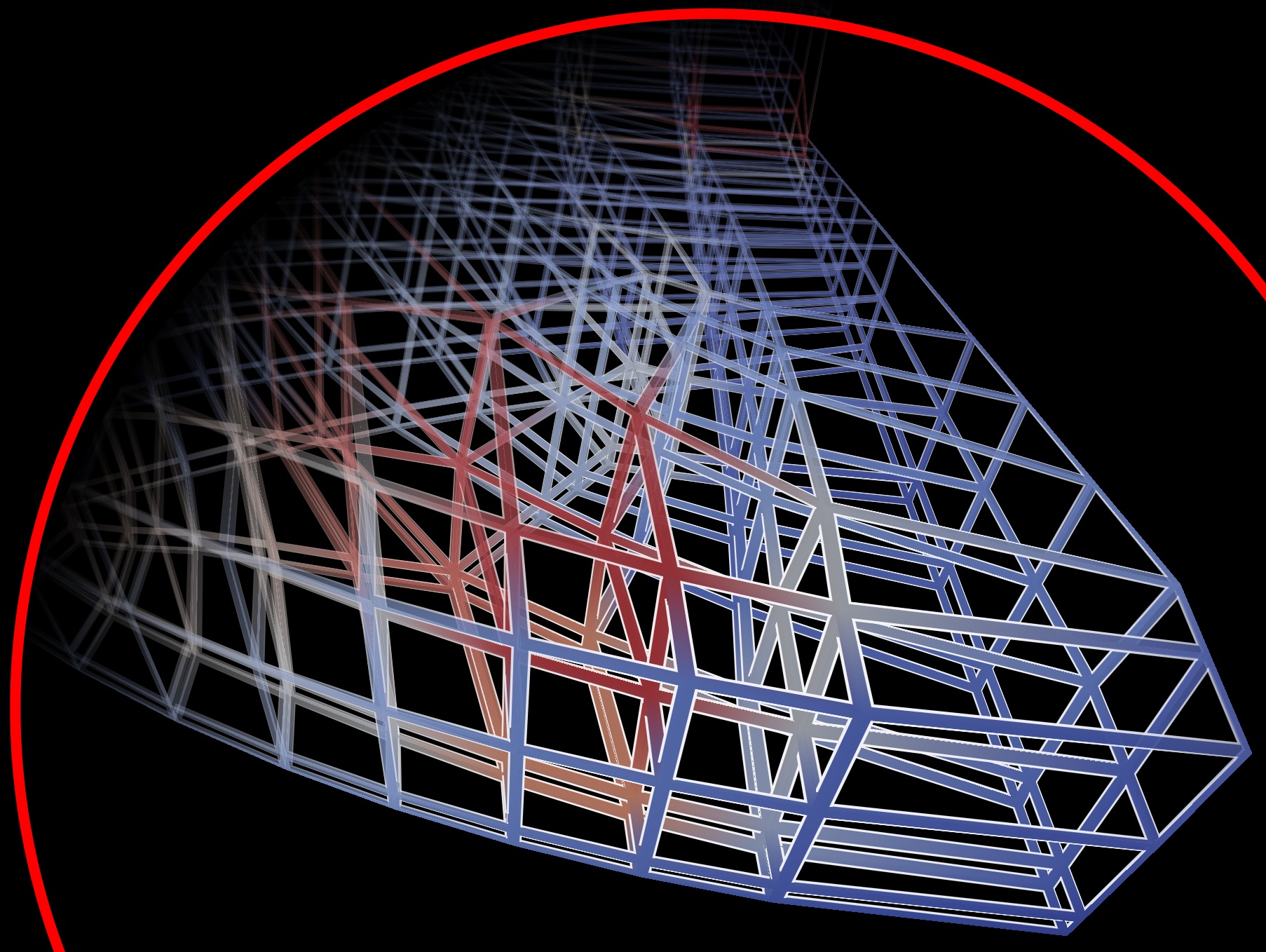}
 \caption{Focus edges are smoothly faded out with increasing distance to the focus center. Left: Edges colored by LoD level. Right: Edges colored by interpolated per-vertex deformation measure. Model fandisk courtesy of \cite{DualSheetMeshing2019}.}
 \label{fig:blendout}
\end{figure}

\begin{figure}[h]
 \centering
 \includegraphics[width=0.48\linewidth]{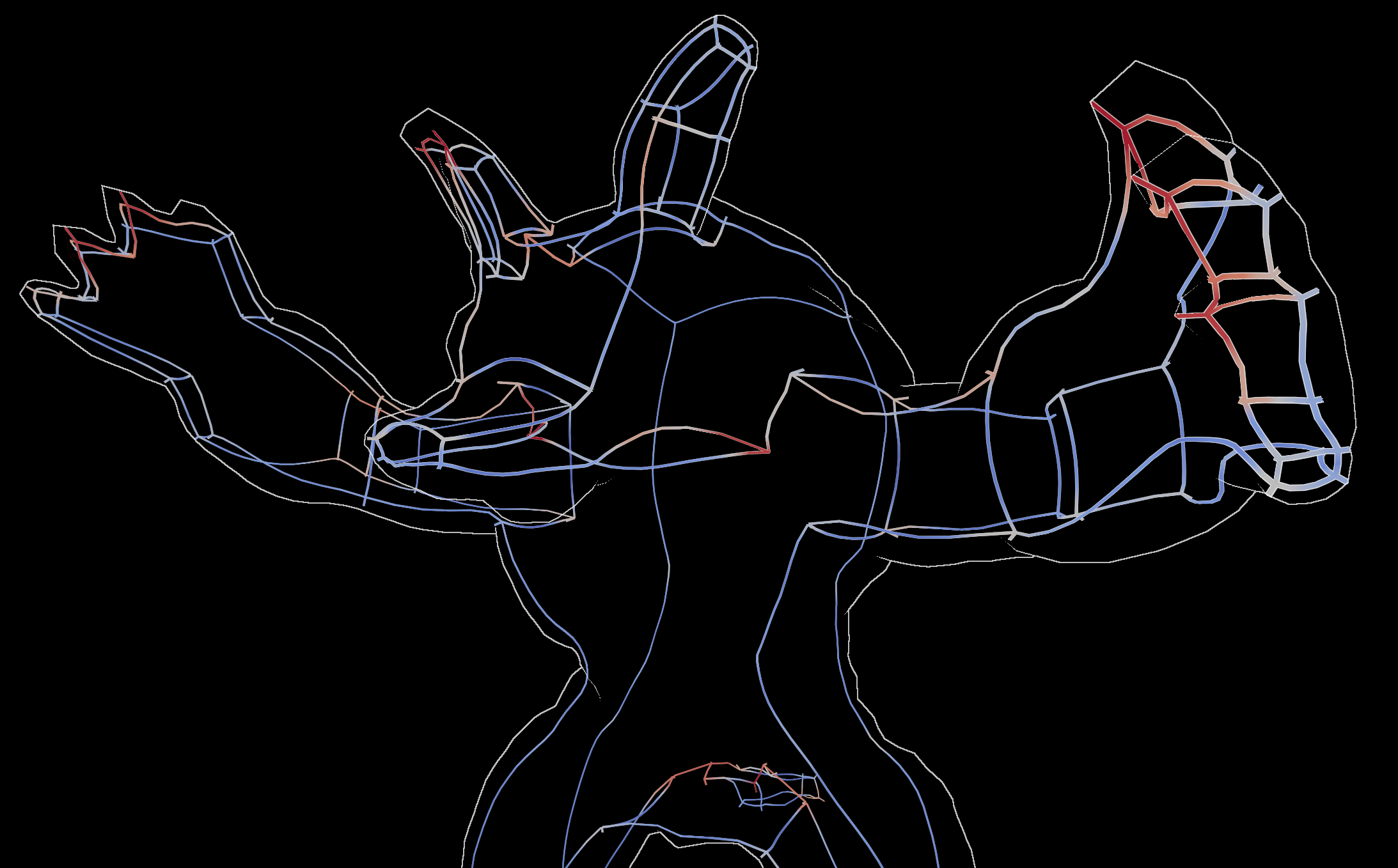}
 \includegraphics[width=0.48\linewidth]{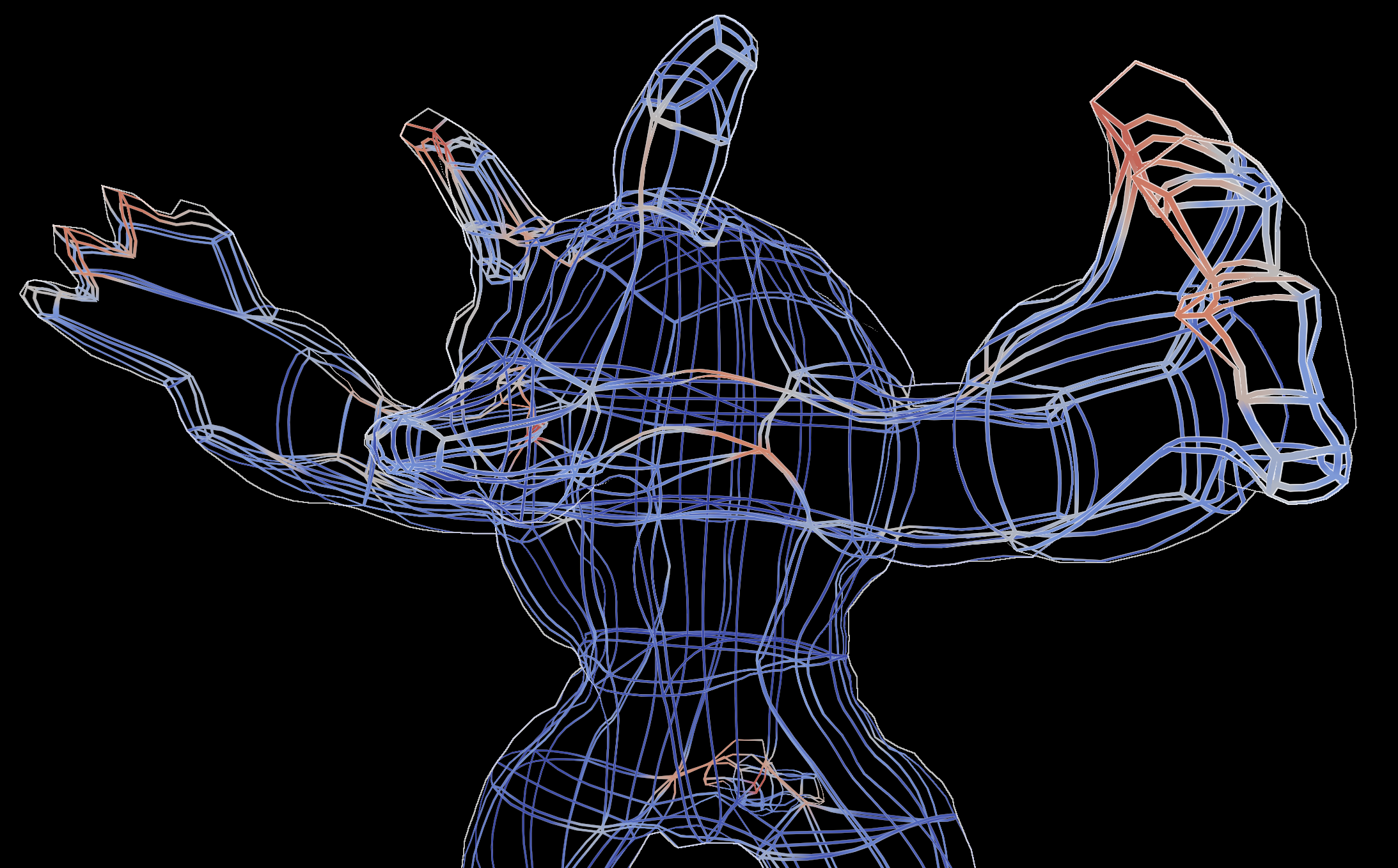}
  \caption{Decreasing $lod$ from 11 (left) to 8 (right) increases the density of contextual edges. No focus selected. Model armadillo courtesy of \cite{EdgeConeRectification2015}.}
 \label{fig:LODvariation}
\end{figure}

The shader renders the focus edges with a thin white depth-dependent halo \cite{Everts:2009:TVCG}. The halo gets thinner with increasing distance to the focus, and the edge colors are slightly darkened to make the edges stand out against the halo (\autoref{fig:blendout}). 
Focus edges blend into contextual edges, which are rendered as simple lines without a halo and colored according to the deformation measure. To maintain certain contextual edges as spatial cues in the focus and context region, the user can interactively select the value of $lod$ (\autoref{fig:LODvariation}). 

\begin{figure}[h]
 \centering
 \includegraphics[width=0.49\linewidth]{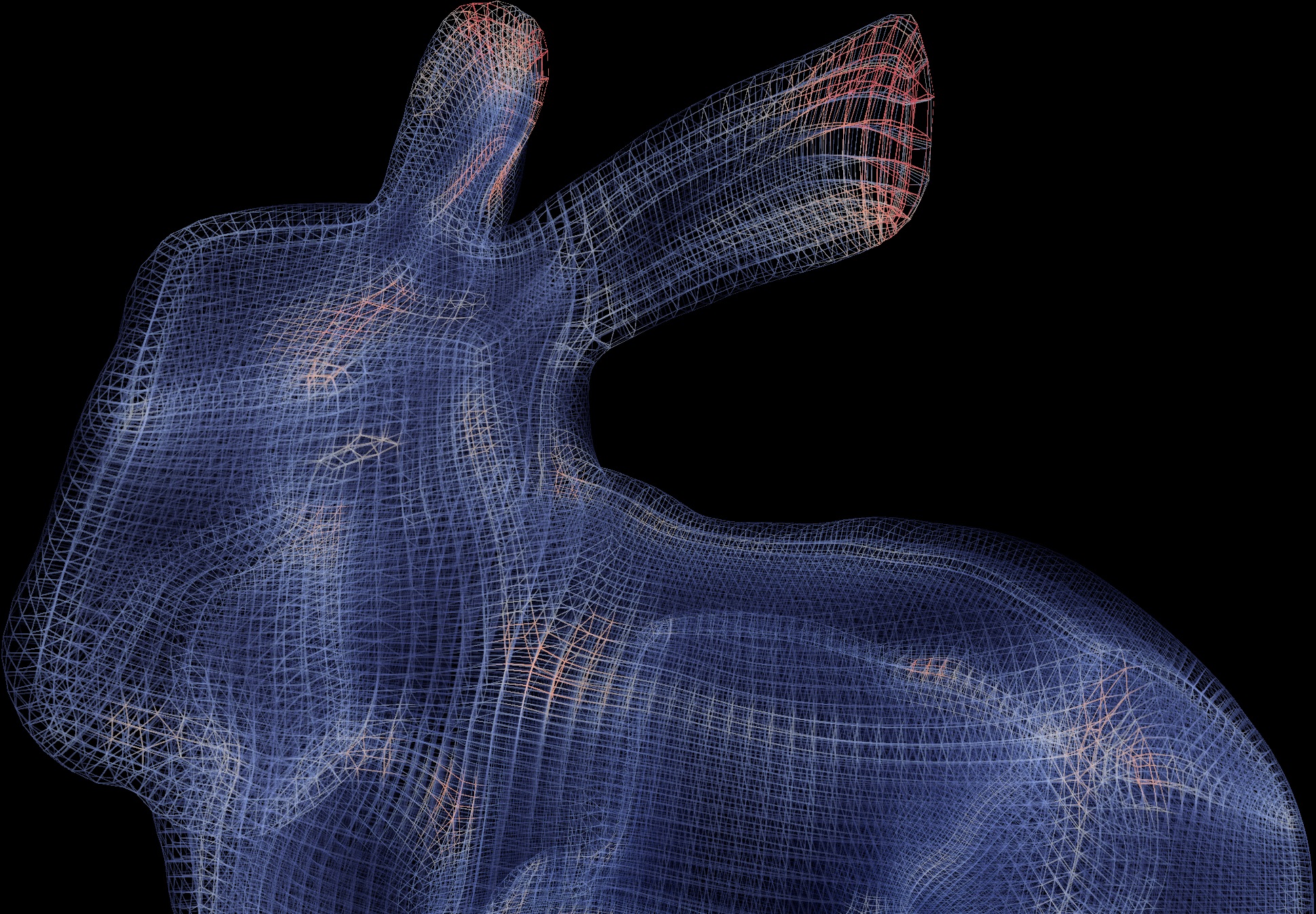}
 \includegraphics[width=0.49\linewidth]{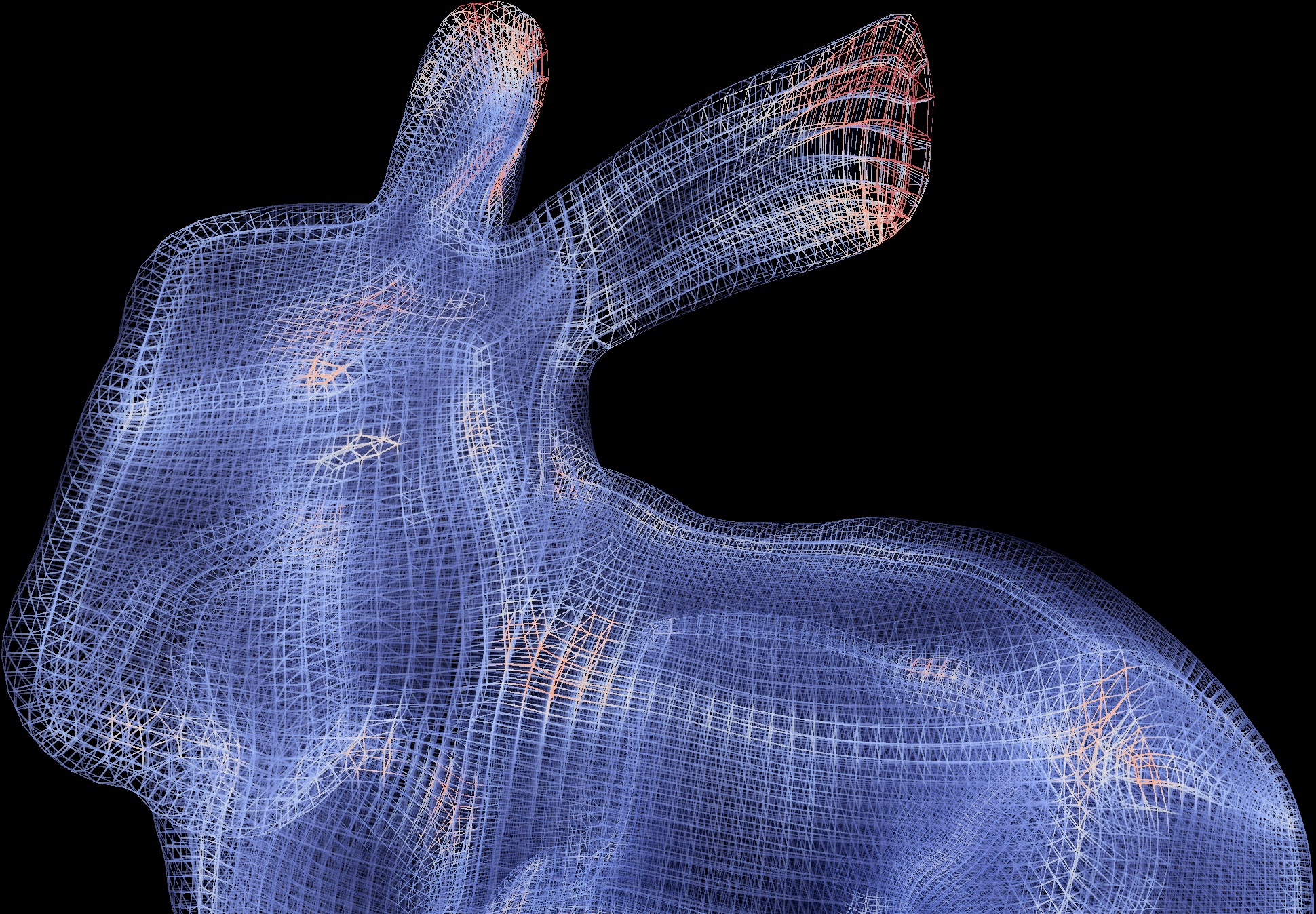}
  \caption{Weakly ($s=1.5$) and strongly ($s=3$) accentuated edges according to ~\autoref{equ:edges-secondary}. No focus selected. Model bunny courtesy of \cite{HexMeshSGP2011}.}
 \label{fig:edges_accentuated}
\end{figure}

Furthermore, fragments in the context that are in close vicinity to an edge, but are not visible at the selected LoD level, are slightly accentuated. If such a fragment doesn't belong to an edge according to ~\autoref{equ:edges-primary}, $s_{e}$ determines how strongly it is emphasized: 
\begin{equation}
\begin{gathered}
s_{e} = d_{edge} \leq \frac{2}{3} \cdot w_{base} \; ? \; s : 1
\end{gathered}
\label{equ:edges-secondary}
\end{equation}
$s_{e}$ is used to enhance the face opacity $\alpha_f$ (see \autoref{eqn:final_compositing}). Since $\alpha_f$ depends on the distance to the focus center (\autoref{equ:face_opacity}), accentuated edges fade out accordingly.  
\autoref{fig:edges_accentuated} demonstrates varying  accentuation of contextual edges by variation of the accentuation strength $s$.  


Both parameters $\alpha_e$ and $s_{e}$ are used to assign the fragment opacity that emphasizes certain edges and smoothly blends between focus edges and contextual edges with increasing distance to the focus center. The edge colors are set via a color table that maps the edge importance values to colors $C_e$ (see Subsection~\ref{sec:fccombined}).   


\subsection{Contextual Volume Rendering} \label{sec:VolRen}

\begin{figure}[h]
 \centering
 \includegraphics[width=0.49\linewidth]{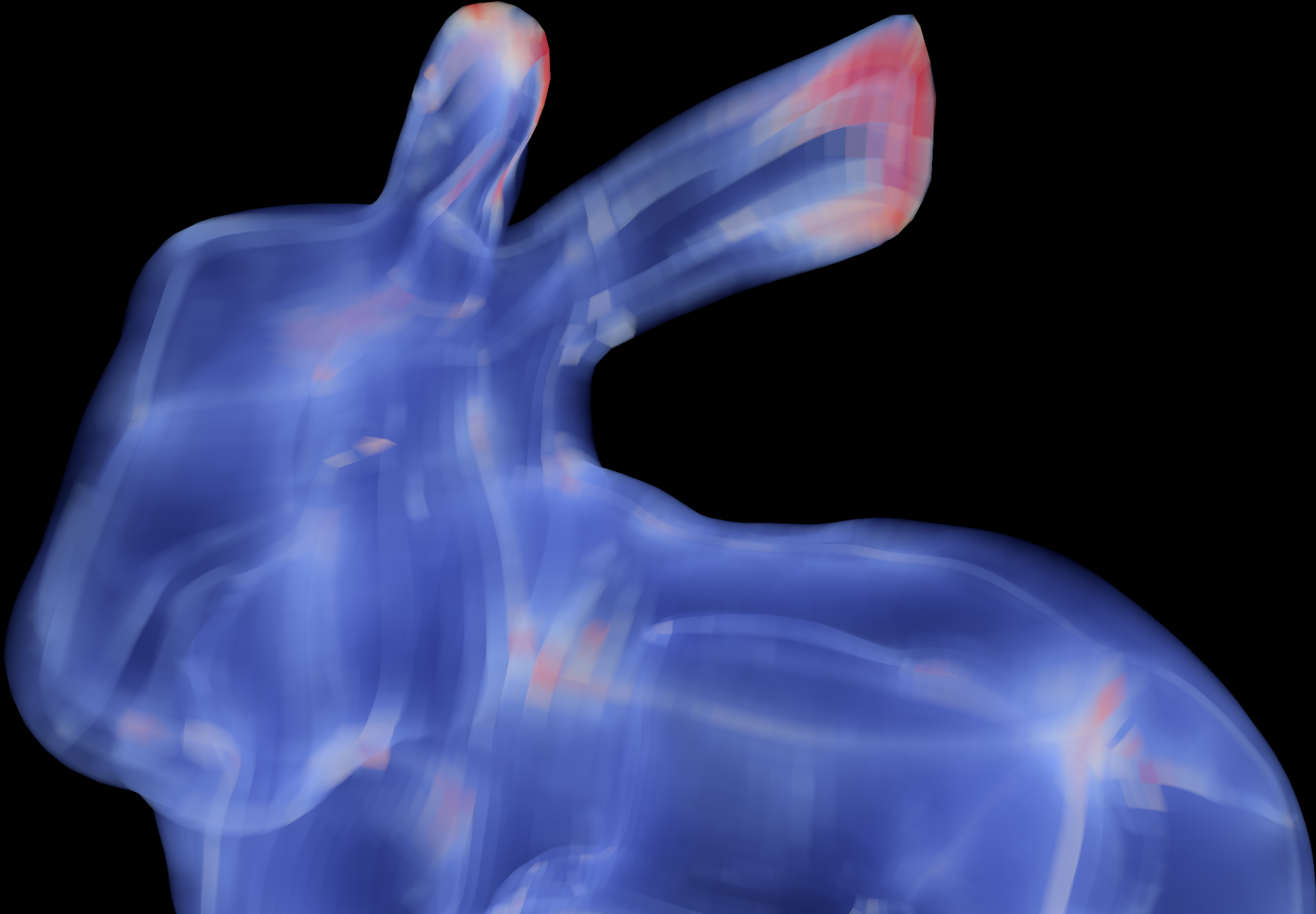}
 \includegraphics[width=0.49\linewidth]{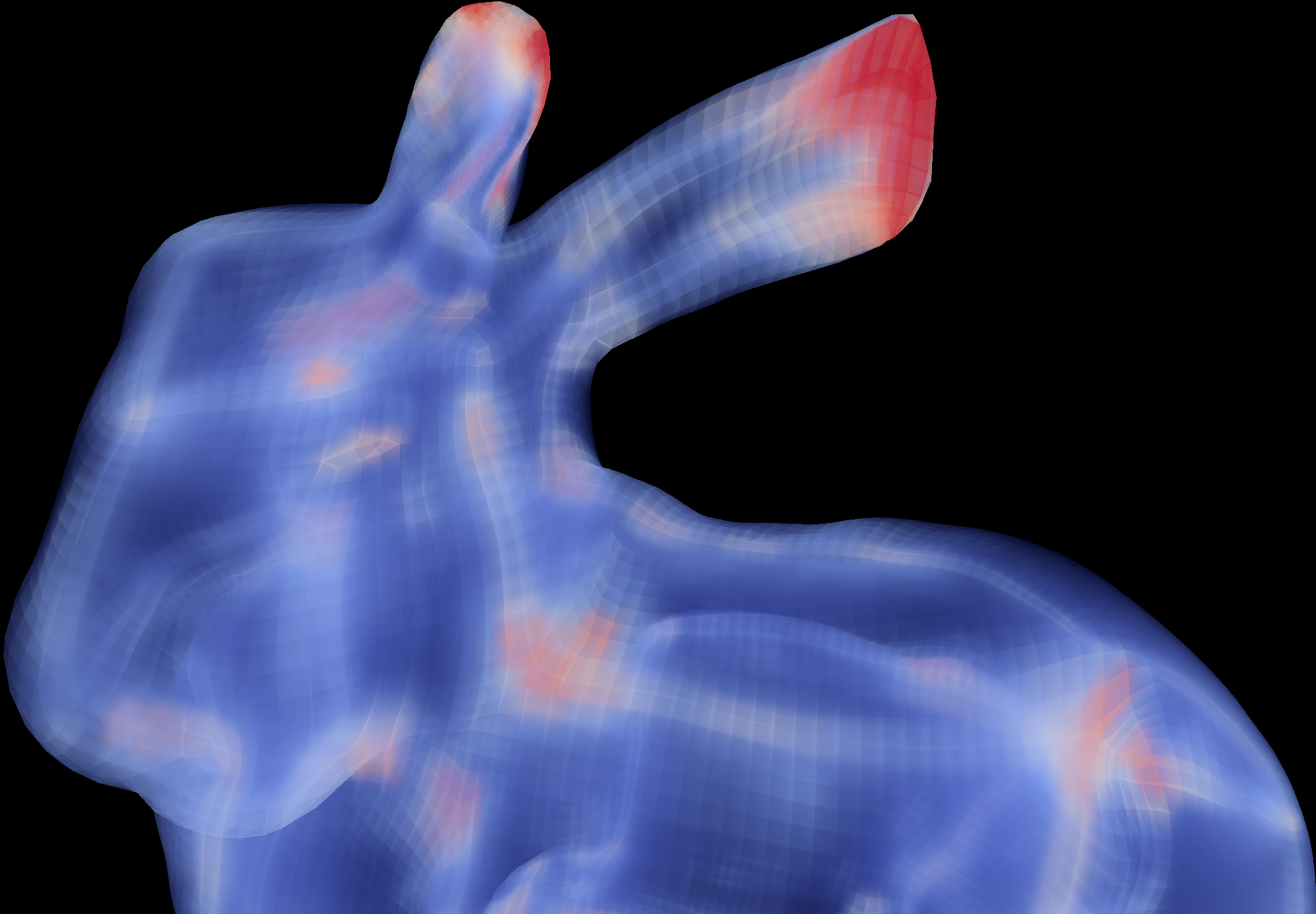}
 \caption{Left: Volume rendering using cell contributions. Right: Face-based volume rendering. The Jacobian ratio (from low to high) is mapped linearly to color (from blue to red) and opacity. Model bunny courtesy of \cite{HexMeshSGP2011}.}
 \label{fig:volume_rendering_types}
\end{figure}

If a fragment is not classified as part of an edge, it is rendered as part of a face to generate a volumetric appearance that hints to important mesh regions. In principle, once the face fragments are rendered and sorted in a fragment list, direct volume rendering using $\alpha$-compositing of cell contributions can be used (\autoref{fig:volume_rendering_types}~(left)). This gives a continuous volumetric appearance, as if the object is filled with a scalar-valued quantity, yet the mesh structure is mostly lost. 

To also accentuate the mesh structure in the context region, we refrain from using direct volume rendering. Instead, the faces are blended in correct visibility order, yet the optical depth through the cells is neglected and face colors are blended using opacities that continually increase with decreasing distance to the focus. I.e., the face opacity $\alpha_f$ is computed by modulating a user selected face opacity $\hat \alpha_f$ using the distance to the focus center and the edge accentuation factor as
\begin{equation}
\alpha_f = \hat \alpha_f \cdot dist^4. 
\label{equ:face_opacity}
\end{equation}

Blending a discrete set of faces generates accentuated jumps in the final colors whenever there is a change in the number of faces falling into adjacent fragments (\autoref{fig:volume_rendering_types} (right)). Increasing opacity artificially increases these jumps in the context region and makes them more noticeably. The face colors $C_f$ are generated by interpolation of per-vertex importance values by the rasterizer. 

\begin{figure*}[hp]
 \centering
 \includegraphics[width=0.48\linewidth]{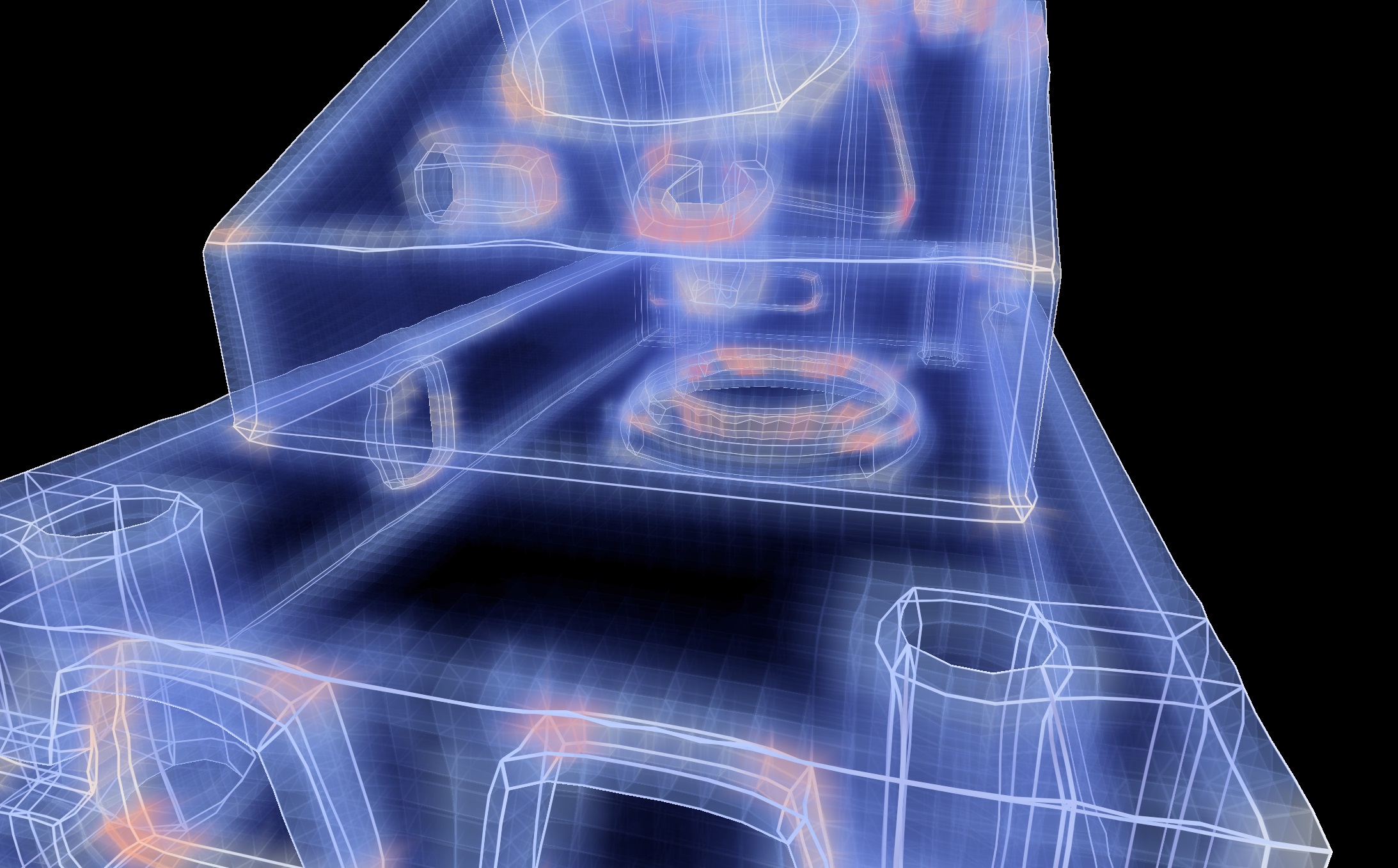}
 \includegraphics[width=0.48\linewidth]{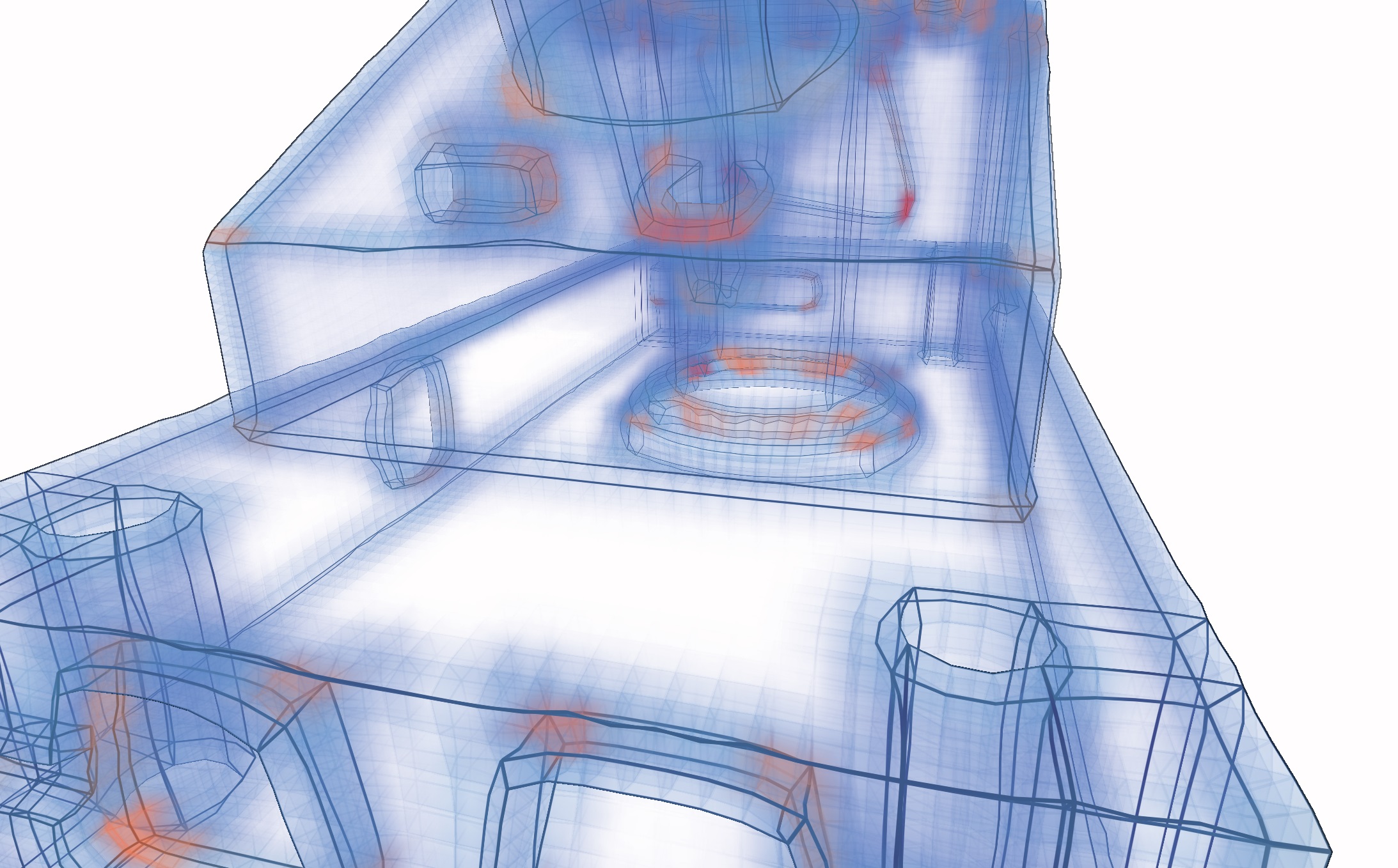}
  \caption{Comparison of black and white background. Model anc101\_a1 courtesy of \cite{HexMeshSGP2011}.}
 \label{fig:BlackWhiteComp}
\end{figure*}

We decided to use a dark background, because the rendering, combined with bold saturated
colors, tends to stand out. A white background shines through and affects the line colors. However, our visualization tool also allows switching to a white background if desired (cf. \autoref{fig:BlackWhiteComp}).

\subsection{Blending Focus and Context} \label{sec:fccombined}
\begin{figure}[h]
 \centering
 \vspace{-0.3cm}
 \includegraphics[width=0.95\linewidth]{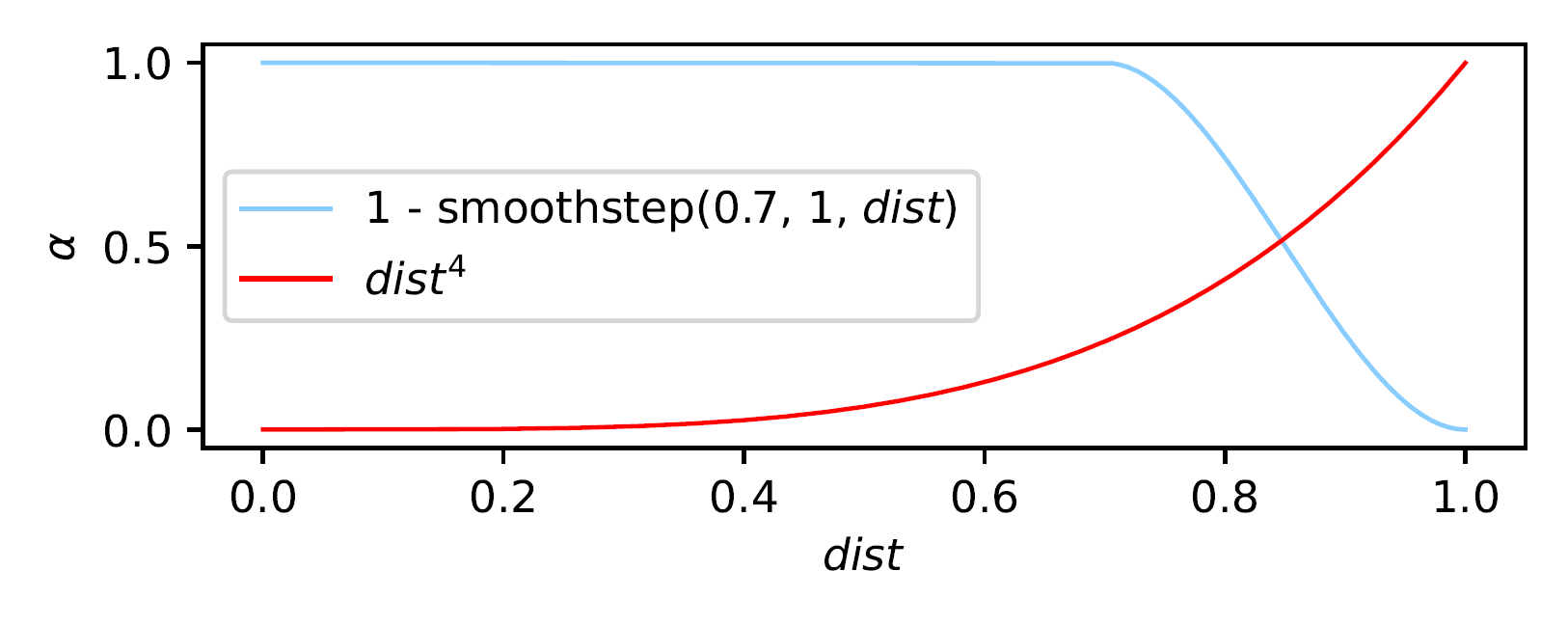}
 \vspace{-0.7cm}
 \caption{Blend factors for focus and context.}
 \label{fig:blending_graphs}
\end{figure}

Each fragment obtains an edge and a face color ($C_e, C_f$), and in addition computes the values $\alpha_e$, $s_{e}$ and $\alpha_f$ according to Equations~\ref{equ:edges-primary},~\ref{equ:edges-secondary} and \ref{equ:face_opacity}. 
The fragment shader blends the edge colors (focus and contextual edges) and face colors (face colors and accentuated lines) according to  
\begin{equation}
\begin{gathered}
\label{eqn:final_compositing}
C = \alpha_e C_e + (1-\alpha_e) s_{e} \alpha_f C_f\\
\alpha = \alpha_e + (1-\alpha_e) s_{e} \alpha_f.
\end{gathered}
\end{equation}
Thus, focus and context information is blended as shown in \autoref{fig:blending_graphs}. Via front-to-back $\alpha$-compositing, all fragments falling into a pixel are finally merged. 

\begin{figure}[h]
 \centering
  \includegraphics[width=\linewidth]{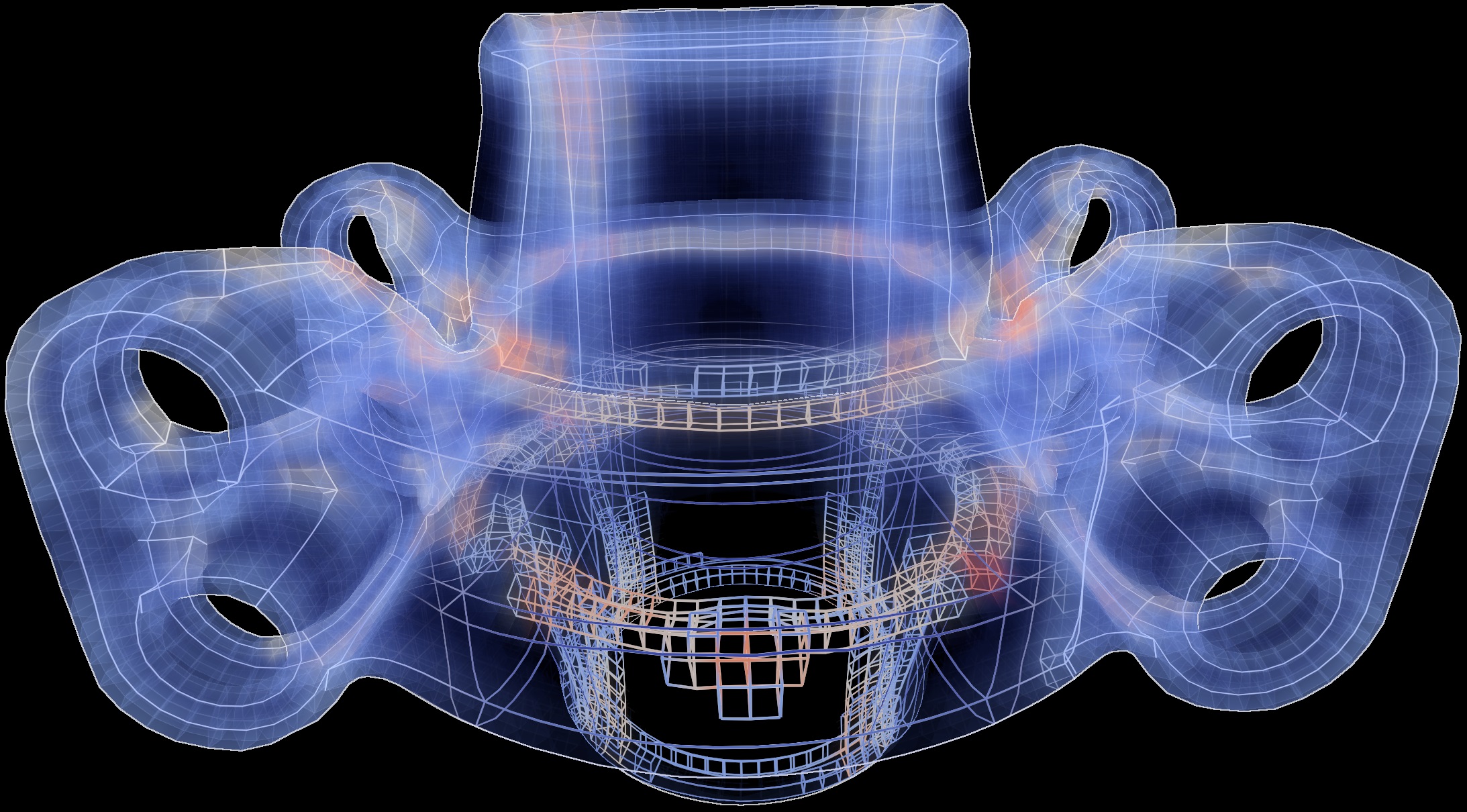}
  \caption{A final mesh rendering showing a smooth transition from the focus edges to the context edges and volumetric representation. Model grayloc courtesy of \cite{AllHex2016}.}
 \label{fig:all_combined}
\end{figure}

\autoref{fig:all_combined} shows the final F+C look. An accentuated edge in the context takes on the color of the face, brightened a little, and its opacity is increased about $50\%$. In addition, white exterior and interior screen-space silhouettes are added to improve the perception of the mesh shape ~\cite{Saito1990,Herzmann1999,raskar1999image}. Therefore, the mesh boundary surface is rendered, and fragments along sharp edges in the depth buffer are emphasized. 

\section{Level of Detail Structure} \label{sec:lod}

\begin{figure}[h]
 \centering
 \includegraphics[width=0.18\linewidth]{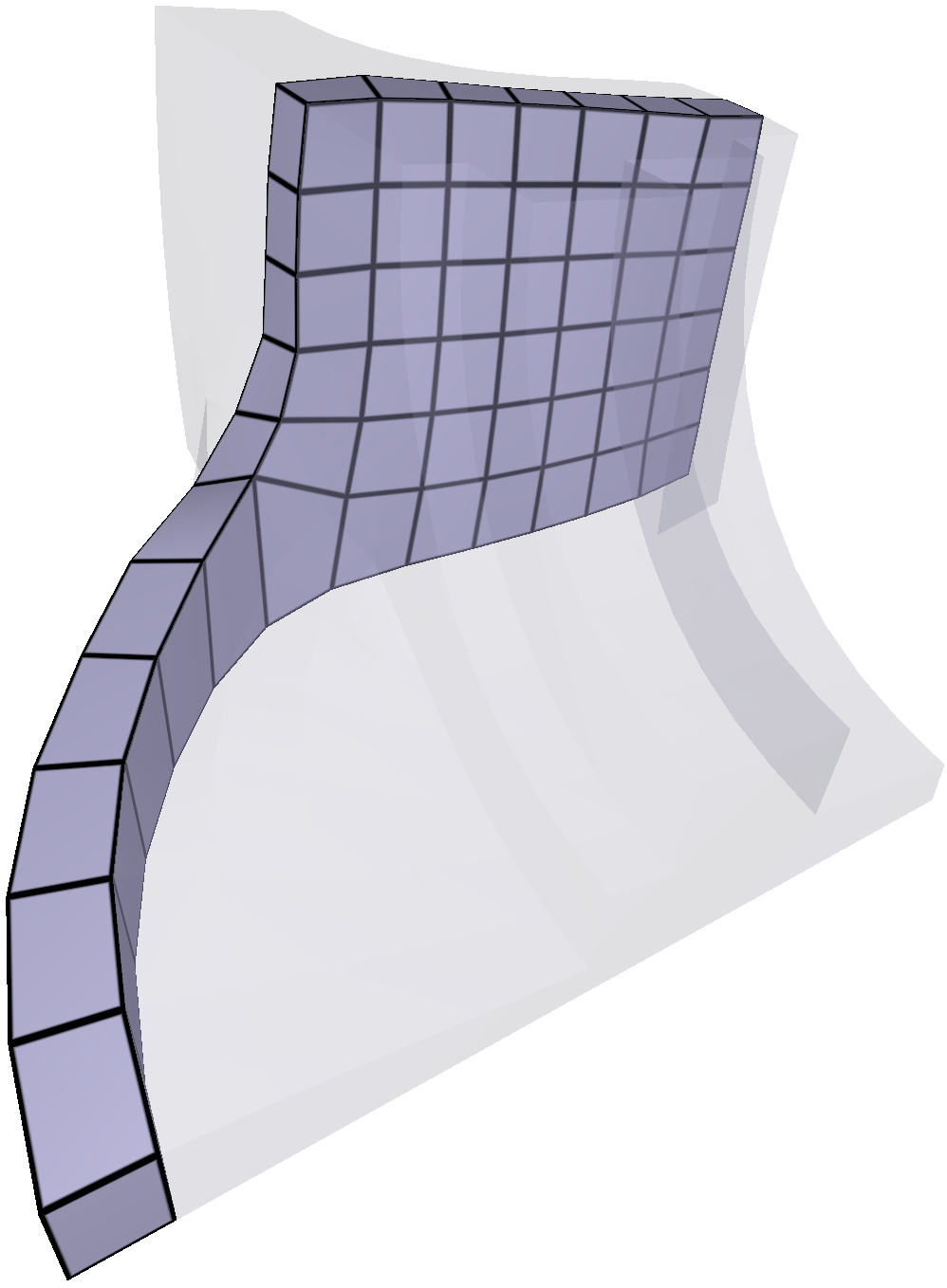}
 \includegraphics[width=0.64\linewidth]{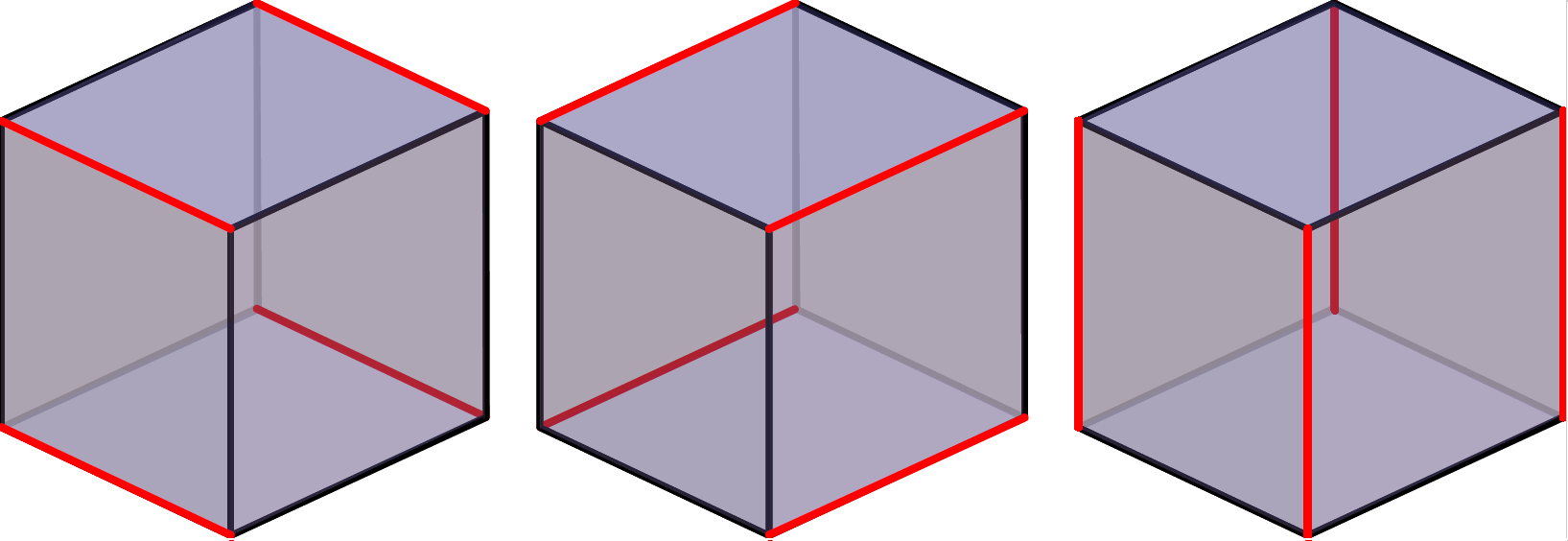}
 \caption{A hexahedral sheet (left), and the three sets of topologically parallel edges (in red) of a hexahedral element \cite{Woodbury:2011}. Model courtesy of~\cite{AllHexMeshing2012}.}
 \label{fig:sheet_overview}
\end{figure}

In the following, we describe the construction of the LoD edge structure for a given hex-mesh using topological simplification. Our approach builds upon the concept of hexahedral sheets. Hexahedral sheets were introduced by Borden et al.~\cite{Borden2002HexahedralSE}, and further formalized by \cite{Woodbury:2011} as a set of hex-elements which are connected to each other via their topologically parallel shared edges. In \autoref{fig:sheet_overview}, we reproduce images from Woodbury et al. to illustrate the relationship between these two topology-based groups. 
In a number of works, the concept of hexahedral sheets has been utilized for hex-mesh construction and simplification \cite{Gao:2015:ACM}, as well as re-meshing \cite{wang2017sheet}. We make use in particular of sheet-based topology simplification, by successively collapsing pairs of neighboring sheets.

We use the approach proposed by \cite{Woodbury:2011} to extract each single sheet: Upon selecting the start edge, all elements incident to the edge are found and added to the sheet (if not done already). For each of the newly added elements, the three edges topologically parallel to the original edge are determined, and the edges are updated with the newly found edges. This process is repeated until there is no new element found. During the extraction of a single sheet, all visited element edges are recorded. Then, an unvisited edge is selected for computing a new sheet until no such edge is left. In this way, the set of sheets covering the entire hex-mesh is extracted. Finally, we define for each sheet a sheet component consisting of all elements belonging to this sheet.



\subsection{Merging Sheet Components}
\begin{figure}[h]
 \centering
 \includegraphics[width=0.3\linewidth]{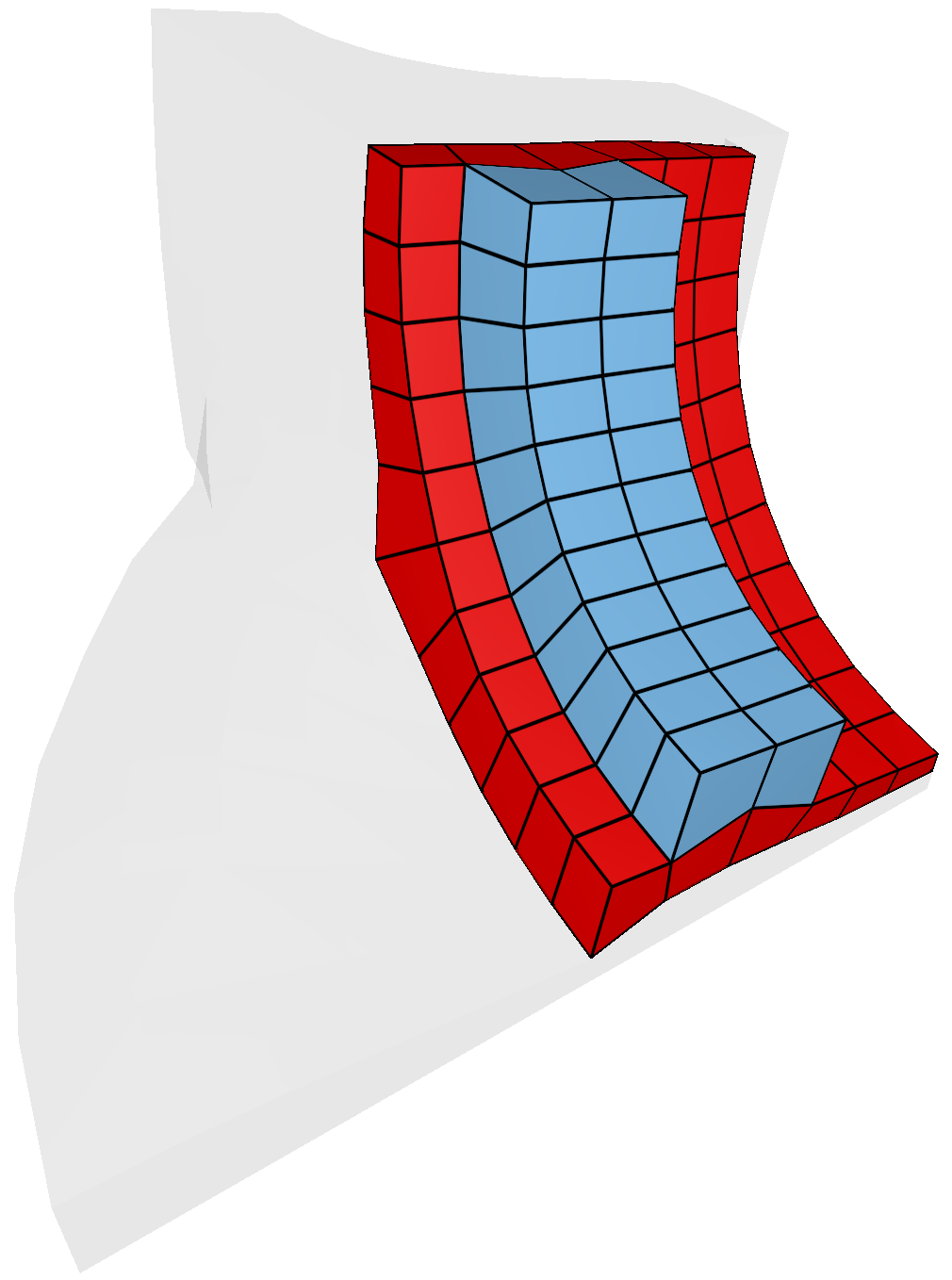}\label{fig:sheet_adjacent}
 \includegraphics[width=0.3\linewidth]{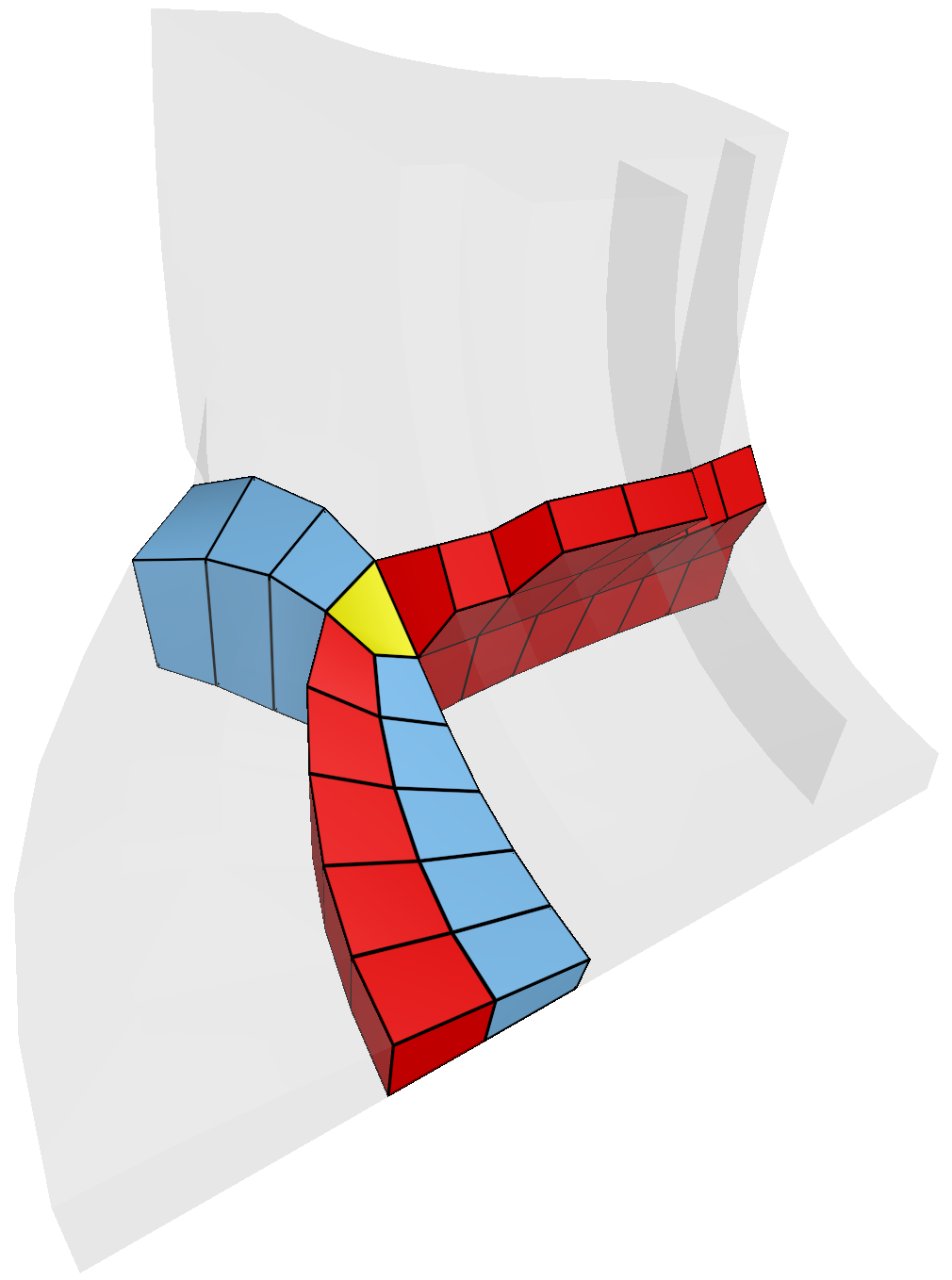}\label{fig:sheet_hybrid}
 \includegraphics[width=0.3\linewidth]{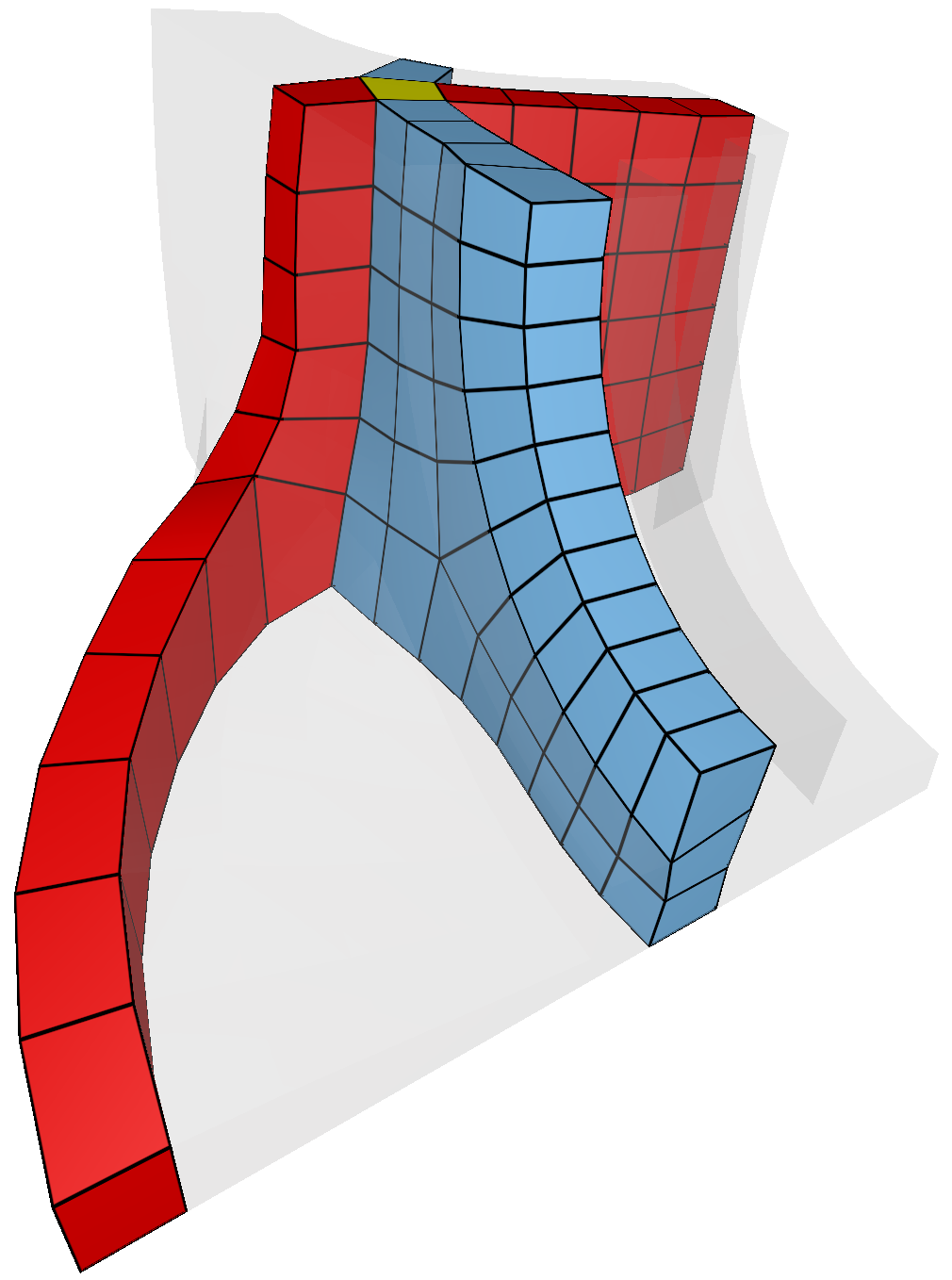}\label{fig:sheet_intersecting}
 \caption{From left to right, the different topological relations (adjacent, hybrid, intersecting) of neighboring hexahedral sheets. Similarity to the constellations by \cite{Xu:2018:TVCG} is intentional. Model courtesy of~\cite{AllHexMeshing2012}.}
 \label{fig:sketch_sheet}
\end{figure}

In an iterative process, pairs of sheet components are merged into a joint component until no components can be merged anymore.   
Therefore, for all pairs of sheet components, their neighborhood relation is classified analogously to the work by \cite{Xu:2018:TVCG} as
\begin{itemize}
    \item adjacent (or tangent), 
    \item intersecting, 
    \item hybrid (i.e., tangent and intersecting),
    \item none.
\end{itemize}
\autoref{fig:sketch_sheet} illustrates the different constellations.
In our design, sheet components are neighbors only if they share at least one boundary face that is no longer on the boundary after merging. 

In addition to the neighborhood relation, for each pair of neighboring components a weight is computed. The weights are used in an iterative merging process to determine the priority of merging for each neighboring component pair.
Building upon \cite{Xu:2018:TVCG}, where the weights consider the percentage of merged boundary faces to the overall number of boundary faces in the two components, the weights are computed as 
\begin{equation}\label{Eqn:adjacencyPriority}
w_{i,j}=\frac{\partial C_i \cap \partial C_j}{\mid \partial C_i \mid  + \mid \partial C_j \mid} \cdot \frac{1}{\mid C_i \mid + \mid C_j \mid}
\end{equation}
Here, $\partial C_i \cap \partial C_j$ is the number of boundary element faces shared by the pair of neighboring components $C_i$ and $C_j$, and $\mid C_i \mid + \mid C_j \mid$ is the number of cells $C_i$ and $C_j$ contain. Different to \cite{Xu:2018:TVCG}, the weights consider the topological size (i.e., the number of cells) for merging to reduce the potential 'jumps' in the LoD structure, i.e., neighboring pairs with smaller topological sizes are favoured at similar ratio between boundary faces. 
Even though we favour a purely topological measure in this work, alternatively one could also opt to use the face areas and cell volumes. 

\begin{figure}[h]
 \centering
 \includegraphics[width=0.9\linewidth] {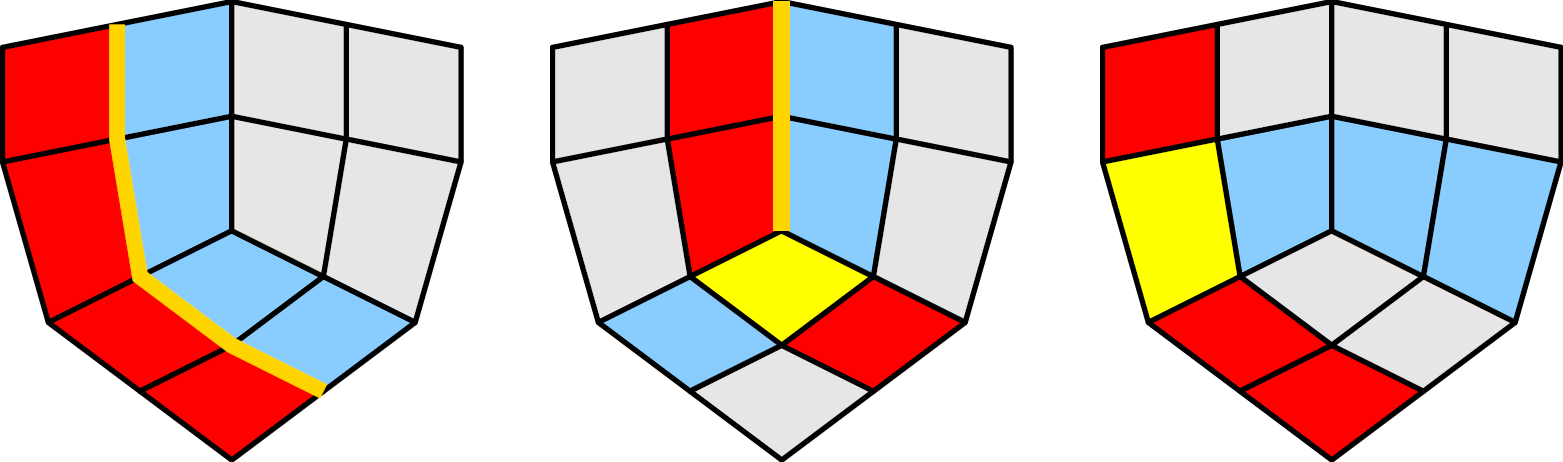}
 \caption{Sheet neighborhoods in a 2D quad mesh. Bold orange lines become invisible after merging. From left to right: Adjacent sheets, hybrid sheets, intersecting sheets. No edges become invisible when sheets intersect.}
 \label{fig:lod_quad_sheet_types}
\end{figure}

The adjacency information is stored in a priority queue, with the weights serving as the priority measure. Pairs of components with highest priority are merged first, yet adjacent sheets always have a higher priority than hybrid sheets, and hybrid sheets always have a higher priority than intersecting sheets.  
During merging, the two matching components are removed from the component queue, and a new component is inserted. The edges on the shared boundary faces of these components are identified and marked as invisible on this level (\autoref{fig:lod_quad_sheet_types}). Then the side element faces of the new component are recomputed, and the adjacency information as well as the priority of neighboring components is updated in the component queue. A next coarser LoD level is established as soon as the number of cells of the merged component is at least more than twice as large as the number of cells of the (merged) components on which the last LoD level starts. The merging process is repeated until only one single component is left. 

An exception to the rules is made for so-called singular edges. Singular or irregular edges are those edges which do not have exactly 2 (on the boundary) or 4 (in the interior) incident cells \cite{Gao:2015:ACM}. These edges form curves which separate the hex-mesh into its regular parts, and they serve as important visual cues regarding the global mesh topology. In particular, valence 1 edges are never set to be invisible, and singular edges of all other valences are only invisible at the coarsest LoD level.

\begin{figure}[t]
 \centering
 \begin{tabular}{cccc}
 \includegraphics[width=0.49\linewidth]{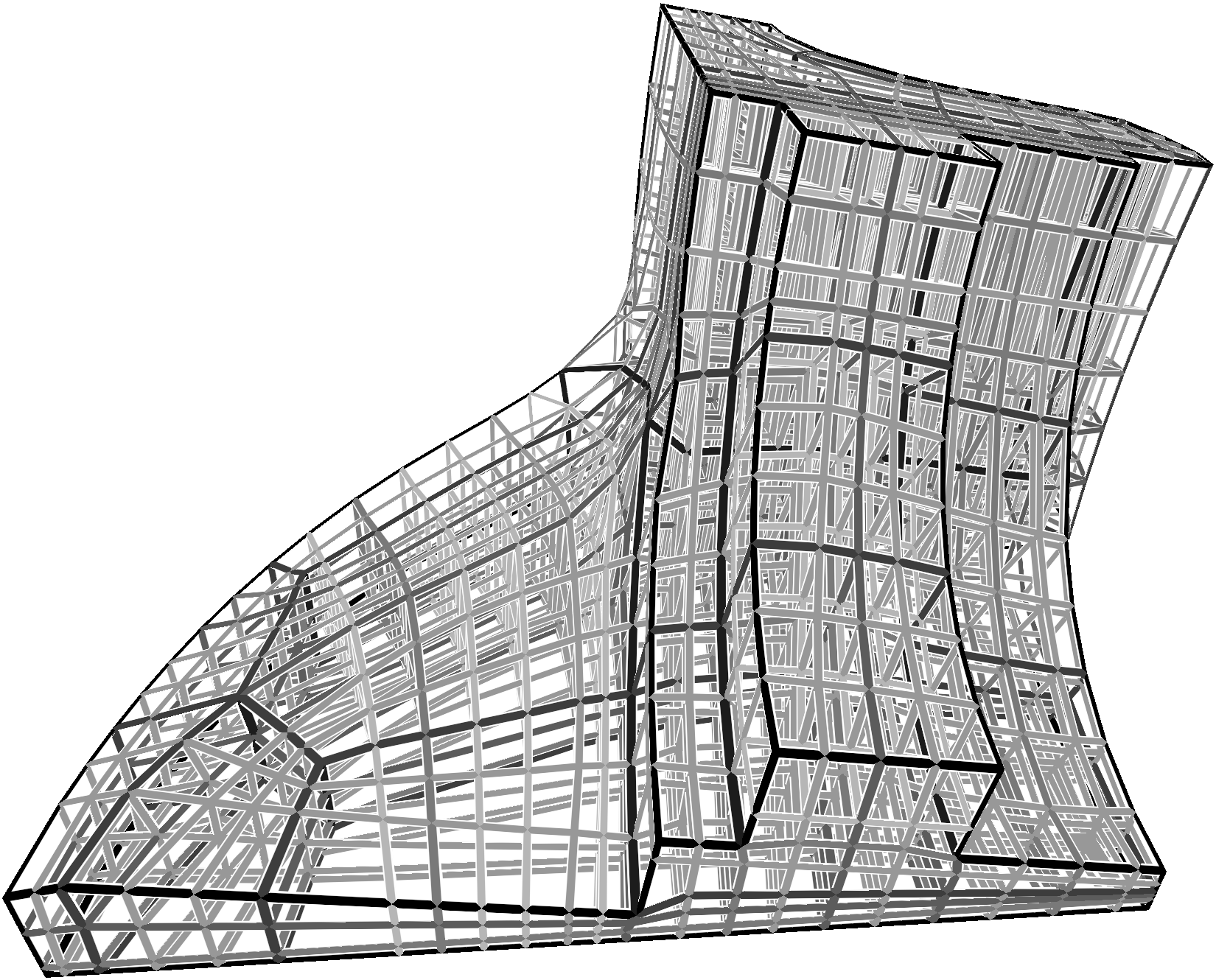} 
 \includegraphics[width=0.49\linewidth]{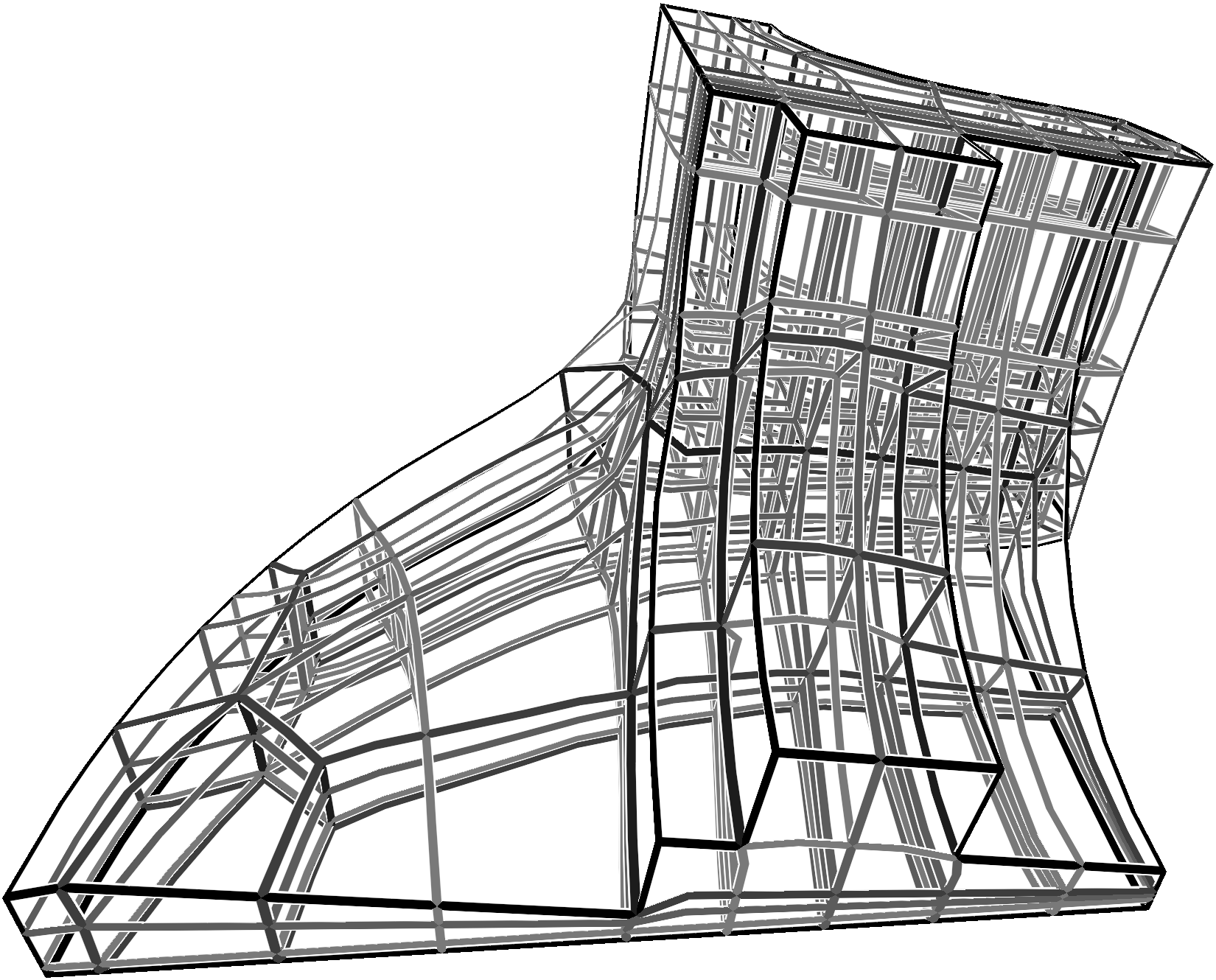}\\ 
 \includegraphics[width=0.49\linewidth]{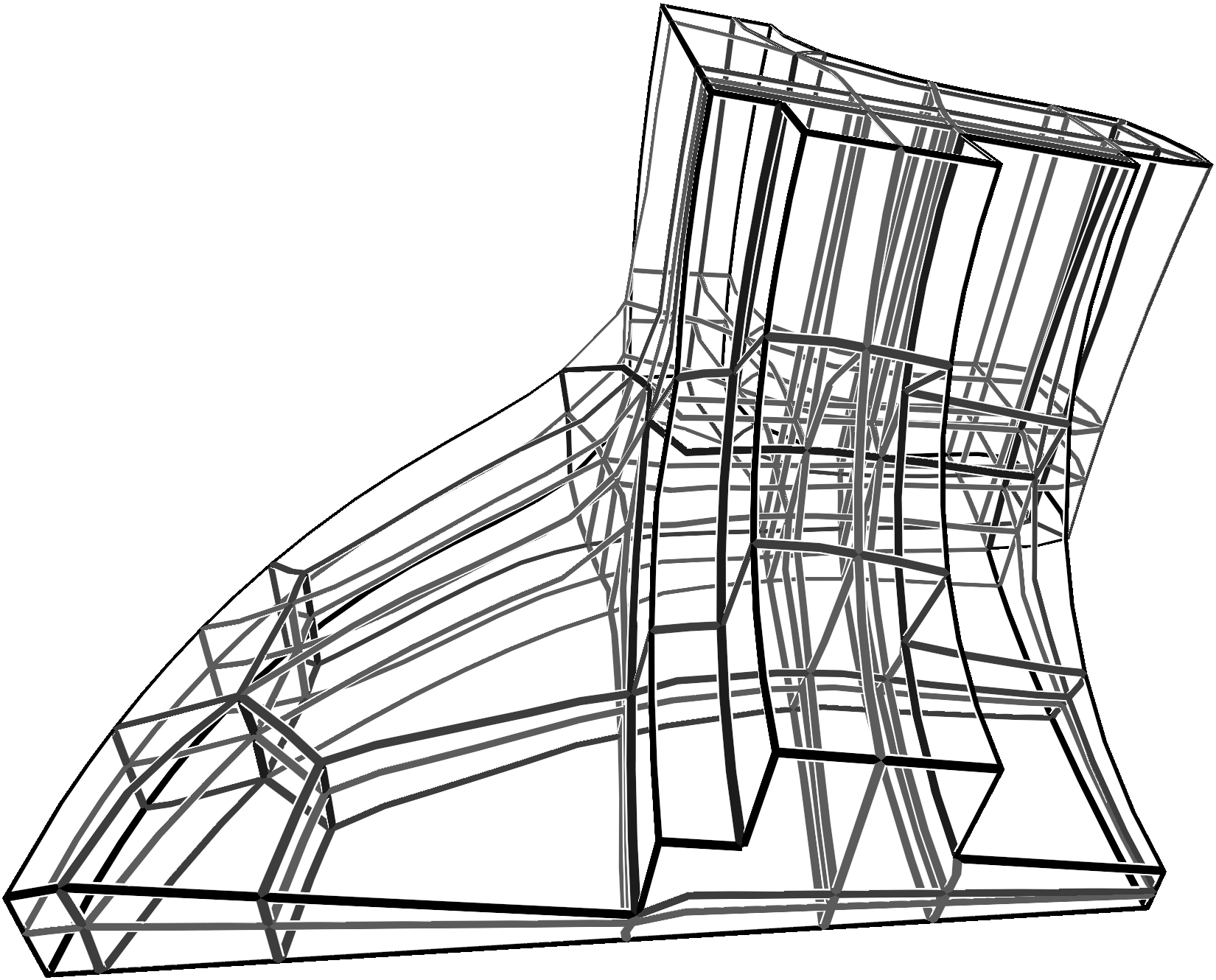} 
 \includegraphics[width=0.49\linewidth]{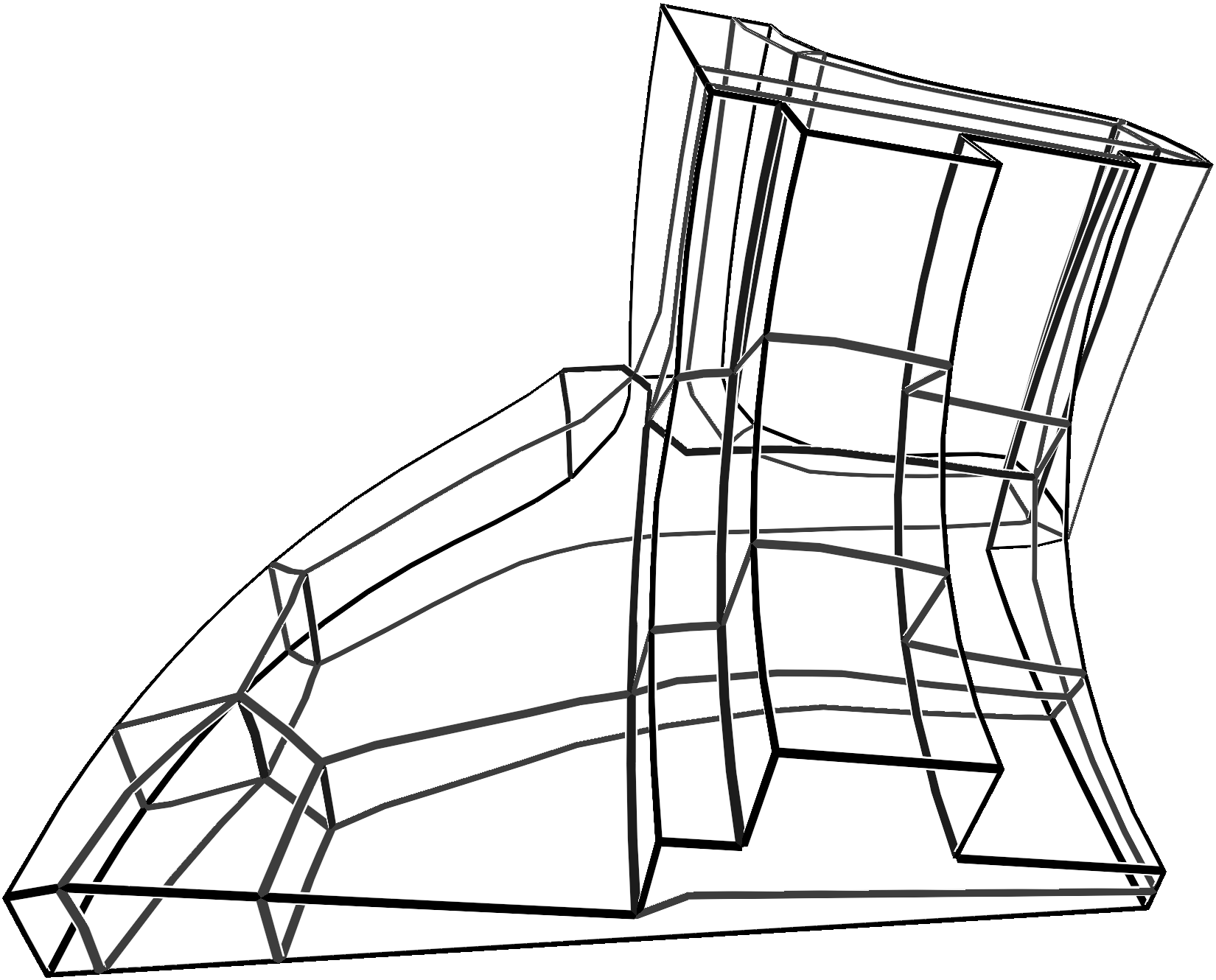}
 \end{tabular}
 \caption{From left to right: LoD levels 0, 2, 3 and 4 of a hex-mesh. Model fandisk courtesy of~\cite{DualSheetMeshing2019}.}
 \label{fig:lod_lines_part}
\end{figure}
\begin{figure}[t]
\centering
\includegraphics[width=0.685\linewidth]{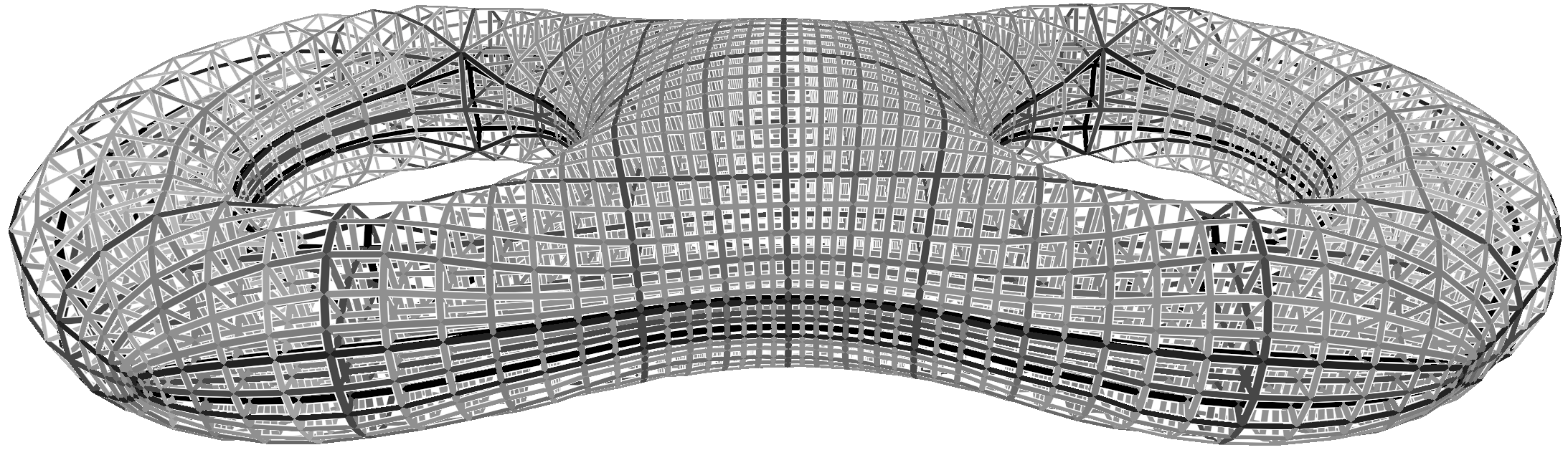} 
\includegraphics[width=0.685\linewidth]{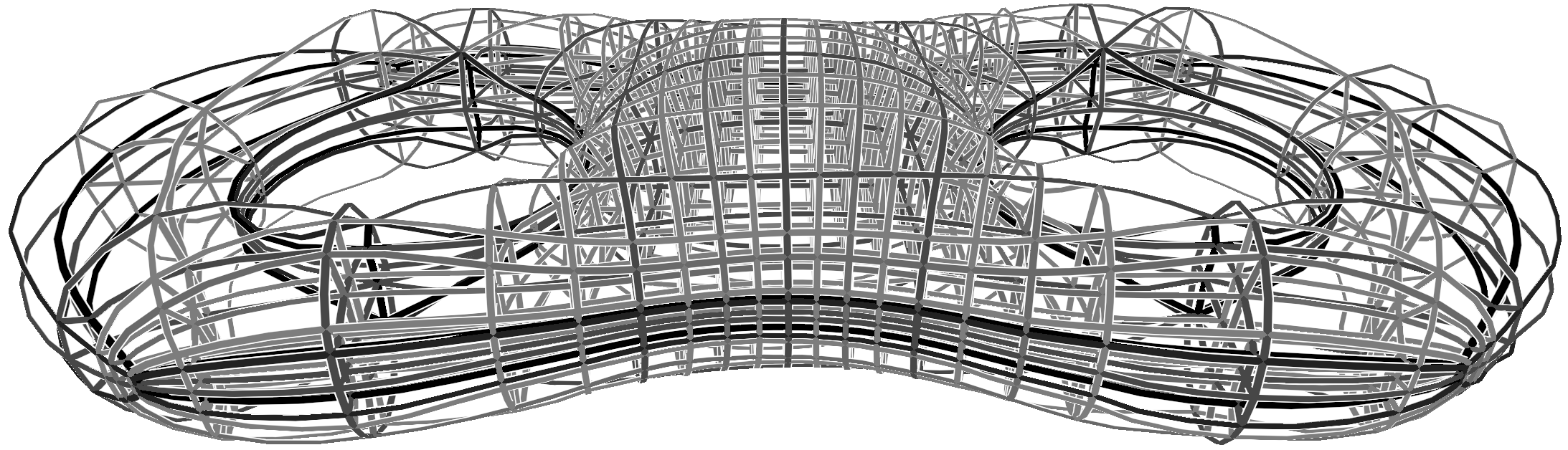} 
\includegraphics[width=0.685\linewidth]{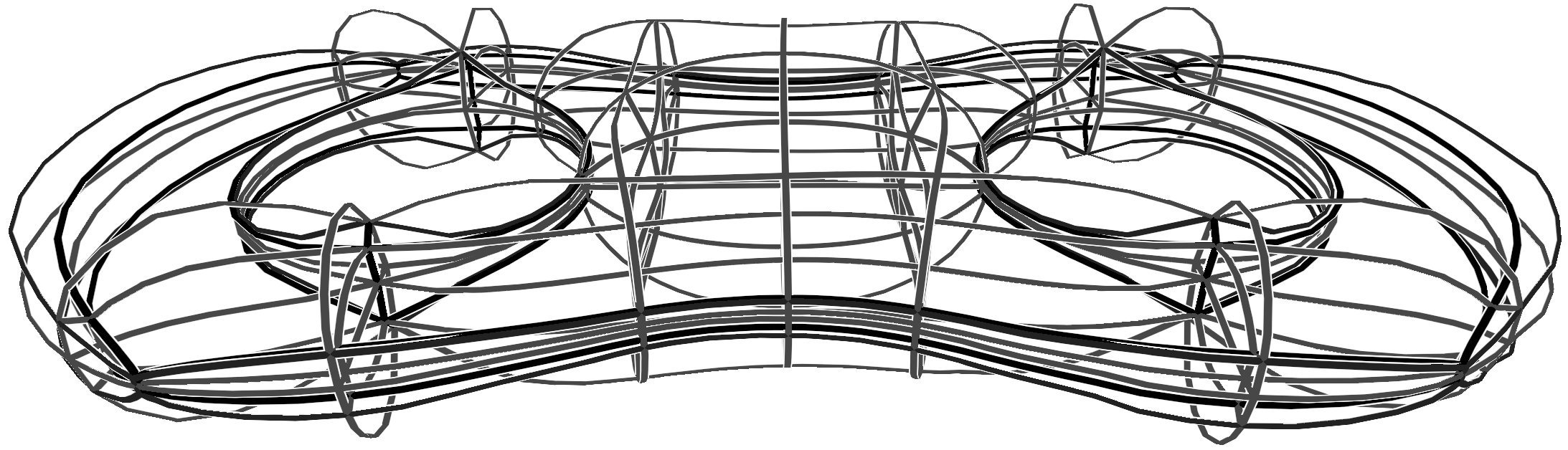} 
\includegraphics[width=0.685\linewidth]{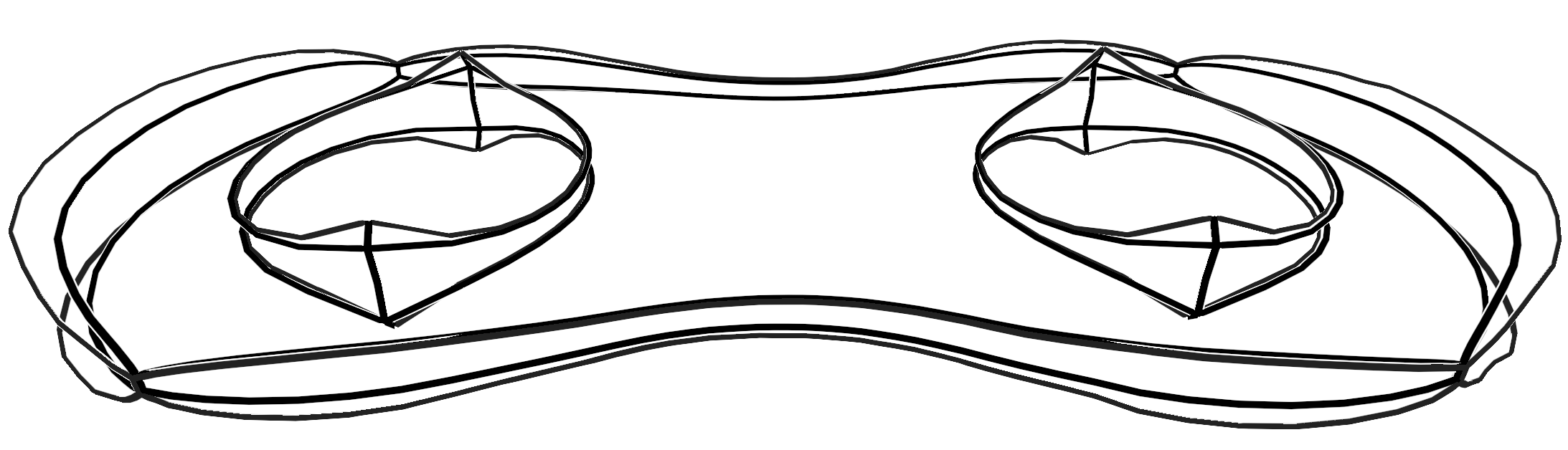}
\caption{From top to bottom: LoD levels 0, 3, 5 and 6 of a hex-mesh. Model eight courtesy of~\cite{SingularityStructureSimplification2019}.}
\label{fig:lod_lines_eight}
\vspace{-0.2cm}
\end{figure}

\autoref{fig:lod_lines_part} and \autoref{fig:lod_lines_eight} show the extracted LoD structures of two hex-meshes. The former shows the model from \autoref{fig:lod}, yet now the edges at different LoD levels, i.e., with $e_{level}$ equal to 0, 2, 3, and 4, are shown separately to better demonstrate the sequence of merging steps. The same representation is used for the latter examples, yet the edges with $e_{level}$ equal to 0, 3, 5 and 6 are shown. In both cases, the greyscale encoding of LoD levels as in \autoref{fig:lod} is used.


\section{GPU Implementation} \label{sec:implementation_details}

Our reference implementation uses the functionality provided by OpenGL 4.5. All data required for rendering is kept on the GPU, so that no CPU-GPU communication is required during rendering and user interaction. Since the fragment shader always performs all computations described in Section~\ref{sec:fcc}, the user can arbitrarily change the size of the focus lens without affecting performance. All constant parameters in 
Equations~\ref{equ:edges-primary},~\ref{equ:edges-secondary} and \ref{equ:face_opacity} are issued via constant shader parameters that can be changed interactively by the user. 

In order to make the single-pass face-based rendering of faces and edges possible, we use \textit{programmable vertex pulling} \cite{VertexPulling2012}. We use a variant called \textit{programmable attribute fetching}, where a fixed-function element array is used for indexed primitive rendering, but all vertex attributes are loaded manually from a dedicated buffer. For each cell face, we create two triangles with shared vertices only between these two triangles. Then, by using the vertex ID the fragment shader computes which face a vertex belongs to, and loads the correct face data. A geometry representation where all vertices are shared between faces is not possible, as vertices need to pass different data to the fragment shader depending on the current face. Thus, the renderer cannot utilize the post-transform cache of indexed vertices between faces, letting the pure geometry throughput fall slightly below the GPU limit.





\begin{figure}[h]
 \centering
 \includegraphics[width=0.9\linewidth]{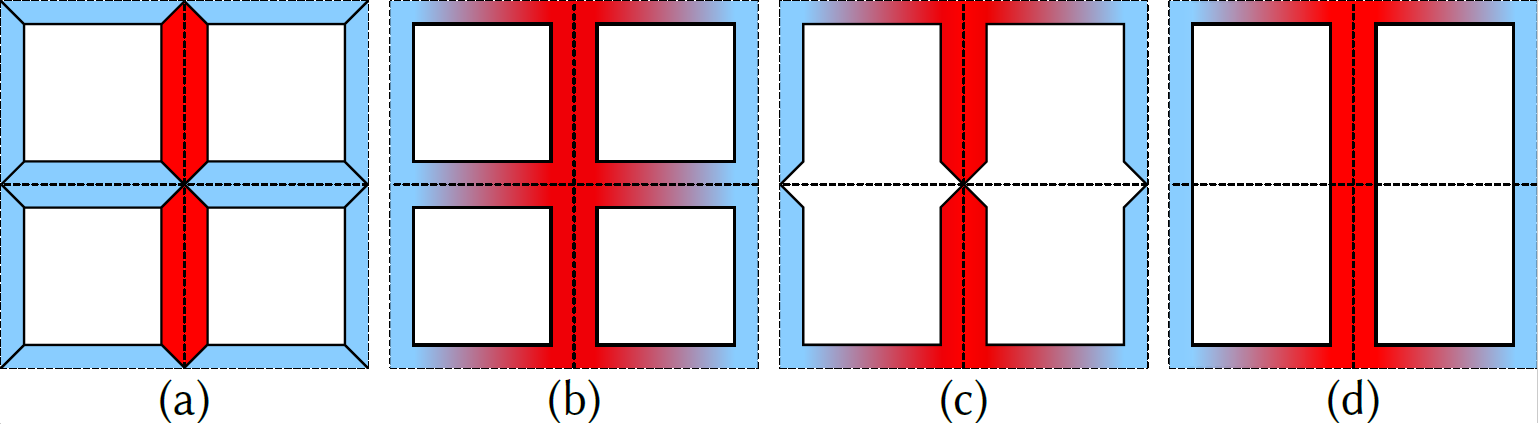}
 \caption{Edge rendering in 2D. (a) Four edges meet in one vertex and form an arrow-like shape. (b) We linearly interpolated colors in order to make the arrow-like shapes disappear. (c) Making edges invisible creates holes. (d) An approach for closing these holes is ignoring lines with a low opacity in the calculation of the closest edge.}
 \label{fig:edge_rendering_arrow}
\end{figure}

The fragment shader uses the vertex positions of all four face corner points to compute the shortest distance to any of the face edges. 
When rendering edges with per-edge constant color, star-shaped patterns occur at edge intersections (\autoref{fig:edge_rendering_arrow}a). Since smoothly interpolated per-vertex colors are rendered, these patterns are hardly visible (\autoref{fig:edge_rendering_arrow}b). Only when two edges intersect and one is not rendered (\autoref{fig:edge_rendering_arrow}c), the pattern is clearly visible. This is avoided by letting the shader ignore edges in the distance calculation which are not visible (\autoref{fig:edge_rendering_arrow}d).


To keep track of the fragments falling into the same pixel, we employ GPU per-pixel linked lists \cite{Yang2010}. All generated fragments are stored in a linked list over all pixels, and a fragment shader sorts these fragments w.r.t. their screen space depth. Here it is assumed that the GPU buffers used for storing the fragments along with a reference to the next neighbor in the global fragment list are large enough. We demonstrate in Section~\ref{sec:results} that even for hex-meshes with a few million elements this is case. 
For scenes with high depth complexity, however, the number of fragments is so large that sorting can become a performance bottleneck. For instance, for the largest hex-mesh used in our experiments about 340 million fragments are generated per frame. 
Therefore, we use a GPU version of priority-queues using a binary tree implementation as search structure ~\cite{Kern:2020:TVCG}, which reduces the time required for sorting to slightly more than half of the overall frame time. 



\section{Results and Analysis}
\label{sec:results}

All our results were rendered on an NVIDIA RTX 2070 SUPER GPU with 8GB of on-chip memory. Only the construction of the LoD hiearchy was performed on the CPU, i.e., a workstation running Ubuntu 20.04 with an AMD Ryzen 9 3900X @3.80GHz CPU and 32GB RAM. We have used different viewport sizes to demonstrate the scalability of the rendering approach in the number of pixels, and in particular to show that even for large meshes and viewports the memory required by per-pixel fragment lists does not exceed the GPU memory. All timings are averages over 128 frames with different camera views where the data sets cover almost all of the screen. The accompanying video shows one of the camera paths we have used to record the performance data. 

\begin{figure*}[hp]
\centering
\includegraphics[height=5.5cm]{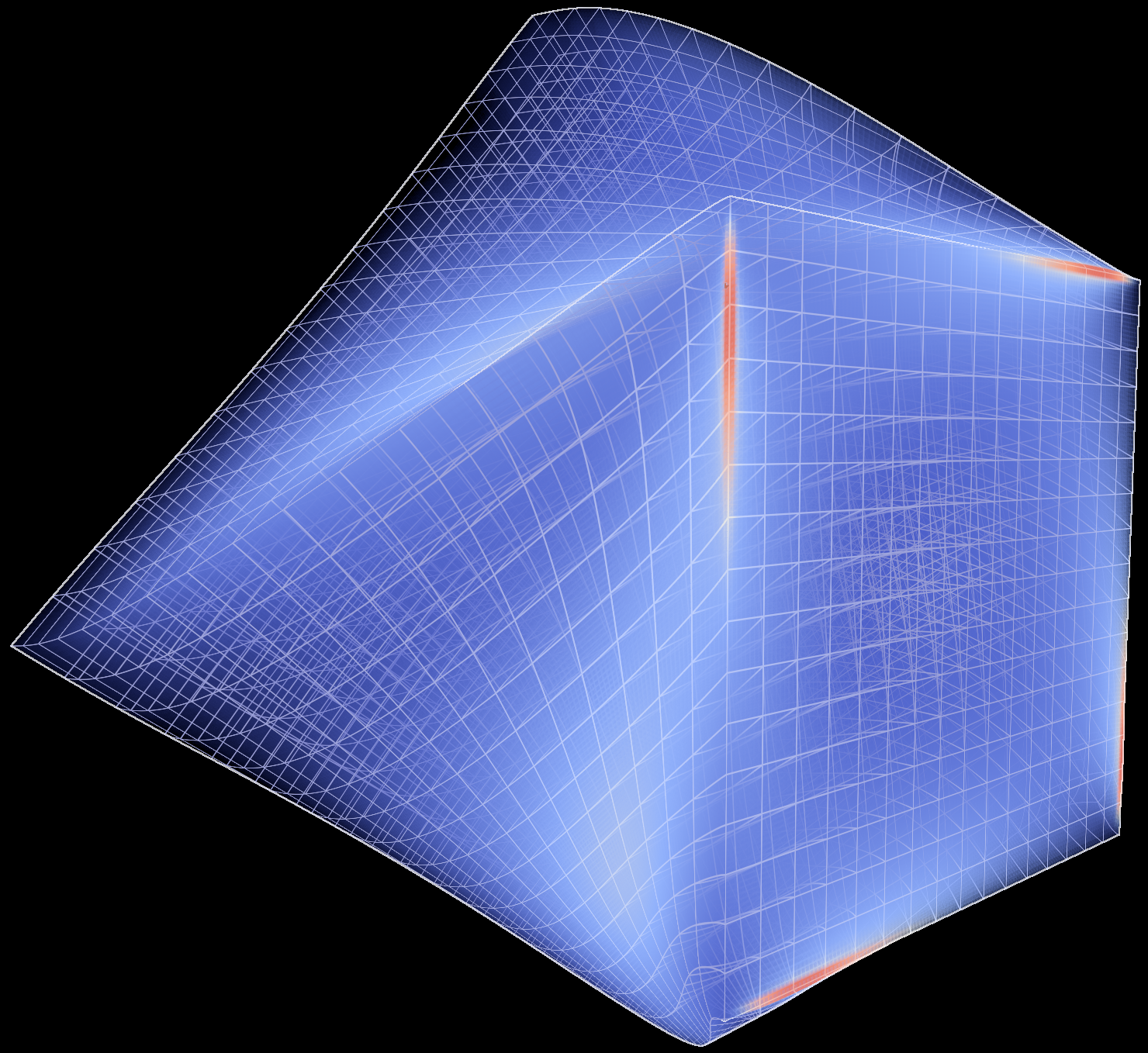}
\includegraphics[height=5.5cm]{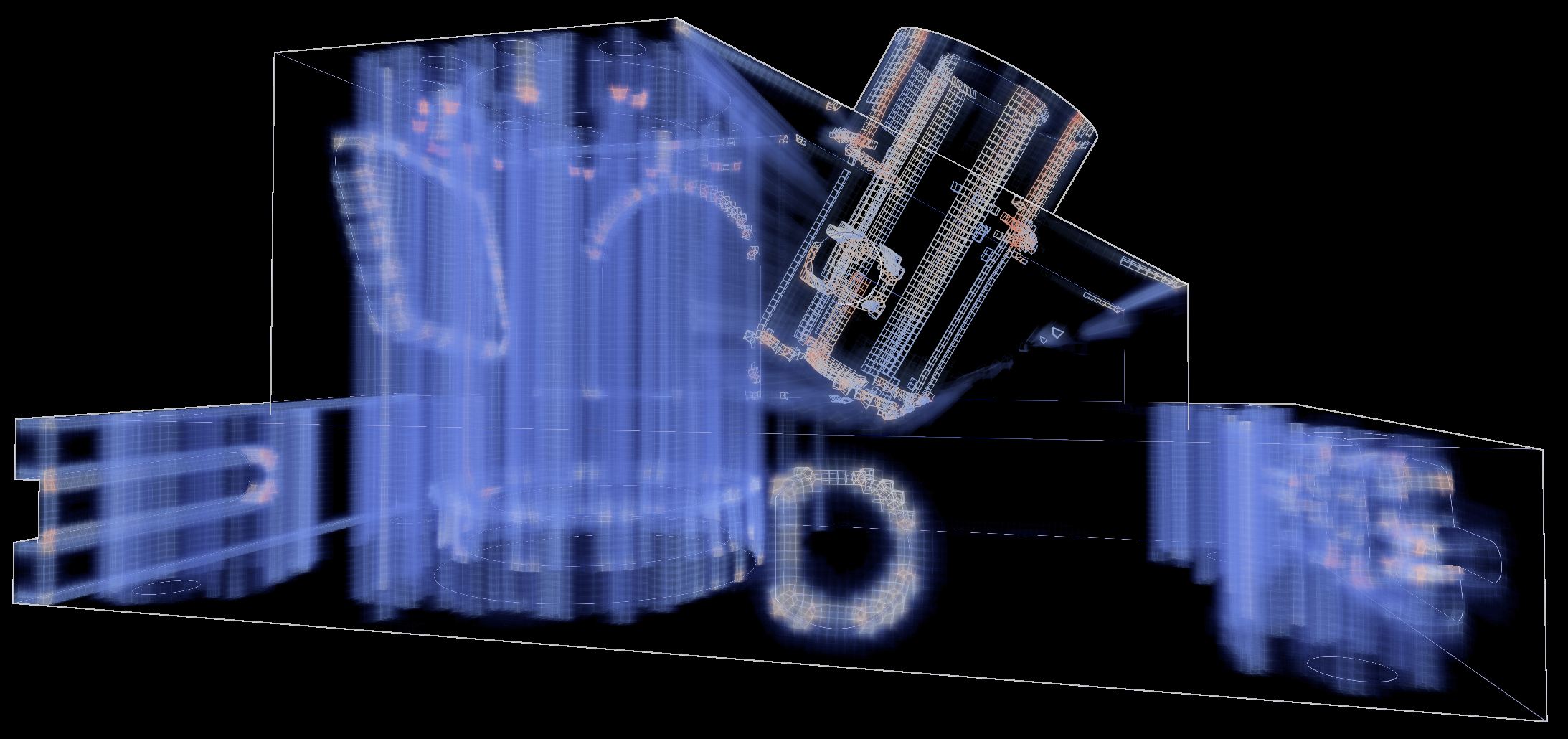}
 \vspace{0.1cm}
 \caption{Left: F+C rendering of cubic128. Right: F+C rendering of example3 reveals mostly elongated sub-structures with high deformations. 
 Model courtesy of \cite{FuzzyClustering2018}.
 }
\label{fig:sidebyside-cubic-example}
\end{figure*}

\autoref{tab:mem-time} lists the number of hex-elements of the test data sets, the GPU memory that is required to store these data sets on the GPU, and the time it requires to build the LoD structure for each data set. We have in particular included the data sets "example3" and "cubic128" (\autoref{fig:sidebyside-cubic-example}) to demonstrate that even large data sets with millions of cells can be stored entirely on the GPU and processed in a short time.
\begin{table}[h]
\small
\begin{center}
\scalebox{0.85}{
\begin{tabular}{cccc}
	\hline
	Data Set&\#Cells&Mesh Buffer Size&LoD Creation Time\\ 
    \hline
    fandisk&1,774&0.9 MiB&0.01s\\
    eight&5,428&2.6 MiB&0.05s\\
    dragon&14,009&6.7 MiB&0.2s\\
    grayloc&24,344&11.2 MiB&0.4s\\
    armadillo&29,935&13.7 MiB&0.5s\\
    dancingchildren&352,93&16.2 MiB&0.9s\\
    anc101\_a1&73,976&33.3 MiB&2.1s\\
    cognit&77,559&35.9 MiB&2.2s\\
    example3&589,040&261.4 MiB&16.3s\\
    cubic128&2,097,152&919.1 MiB&23.0s\\
    \hline
\end{tabular}}
\end{center}
\caption{Data set statistics. Model fandisk courtesy of~\cite{DualSheetMeshing2019}, eight courtesy of~\cite{SingularityStructureSimplification2019}, dragon, armadillo and dancingchildren courtesy of~\cite{EdgeConeRectification2015}, grayloc courtesy of~\cite{AllHex2016}, anc101\_a1 courtesy of~\cite{HexMeshSGP2011}, cognit courtesy of~\cite{Huang2014:LCO}, example3 courtesy of~\cite{FuzzyClustering2018}, model cubic\_128 is a twisted Cartesian grid of size $128^3$.}
\label{tab:mem-time}
\vspace{-0.2cm}
\end{table}

\autoref{tab:perf-mem} provides a performance statistics, distinguishing between the fragment shader used to determine the focus+context (F+C) look and the shader that sorts and blends the fragments in the per-pixel fragment list. In addition, the memory requirements of per-pixel fragment lists are given. Even for the largest data set, interactive frame rates can be achieved, and only at the largest viewport size the frame rate drops slightly below full interactivity. In all experiments, the fragment shader consumes the vast amount of the total frame time. The time for resolving the per-pixel fragment lists is between $43\%$ and $72\%$ of the total rendering time, and it is dependent on the depth complexity of the data set, i.e., the number of cells falling into the single pixels. It can be seen that going further beyond a few millions of elements can exceed the available GPU memory. This problem can be addressed by subdividing the screen into parts and rendering to each part separately. Since this approach requires to process each cell multiple times in the geometry processing stage and the rasterizer but does not increase the number of fragment shader operations, only a marginal overhead can be expected.

\begin{table}[h]
\small
\begin{center}

\scalebox{0.85}{
\begin{tabular}{cccccc}
	\hline
	Data Set&Viewport&FPS&F+C&PPFL&Mem. PPFL\\
    \hline
    \multirow{3}{*}{grayloc}&1280x720&154 FPS&2.2ms&4.3ms&0.21 GiB\\
    &1920x1080&85 FPS&3.9ms&7.8ms&0.47 GiB\\
    &2560x1440&52 FPS&6.5ms&12.7ms&0.84 GiB\\
    \multirow{3}{*}{anc101\_a1}&1280x720&96 FPS&4.1ms&6.3ms&0.37 GiB\\
    &1920x1080&51 FPS&6.7ms&12.9ms&0.83 GiB\\
    &2560x1440&32 FPS&11.0ms&20.3ms&1.48 GiB\\
    \multirow{3}{*}{cognit}&1280x720&127 FPS&3.2ms&4.6ms&0.20 GiB\\
    &1920x1080&74 FPS&4.6ms&8.8ms&0.45 GiB\\
    &2560x1440&49 FPS&6.6ms&13.8ms&0.80 GiB\\
    \multirow{3}{*}{example3}&1280x720&44 FPS&13.0ms&9.8ms&0.64 GiB\\
    &1920x1080&26 FPS&17.8ms&20.6ms&1.45 GiB\\
    &2560x1440&17 FPS&23.6ms&33.6ms&2.57 GiB\\
    \multirow{3}{*}{cubic128}&1280x720&12 FPS&37.3ms&46.9ms&1.19 GiB\\
    &1920x1080&7 FPS&45.8ms&101.5ms&2.68 GiB\\
    &2560x1440&-&-&-&4.8 GiB*\\
    \hline
\end{tabular}}

\end{center}
\caption{Performance statistics for selected data sets at different viewport sizes: Frames per second (FPS), times required by the F+C fragment shader (F+C) and the shader that sorts and blends the fragments in the per-pixel fragment lists (PPFL), and the memory consumed by the fragment lists (Mem. PPFL). Buffer sizes are capped at 4 GiB due to OpenGL buffer restrictions. 
}
\label{tab:perf-mem}
\vspace{-0.5cm}
\end{table}

\begin{figure*}[hp]
 \centering
 \includegraphics[height=5.55cm]{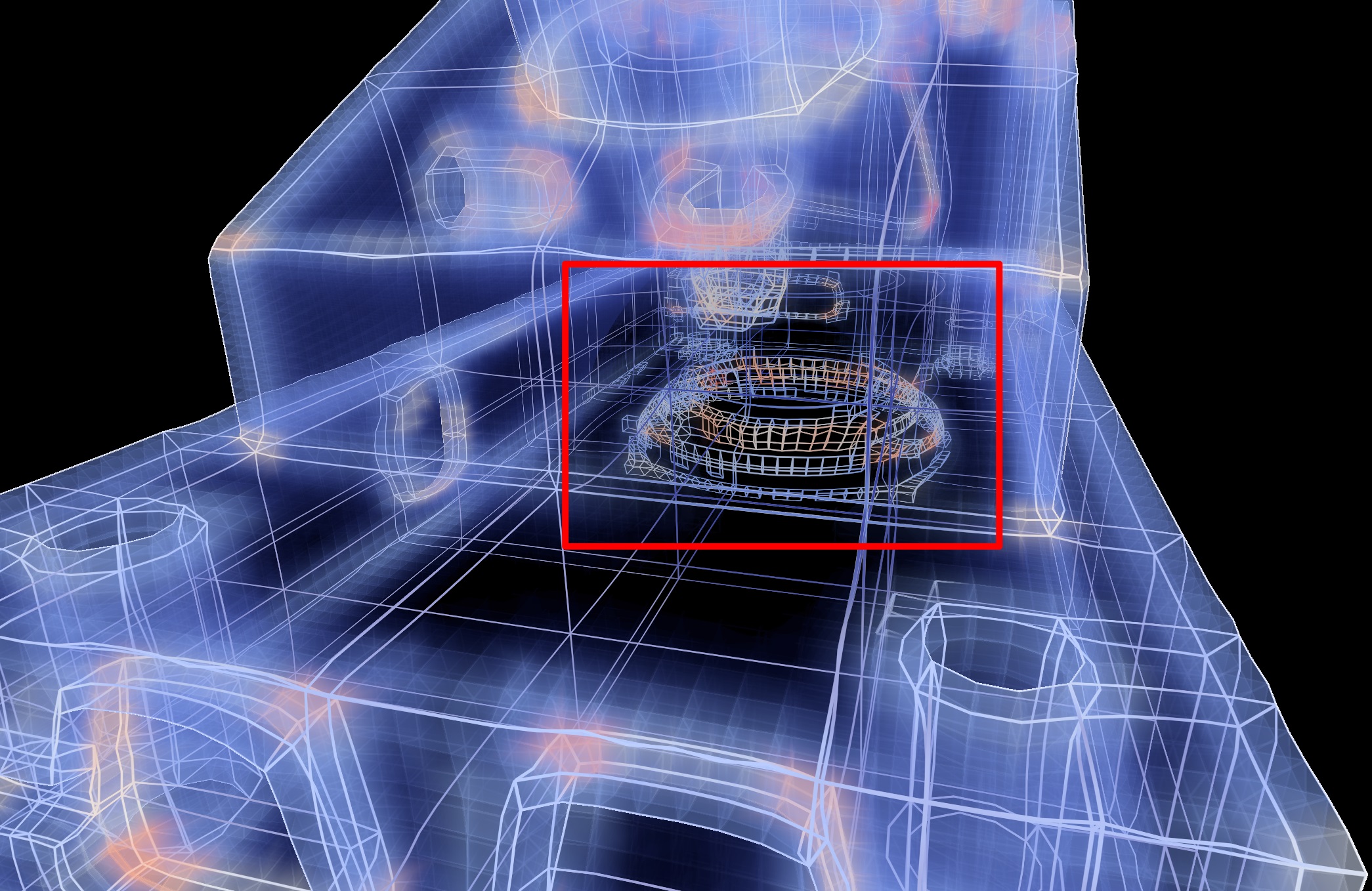}
 \includegraphics[height=5.55cm]{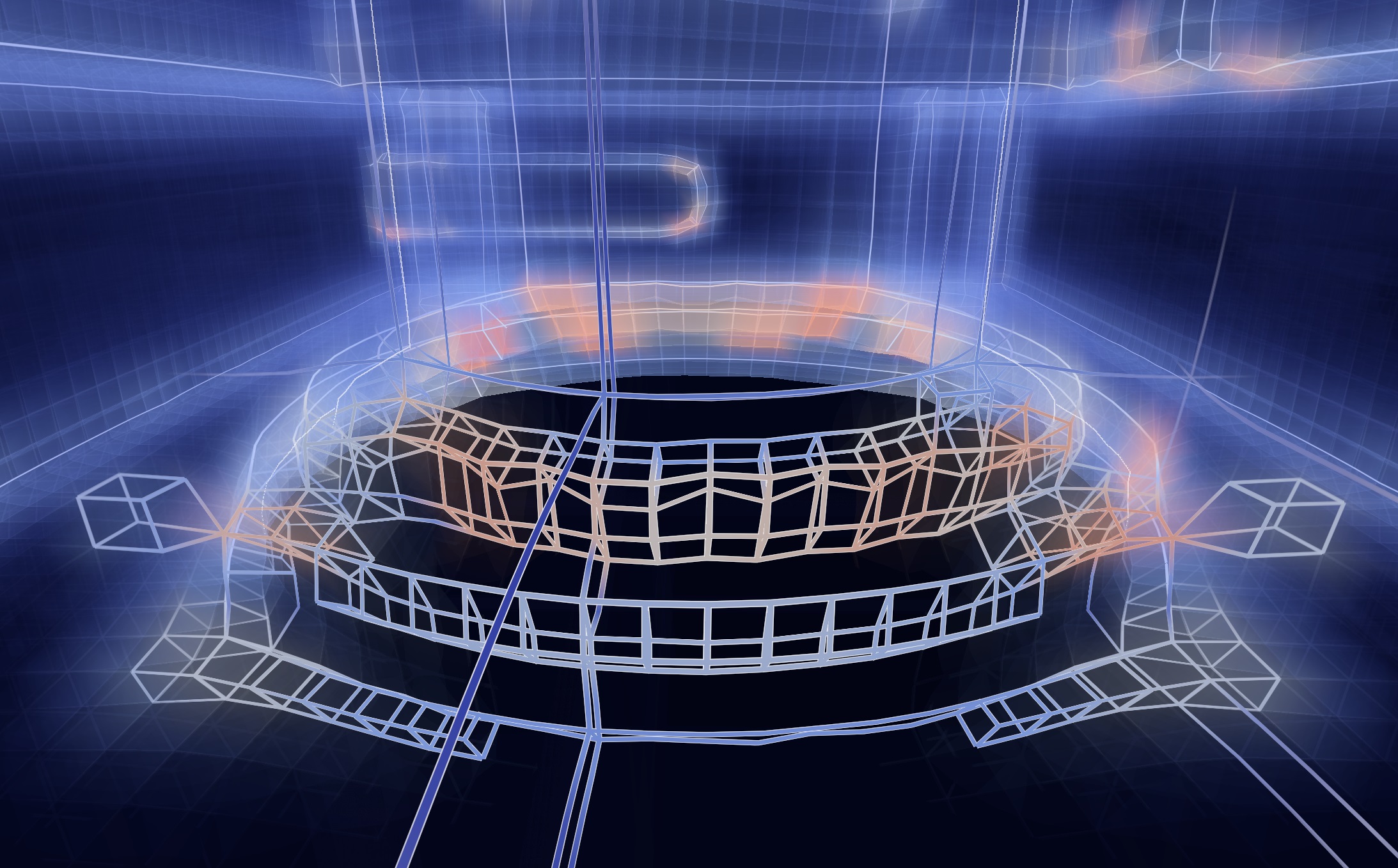}
 \vspace{0.1cm}
  \caption{The first F+C view shows a selected mesh sub-structure with highly deformed cells (framed region) in its global surrounding. Zoom-in and focus size adjustment enables a fine granular cell analysis. Surrounding cells with high deformation are still present in the context. Model anc101\_a1 courtesy of \cite{HexMeshSGP2011}.
  }
 \label{fig:sidebyside-anc}
\end{figure*}

In the following, we show results of interactive visual inspections of some of the test data sets using the proposed F+C renderer. In all examples, the per-cell scaled Jacobian ratio is mapped to color (from blue to red) and opacity (from 0 to 1).
\autoref{fig:sidebyside-anc} and \autoref{fig:sidebyside-grayloc} show the use of F+C rendering to obtain an overview of the spatial locations of regions with highly deformed cells, and to select a particular focus region for a more detailed analysis. One can see that due to the combination of contextual lines with volumetric face-based rendering and accentuated edges, the user quickly understands the basic structure of the mesh and its subdivision into multiple regular sheet components. Once a focus region is selected, a detailed analysis of the cells in that region is performed via close-up views and interactive navigation. During inspection, LoD levels, transfer functions for edge and face colors and opacities, as well as edge thickness can be changed interactively to enhance the visual representation.

\begin{figure*}[hp]
 \centering
 \includegraphics[width=0.24\linewidth]{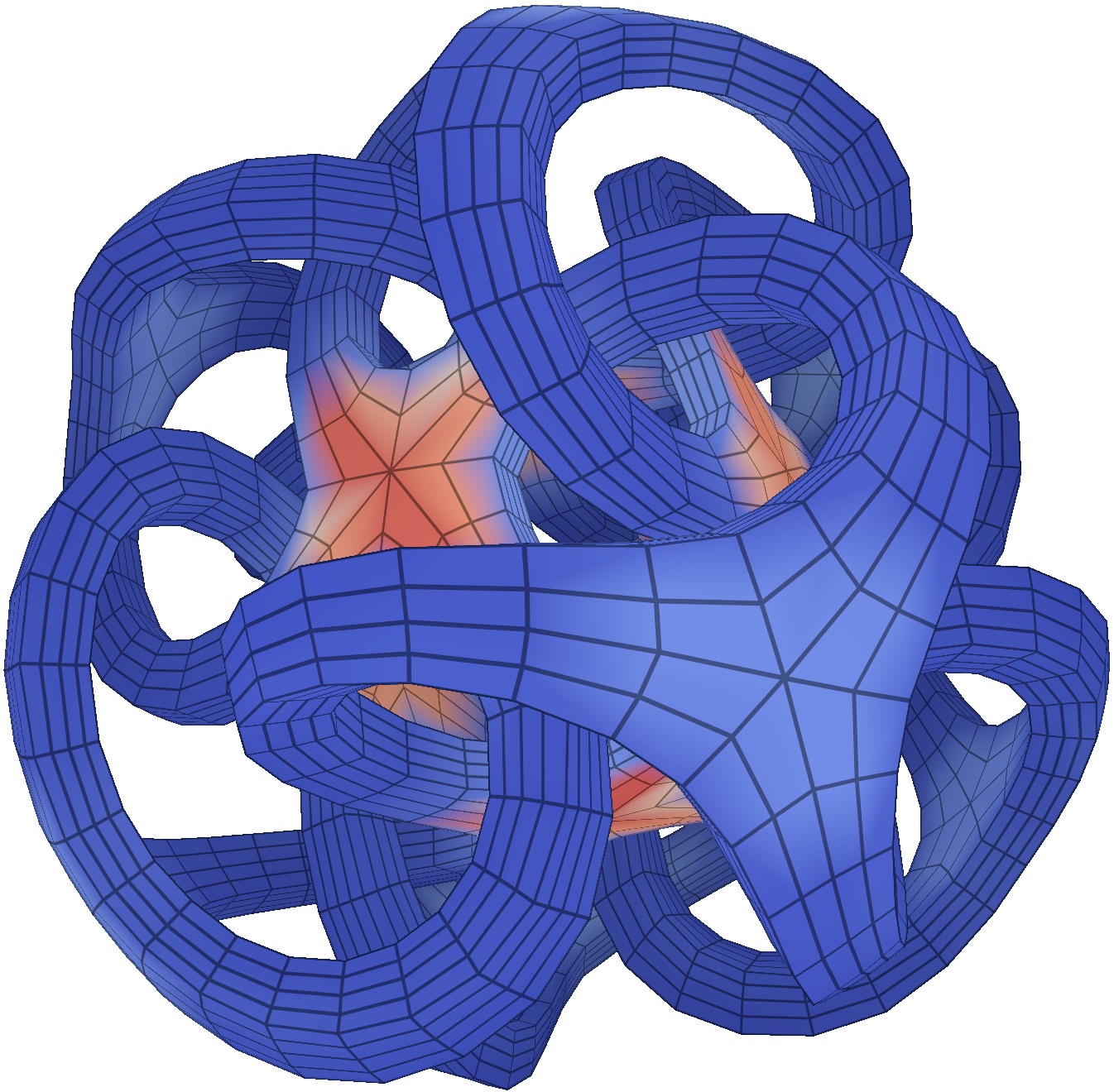} \hspace{2cm}
 \includegraphics[width=0.24\linewidth]{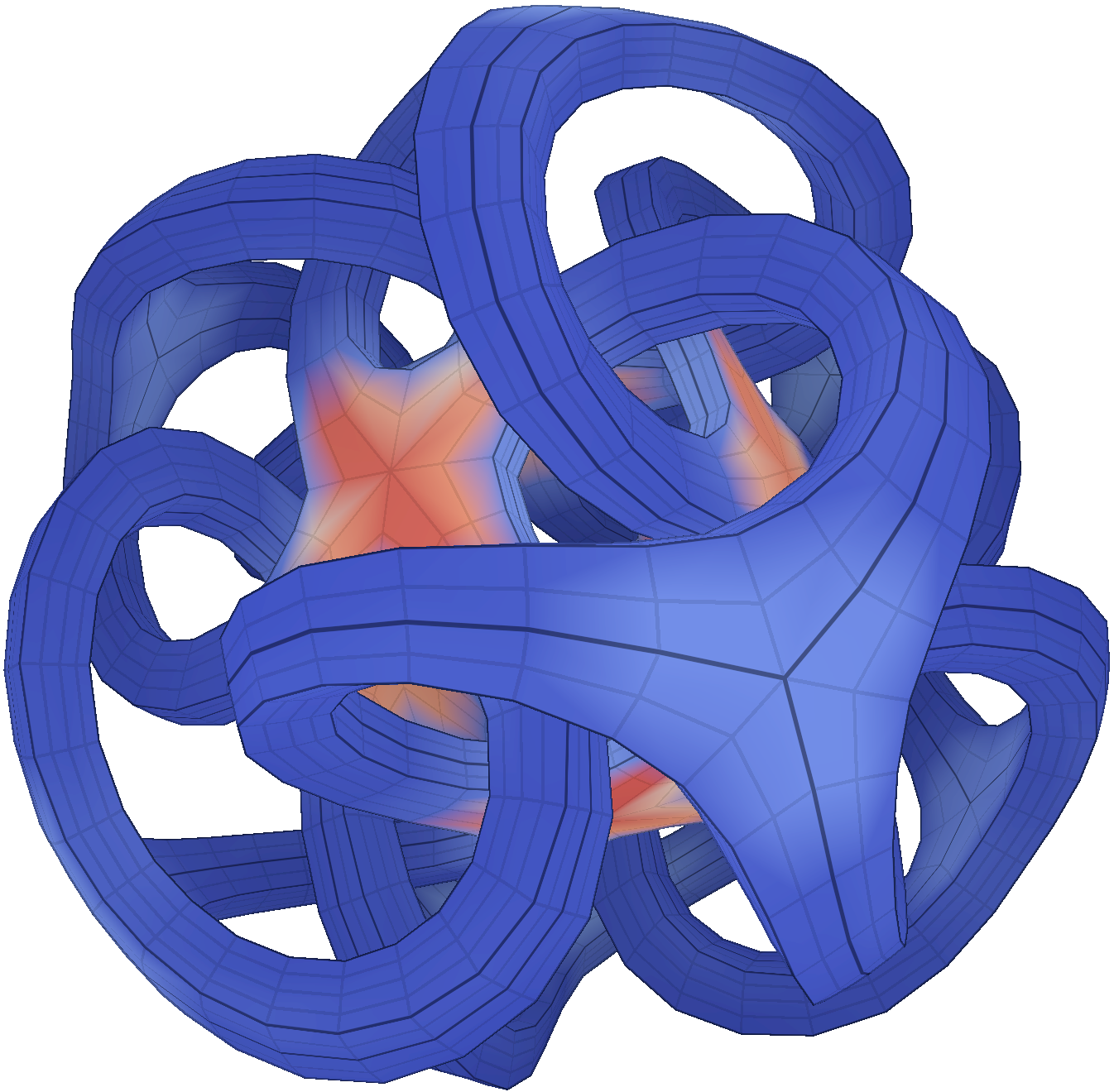}
  \caption{From left to right: Surface rendering with wireframe edges. Same as before, and edge colors are modulated by their LoD to only slightly accentuate edges belonging to a low level. Model metatron courtesy of \cite{LoopyCuts2020}.}
 \label{fig:Comp3}
\end{figure*}

\begin{figure*}[!h]
 \centering
 \includegraphics[height=5.6cm]{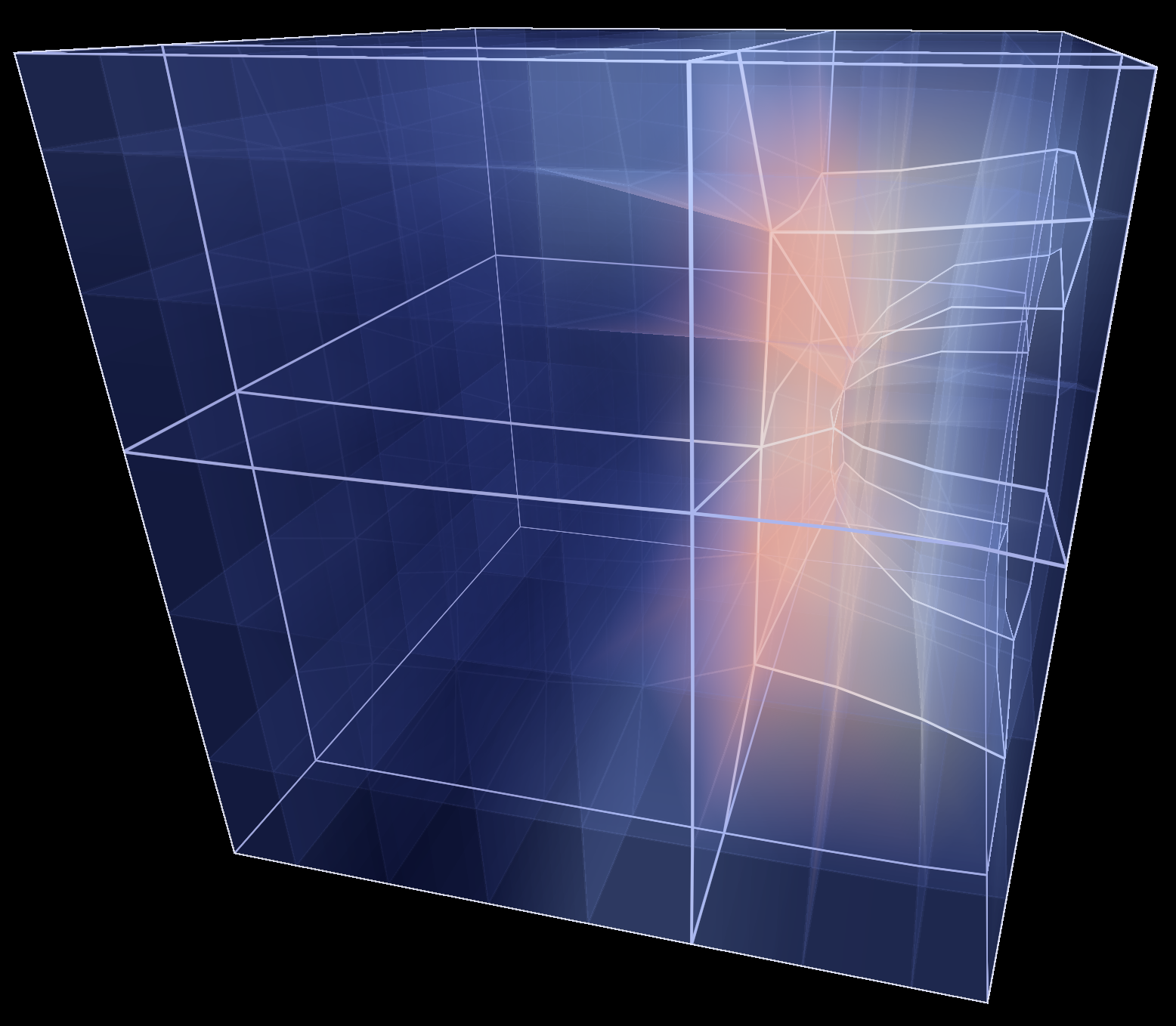}
 \includegraphics[height=5.6cm]{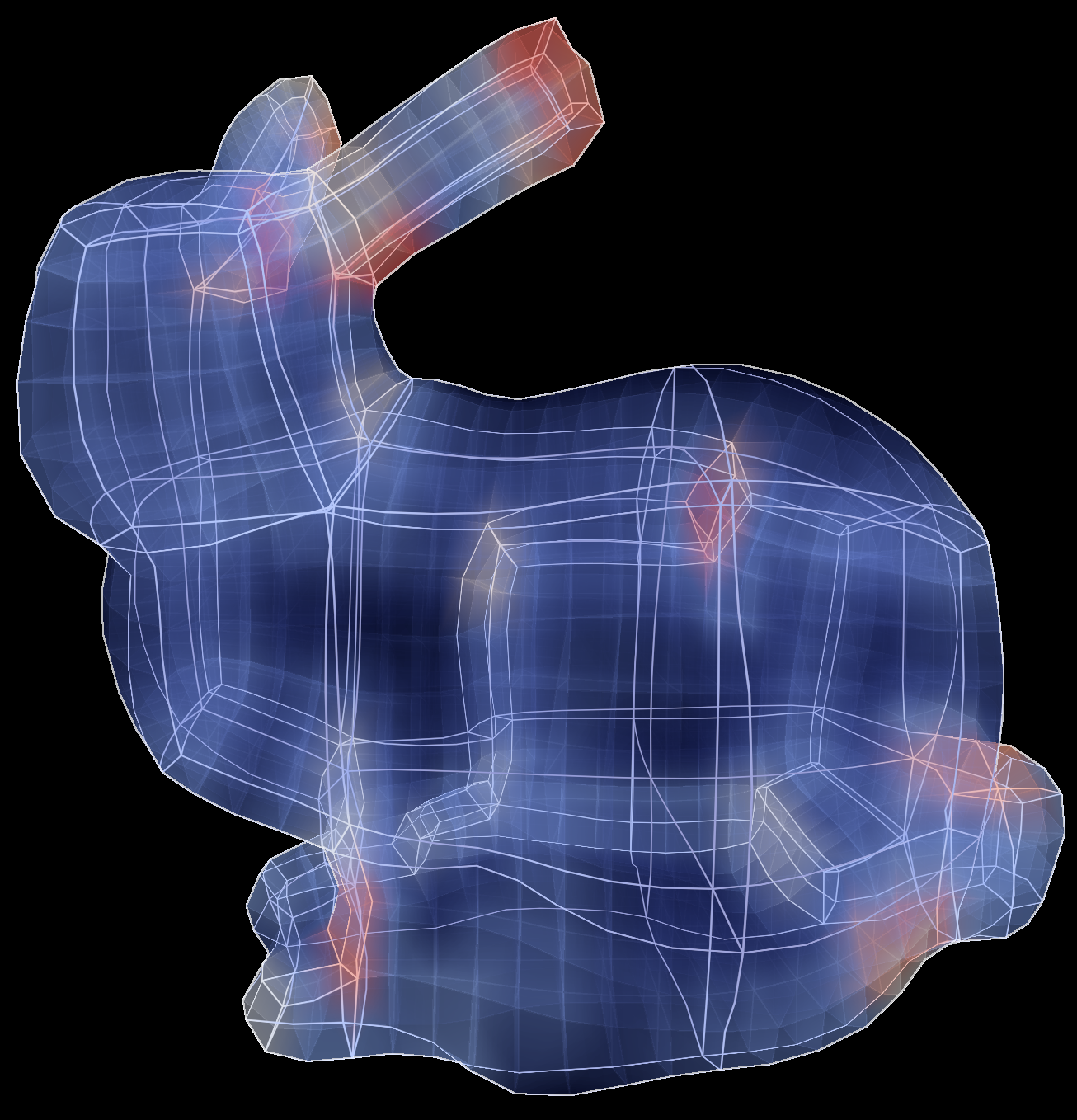}
 \includegraphics[height=5.6cm]{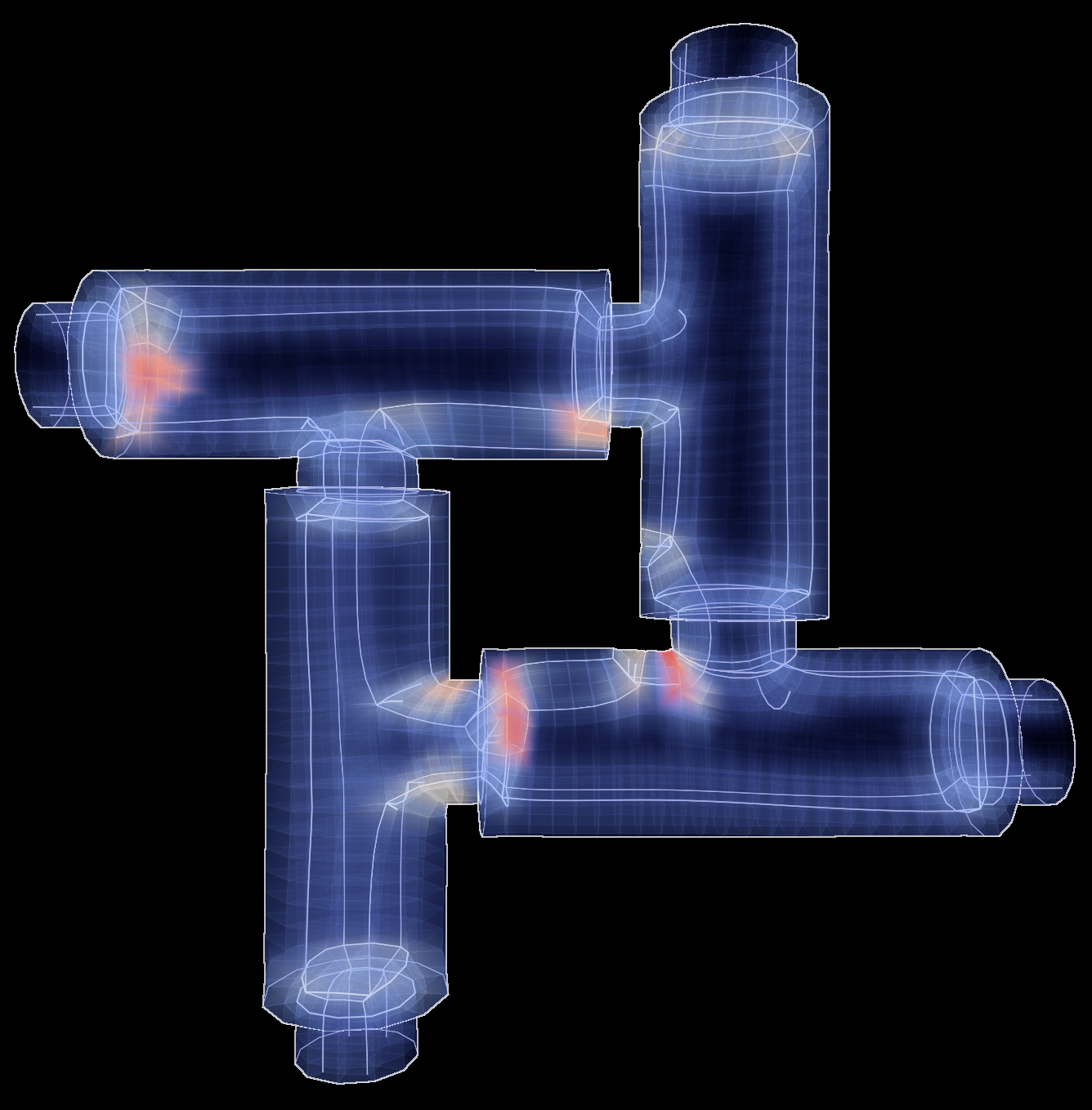}
  \caption{Contextual volume rendering visualizations of LoopyCuts-based models. Models courtesy of \cite{LoopyCuts2020}.}
 \label{fig:LoopyCuts}
\end{figure*}

\begin{figure*}[h]
 \centering
 \includegraphics[height=4.2cm]{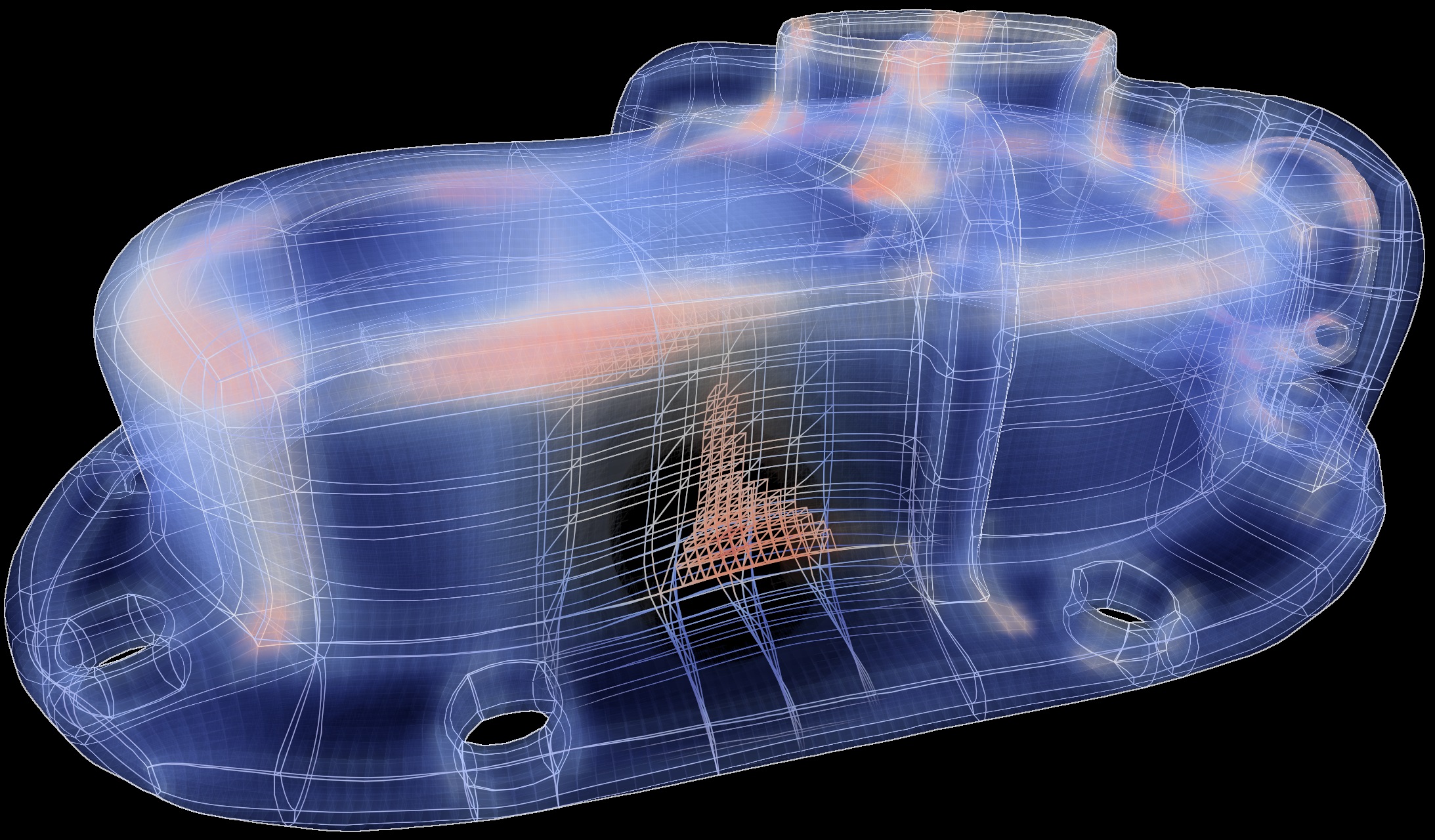}
 \includegraphics[height=4.2cm]{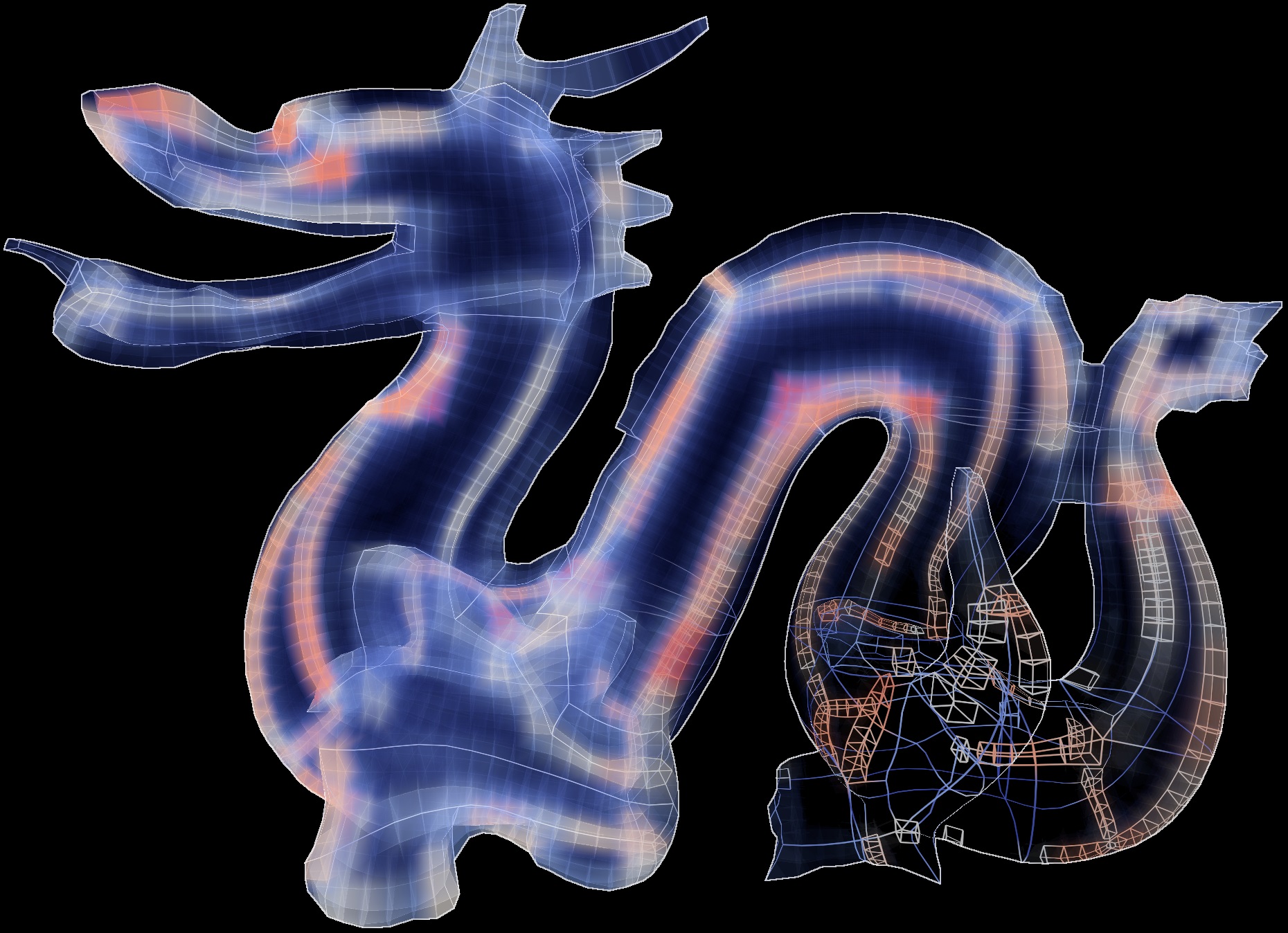}
 \includegraphics[height=4.2cm]{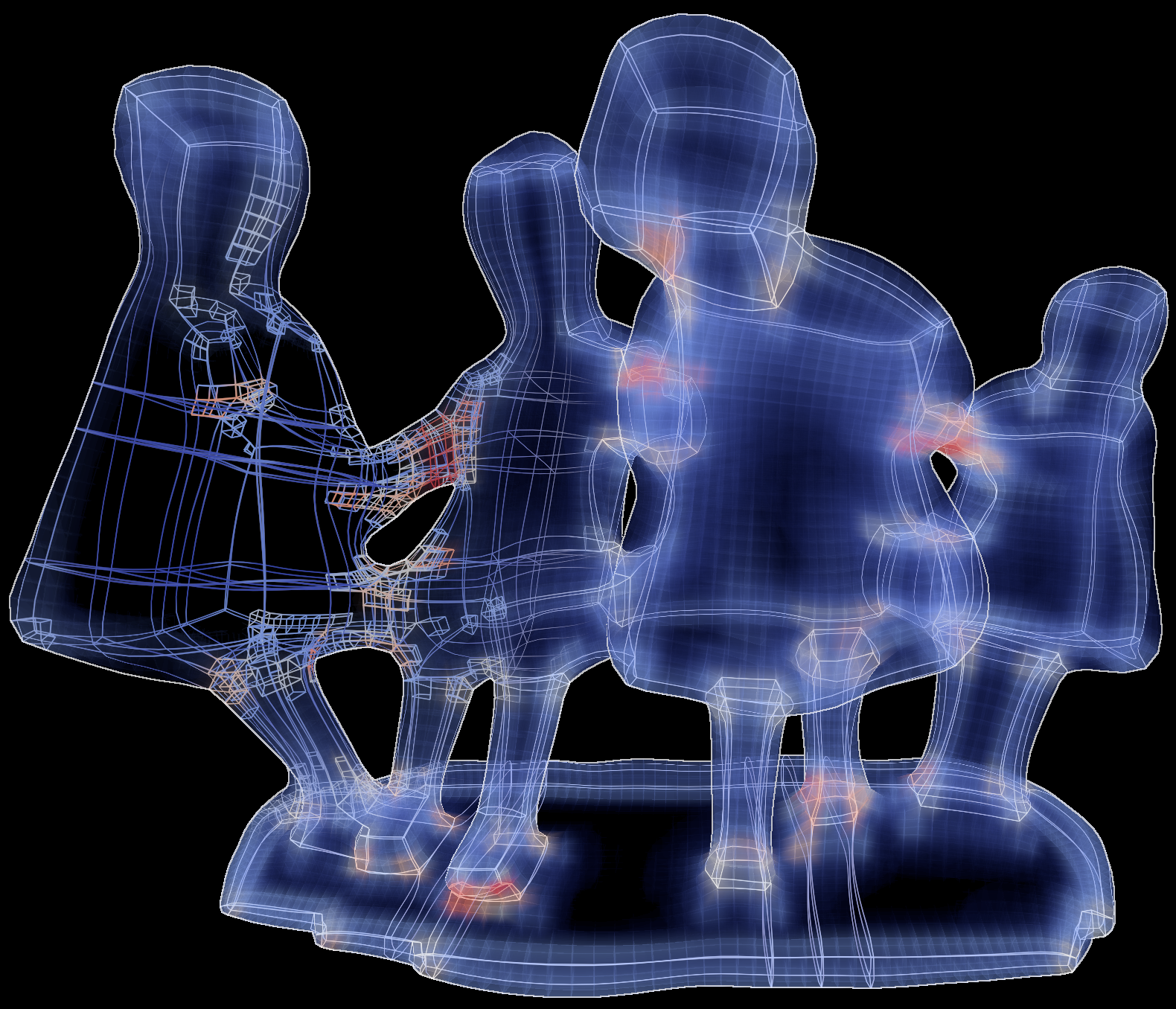}
  \caption{F+C renderings of different meshes. Models cognit, dragon and dancingchildren courtesy of \cite{Huang2014:LCO, EdgeConeRectification2015}.
  }
 \label{fig:dancingchildren}
\end{figure*}

\autoref{fig:sidebyside-cubic-example} (left) shows a deformed Cartesian grid comprised of $128^3$ cells. The deformed grid is created by performing a Finite Element analysis with a specific boundary condition to let the mesh twist. High deformations occur in the orange regions, yet the cells are so small that the mesh structure cannot be seen. Via the edges from a selected coarse LoD level, the basic mesh structure is preserved, and the user can now zoom at a high deformation region and use focus rendering to investigate the deformations in more detail. \autoref{fig:sidebyside-cubic-example} (right) shows a rendering of a hex-mesh that was generated via the method from \cite{FuzzyClustering2018}.
As can be seen, the meshing approach creates many singular edge columns, i.e., cells with higher deformations are laid out along straight vertical structures, while the remaining parts of the mesh show almost zero deformation. The focus view reveals the structure of the cells with a deformation larger than a selected threshold in the selected region.
In \autoref{fig:dancingchildren} we show further results of F+C rendering. In \autoref{fig:Comp3} and \autoref{fig:LoopyCuts}, more applications of our approach can be seen.

\subsection{Evaluation}

\begin{figure}[h]
 \centering
 \includegraphics[width=0.45\linewidth]{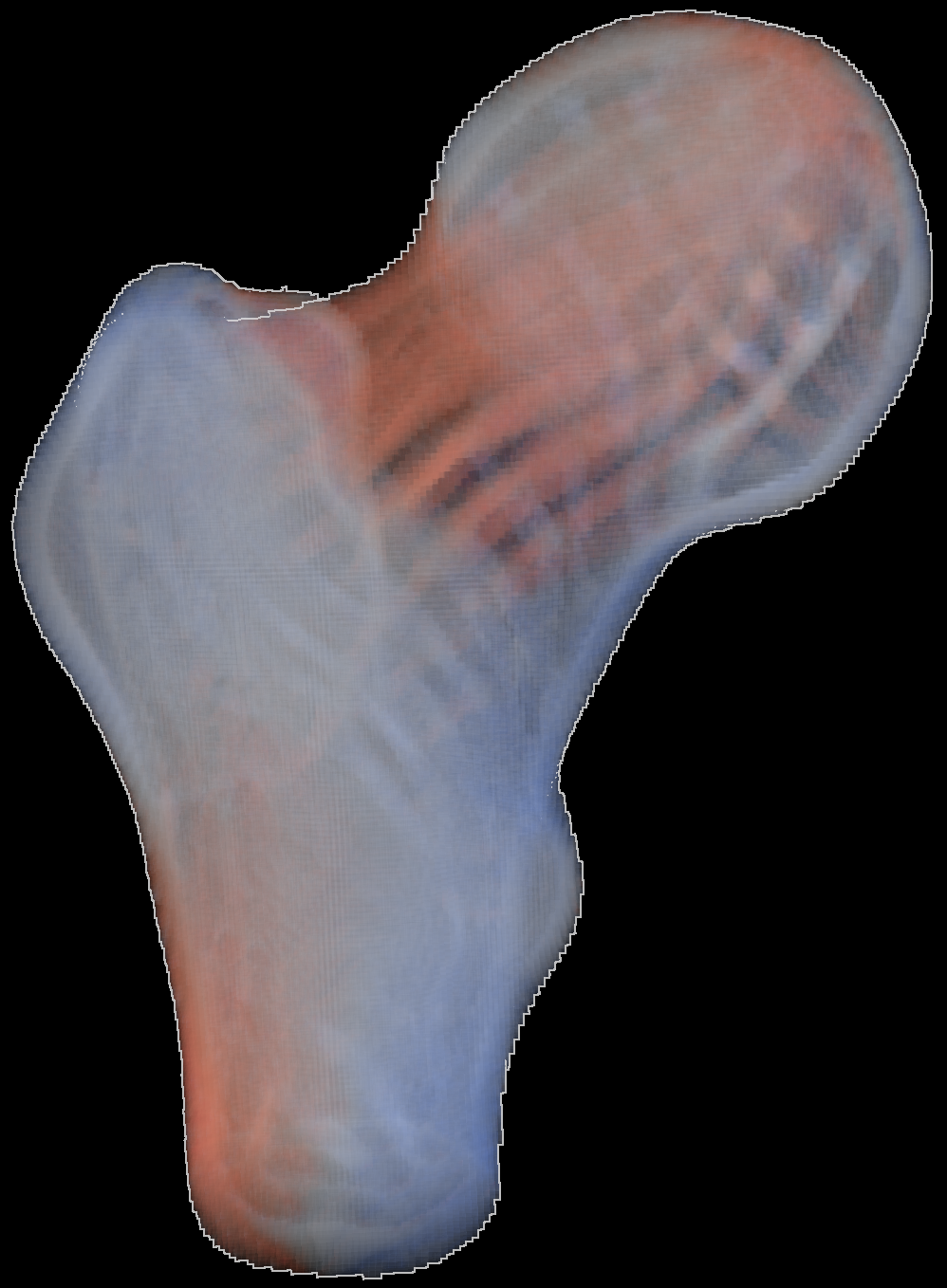}
  \caption{Contextual volume rendering is used to visualize a scalar field that is given at the vertices of Cartesian grid. Values represent the stress anisotropy of a femur model for a certain load condition.}
 \label{fig:femur}
\end{figure}

\begin{figure*}[hp]
 \centering
 \includegraphics[width=0.48\linewidth]{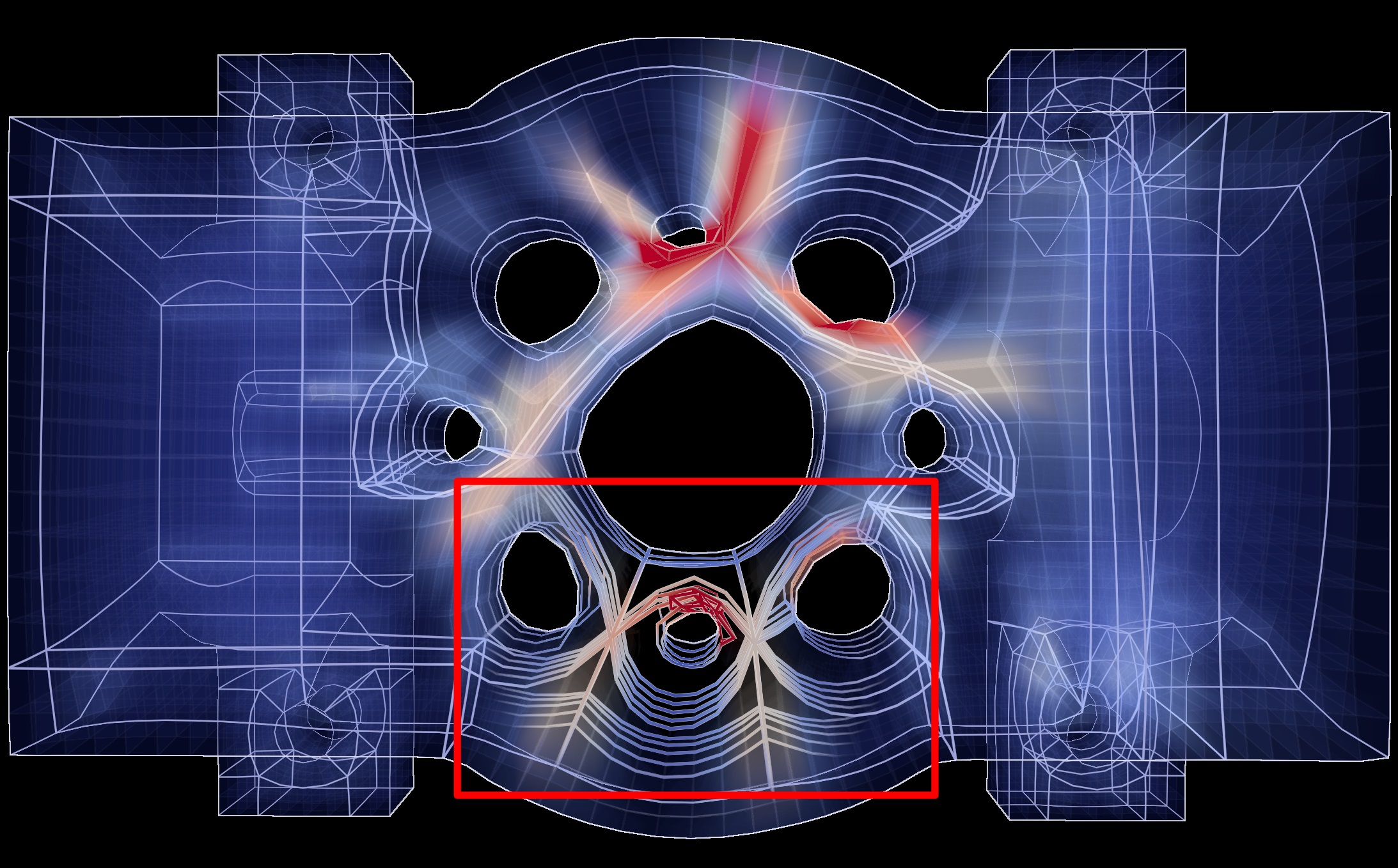}
 \includegraphics[width=0.48\linewidth]{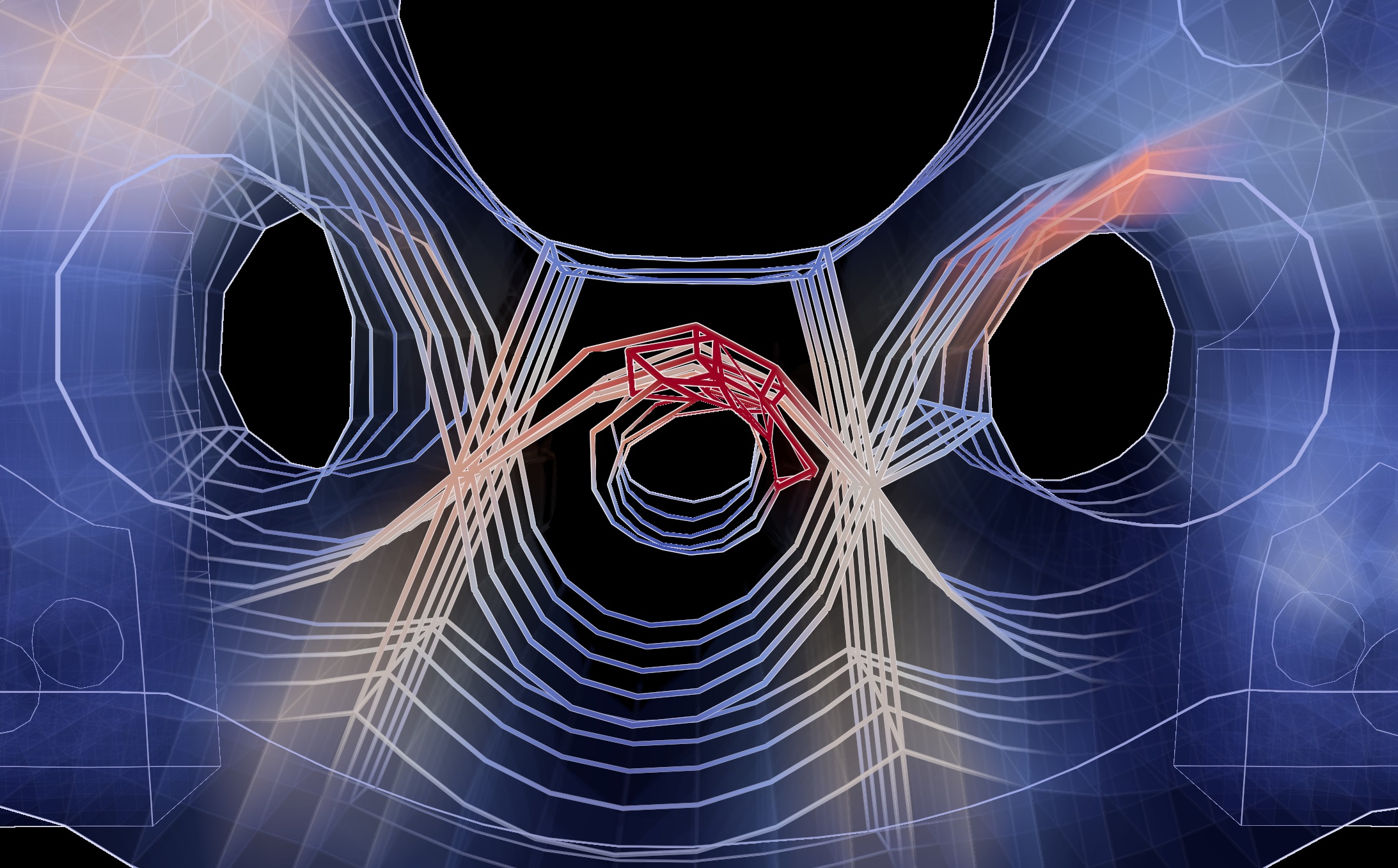}
 \includegraphics[width=0.48\linewidth]{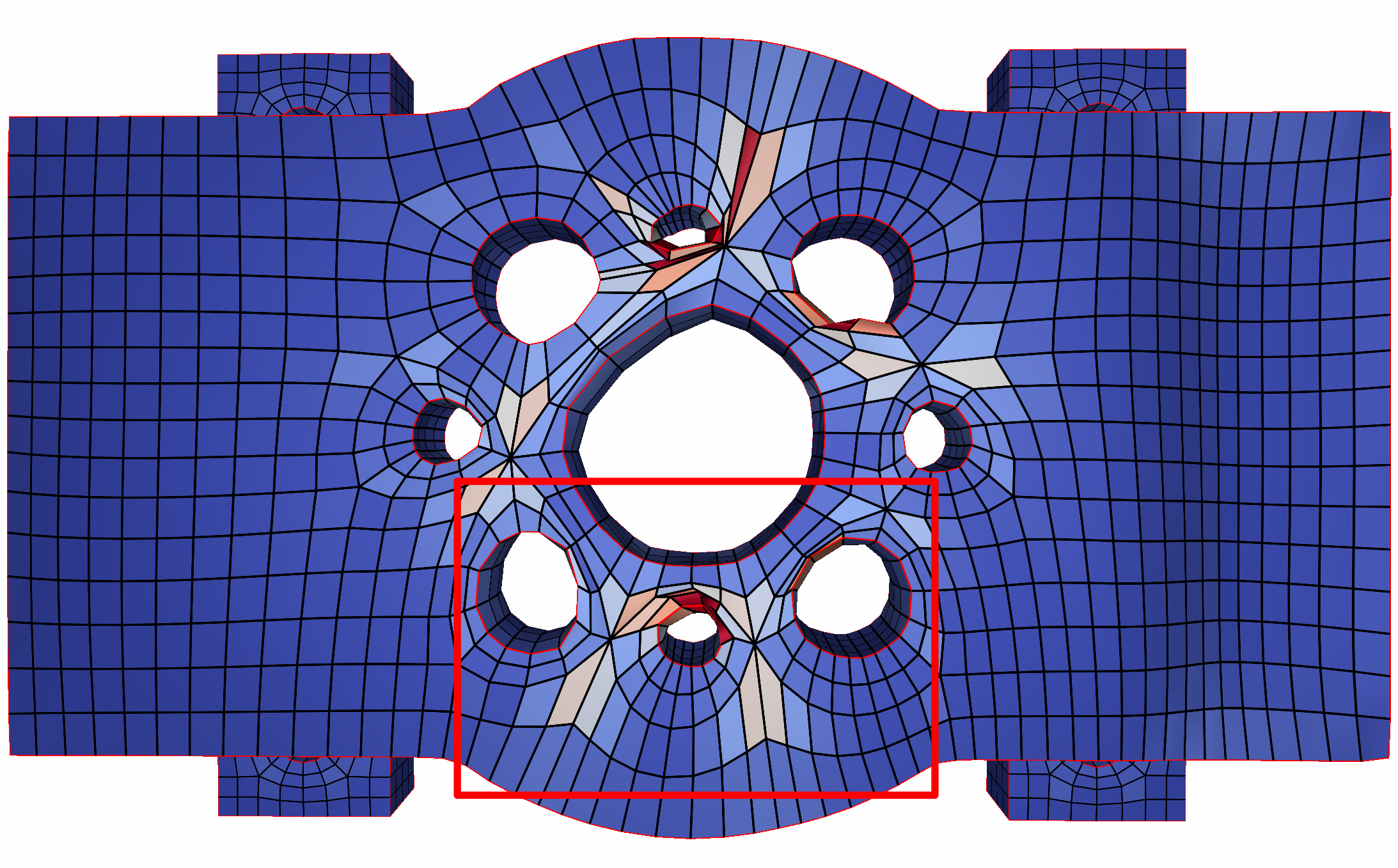}
 \includegraphics[width=0.48\linewidth]{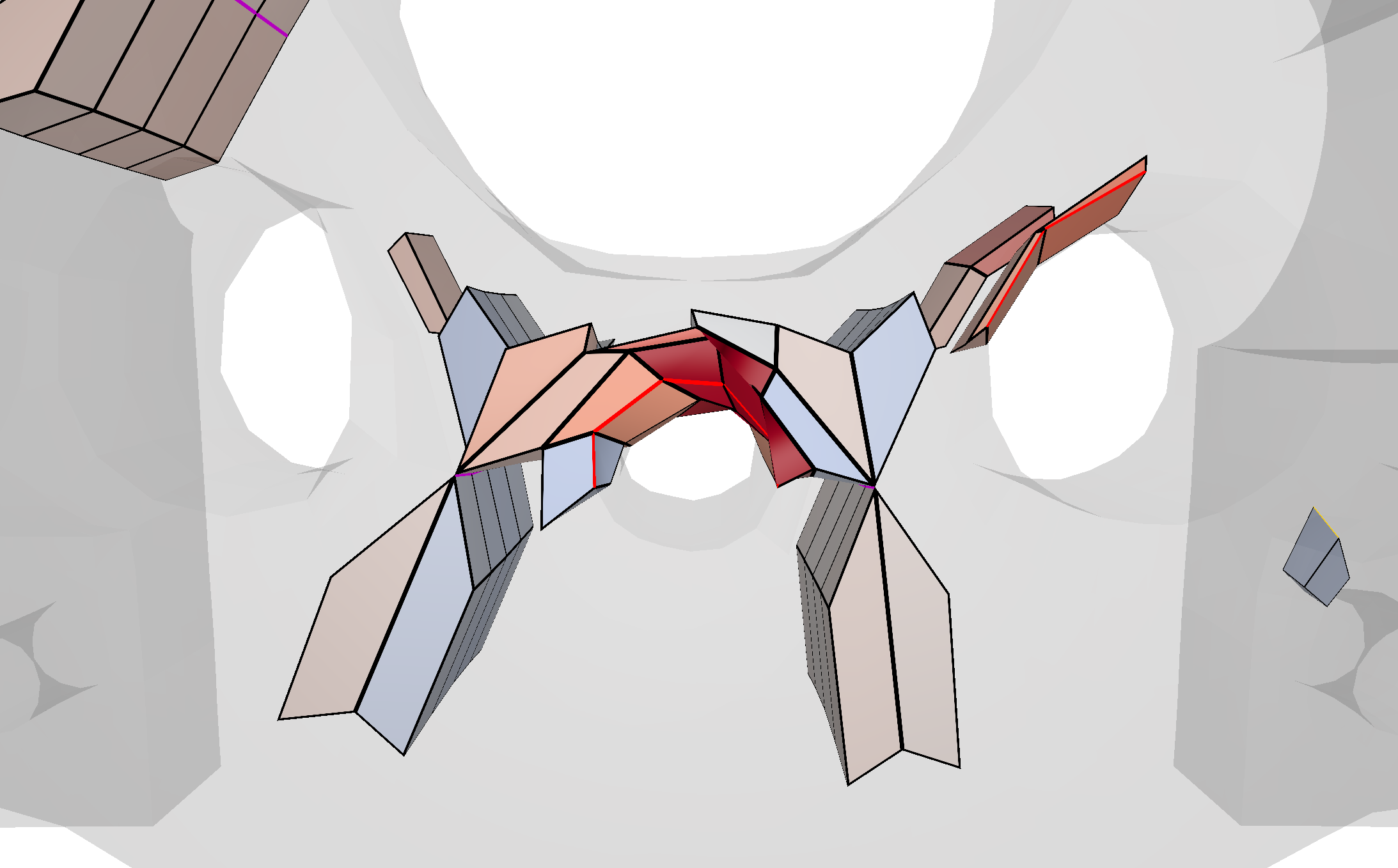}
  \caption{Top: F+C visualization. Bottom: Same model and views with opaque surface rendering and quality-based cell filtering. Model motor\_tail courtesy of \cite{LoopyCuts2020}.}
 \label{fig:Comp4}
\end{figure*}

\begin{figure*}[hp]
 \centering
 \includegraphics[width=0.19\linewidth]{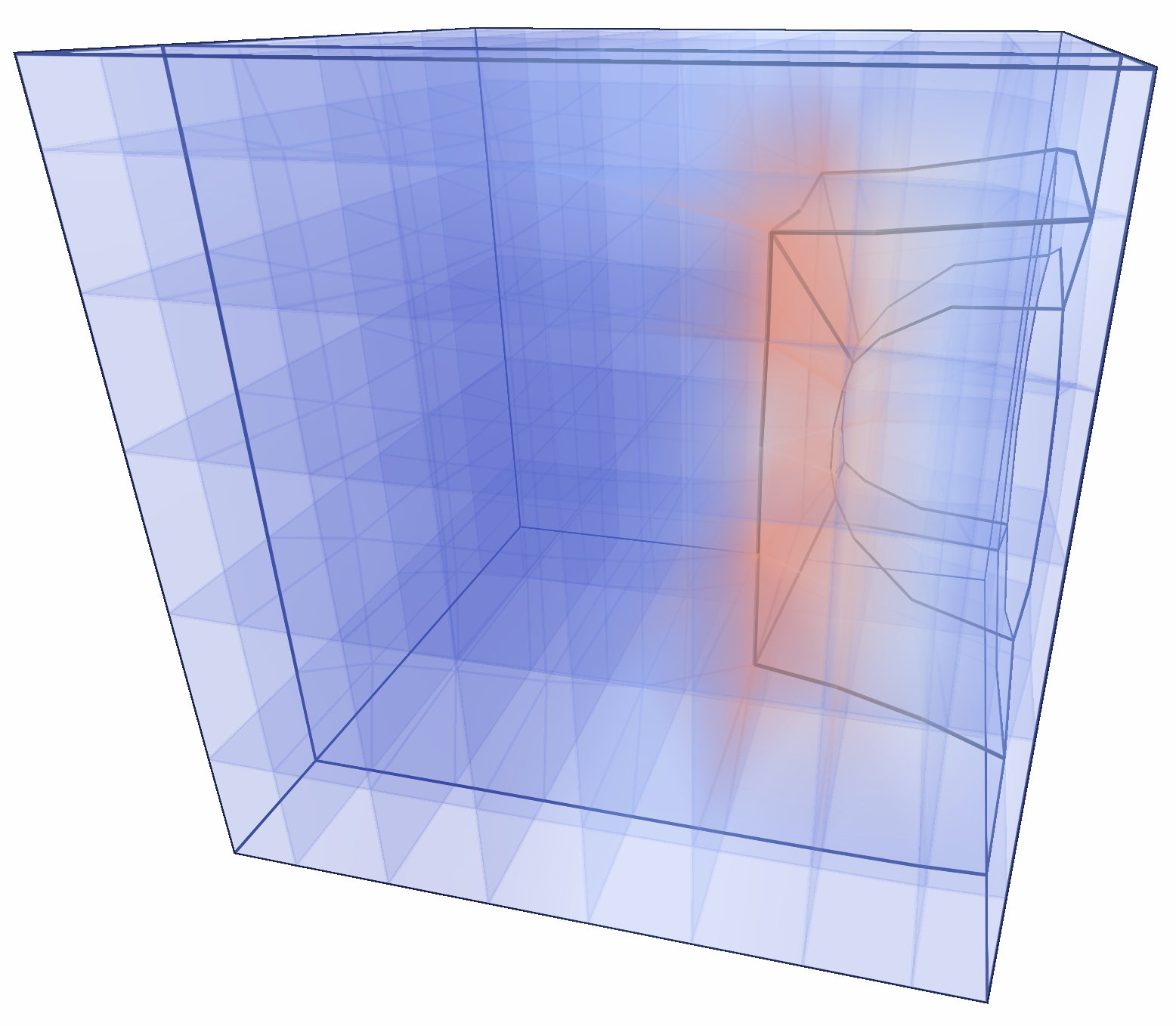}
 \includegraphics[width=0.19\linewidth]{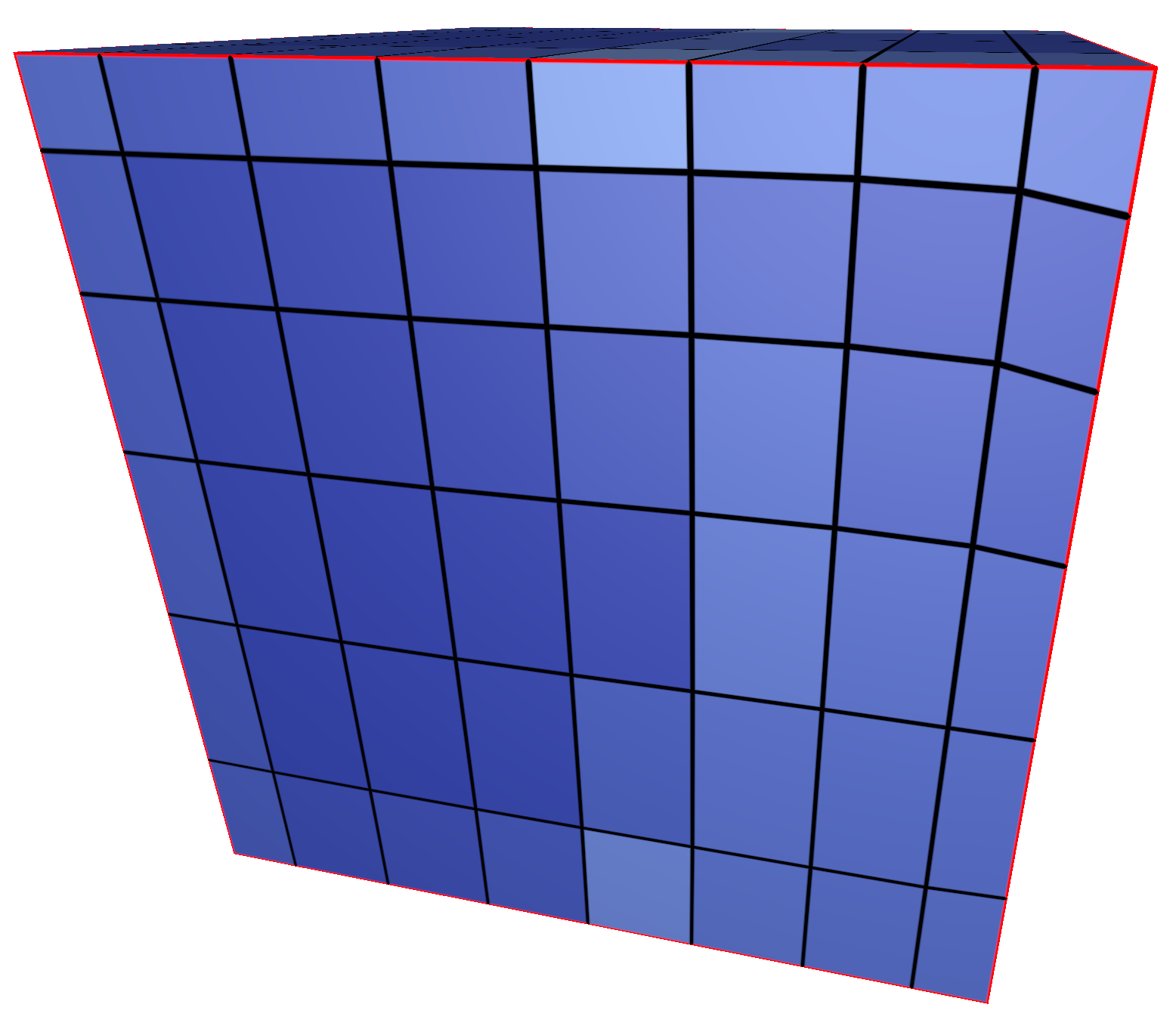}
 \includegraphics[width=0.19\linewidth]{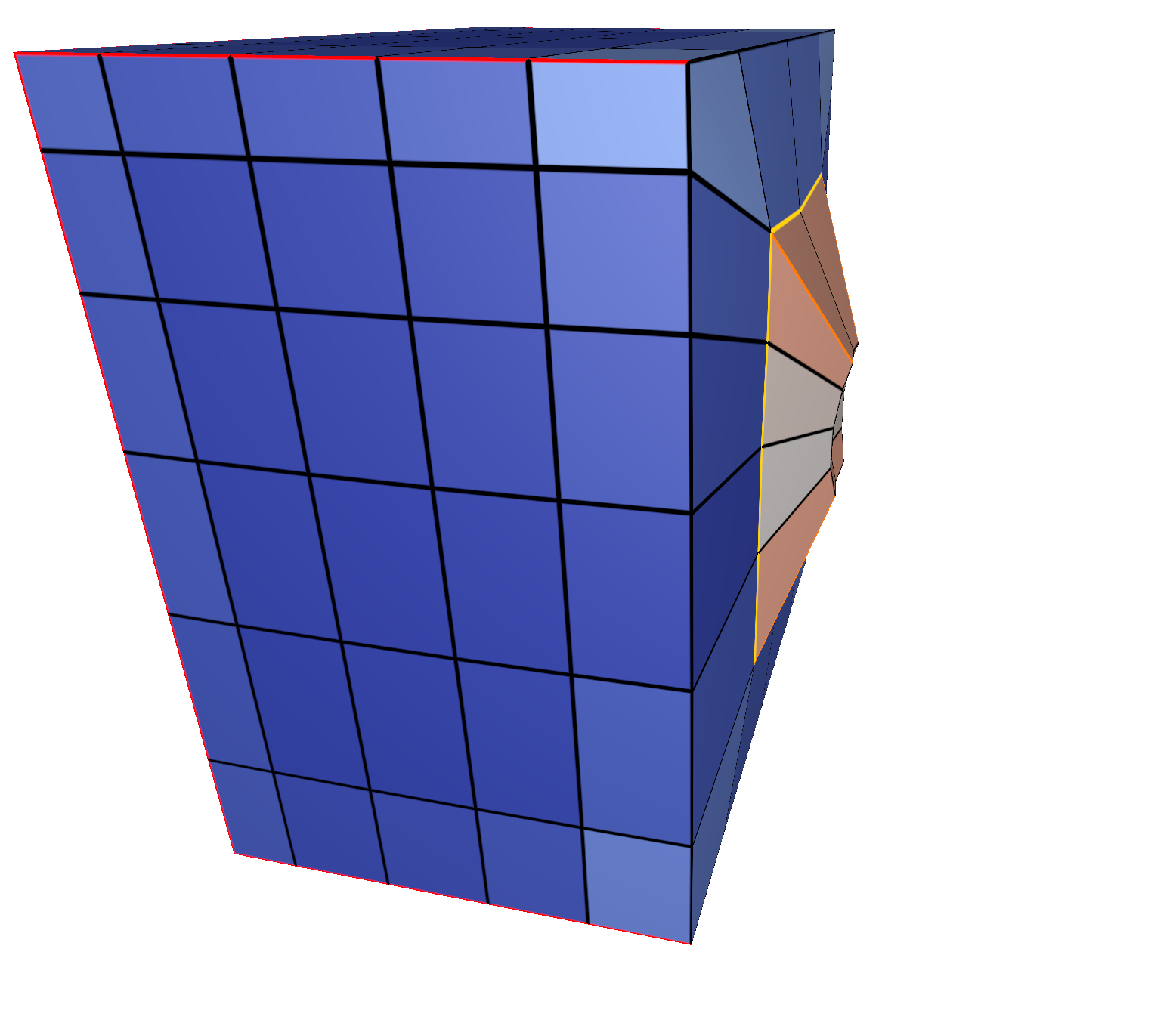}
 \includegraphics[width=0.19\linewidth]{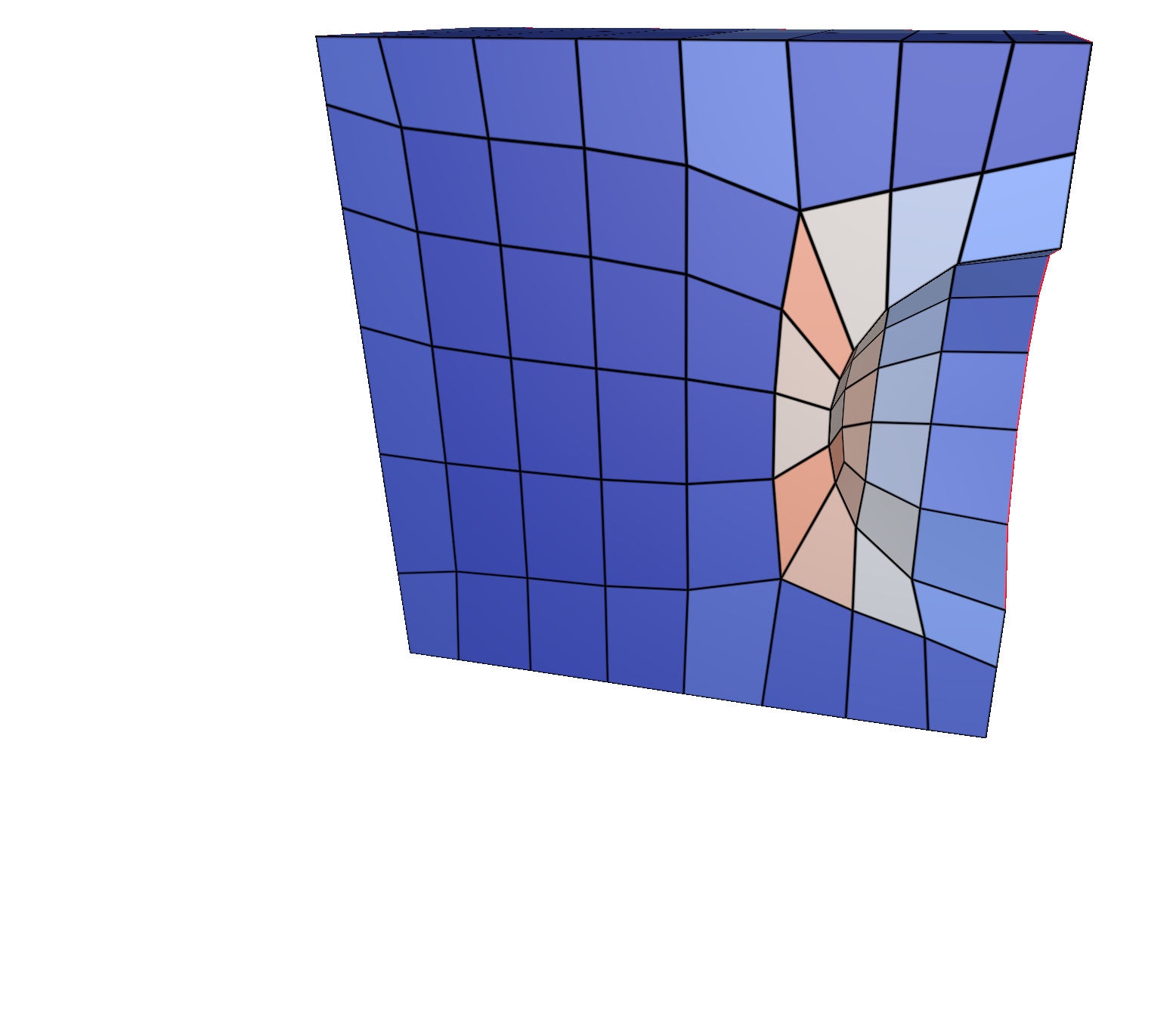}
 \includegraphics[width=0.19\linewidth]{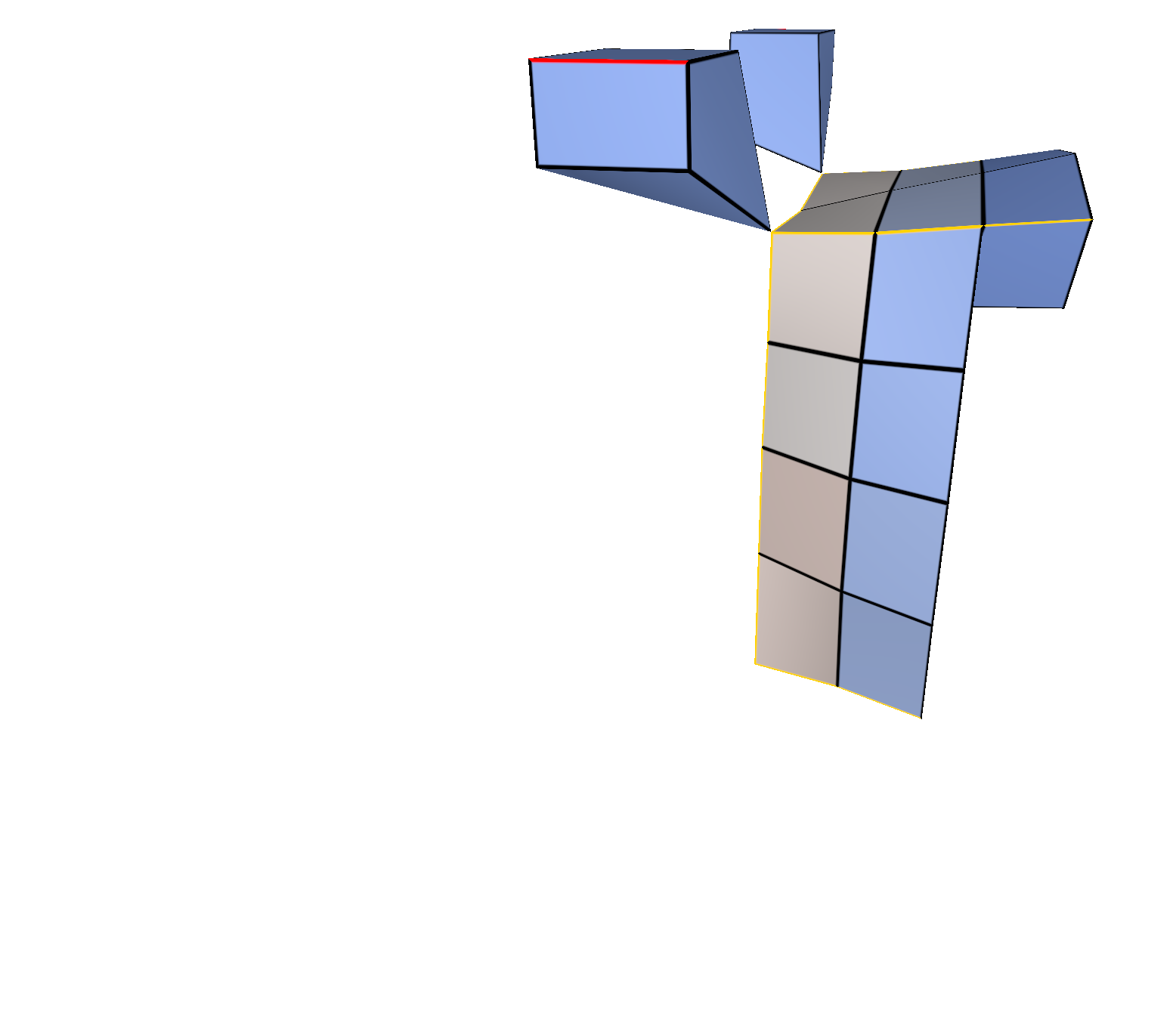}
  \caption{From left to right: Contextual volume rendering (ours). Fully opaque surface rendering. Slicing with oblique slicing plane. Additional slicing plane removing front elements. Quality-based cell filtering.
  Opaque surface rendering requires multiple operations to reveal the interesting mesh structures, and context information is often lost. 
  Model cube\_carved courtesy of \cite{LoopyCuts2020}.}
 \label{fig:Comp2}
\end{figure*}

\begin{figure*}[hp]
 \centering
 \includegraphics[height=5.5cm]{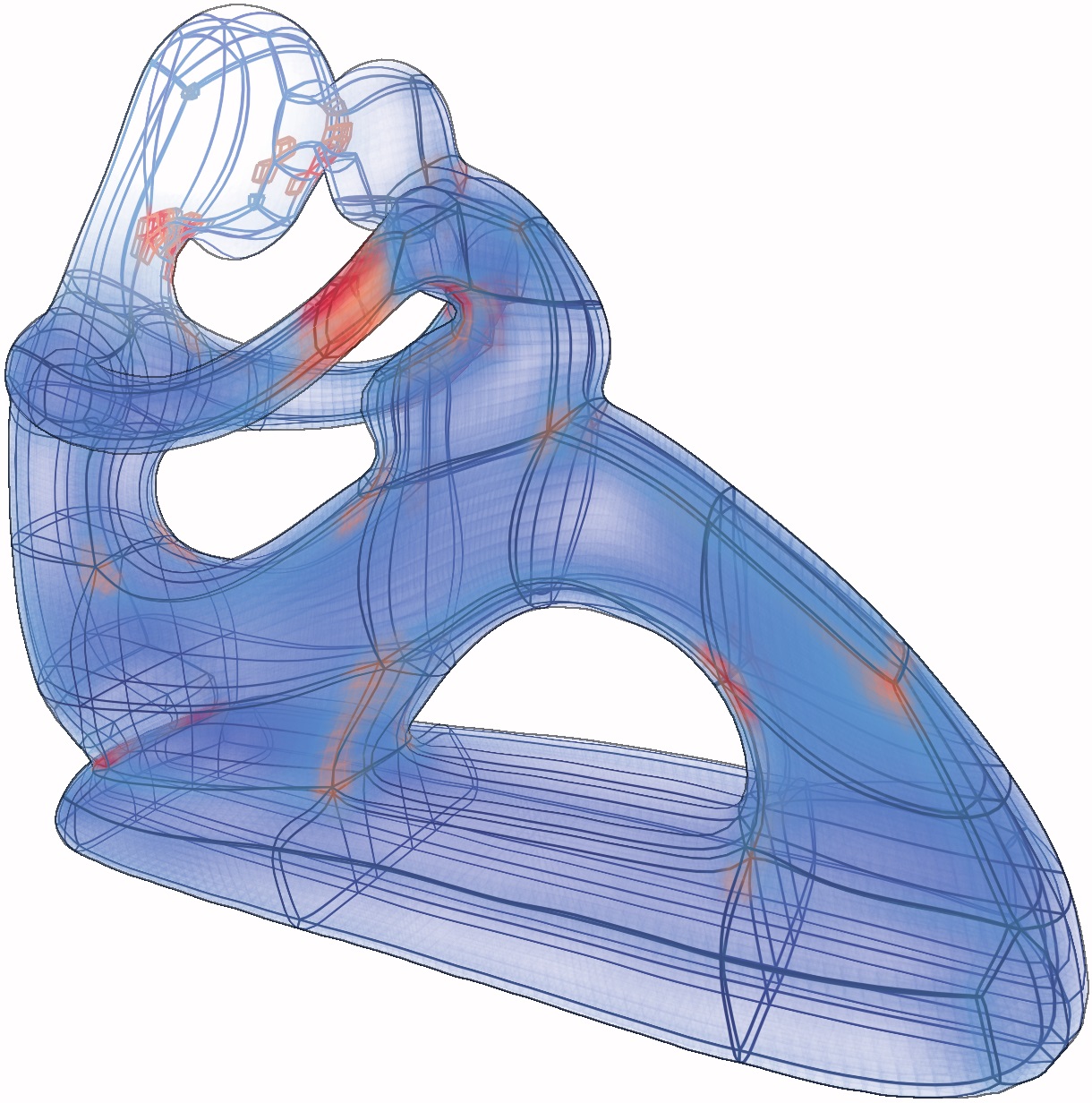}
 \hspace{0.2cm}
 \includegraphics[height=5.5cm]{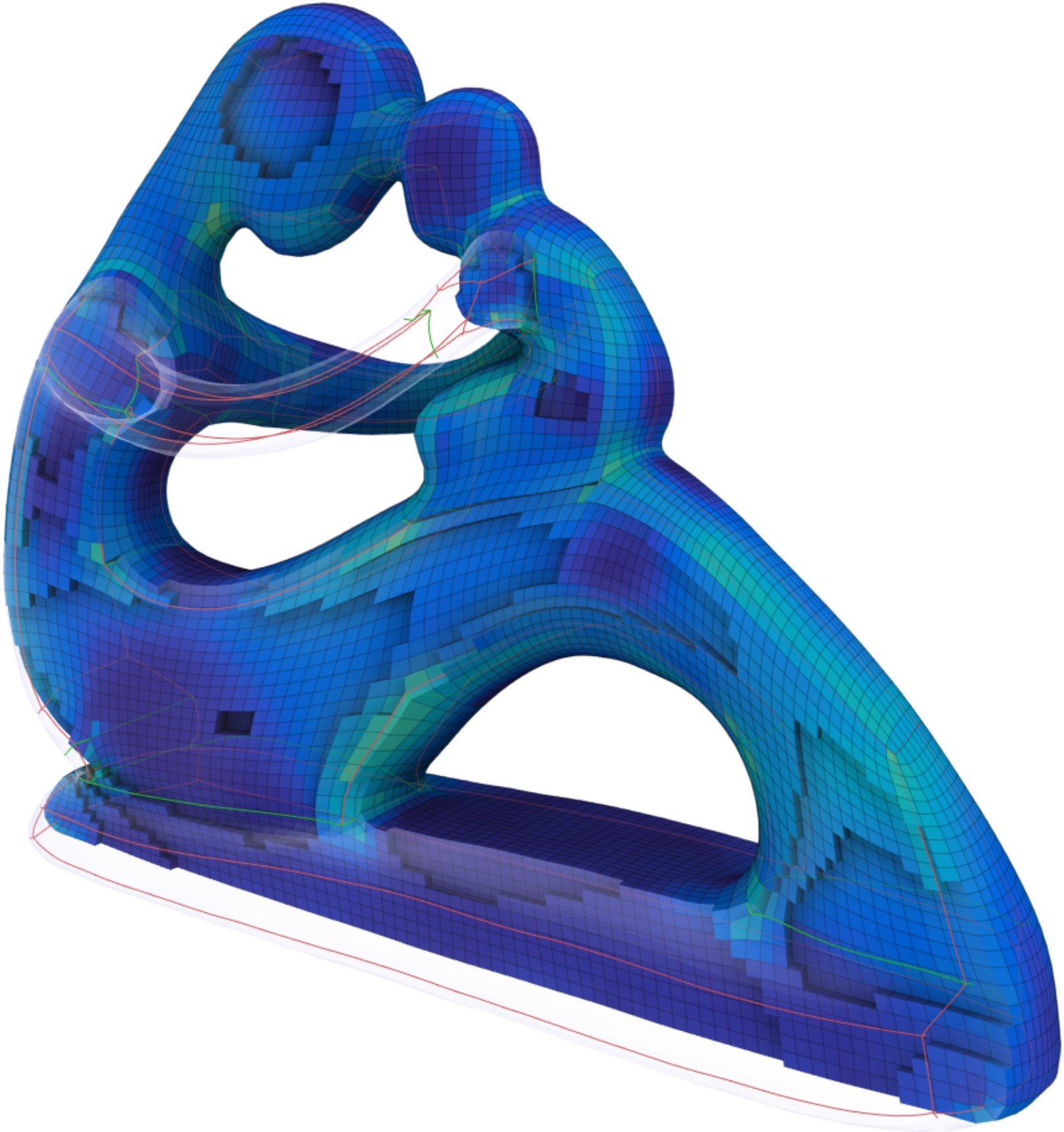}
 \hspace{0.2cm}
 \includegraphics[height=5.5cm]{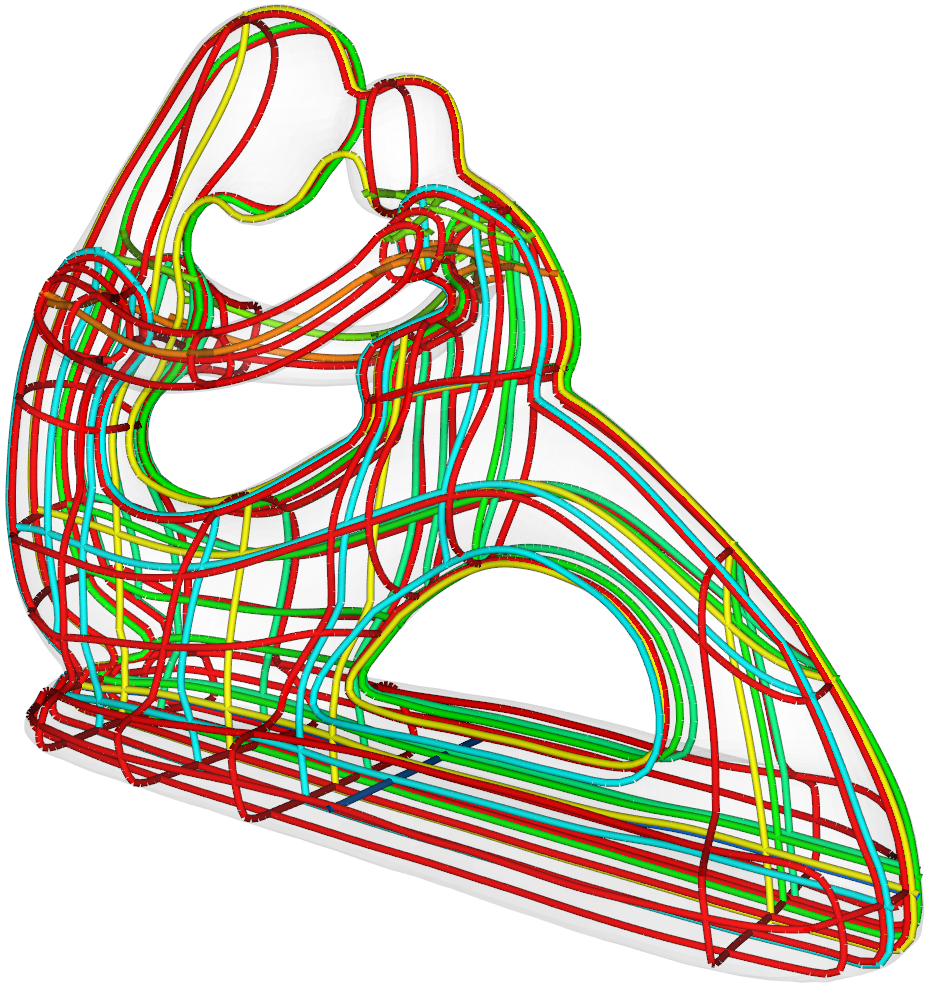}
  \caption{Comparison of different rendering techniques. From left to right: F+C visualization with focus on woman's head (ours). Visualization in HexaLab \cite{Bracci:2019} with slicing. Main sheet visualization by \cite{Xu:2018:TVCG}. Rightmost image courtesy of \cite{Xu:2018:TVCG}. Model fertility courtesy of \cite{Polycut2013}.}
 \label{fig:CompAll}
\end{figure*}

\begin{figure*}[hp]
 \centering
 \includegraphics[width=0.34\linewidth]{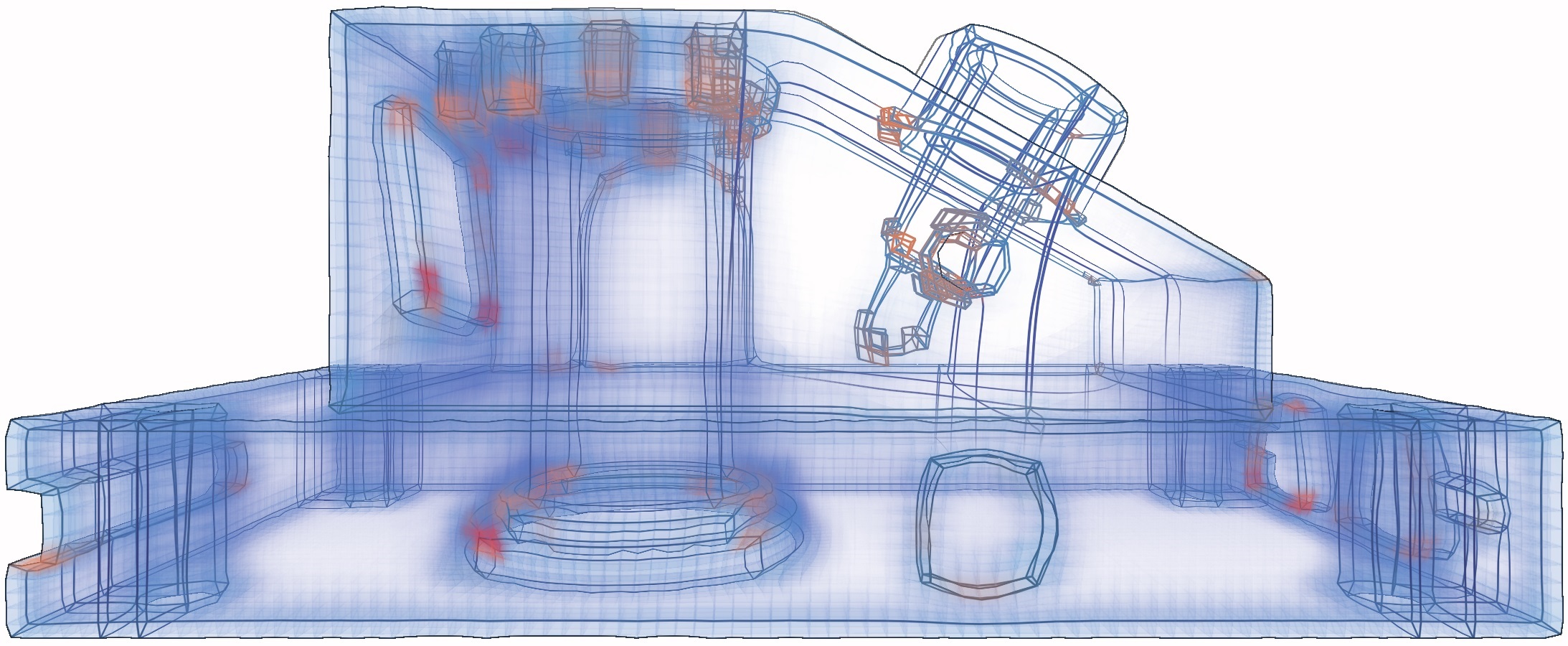}
 \includegraphics[width=0.34\linewidth]{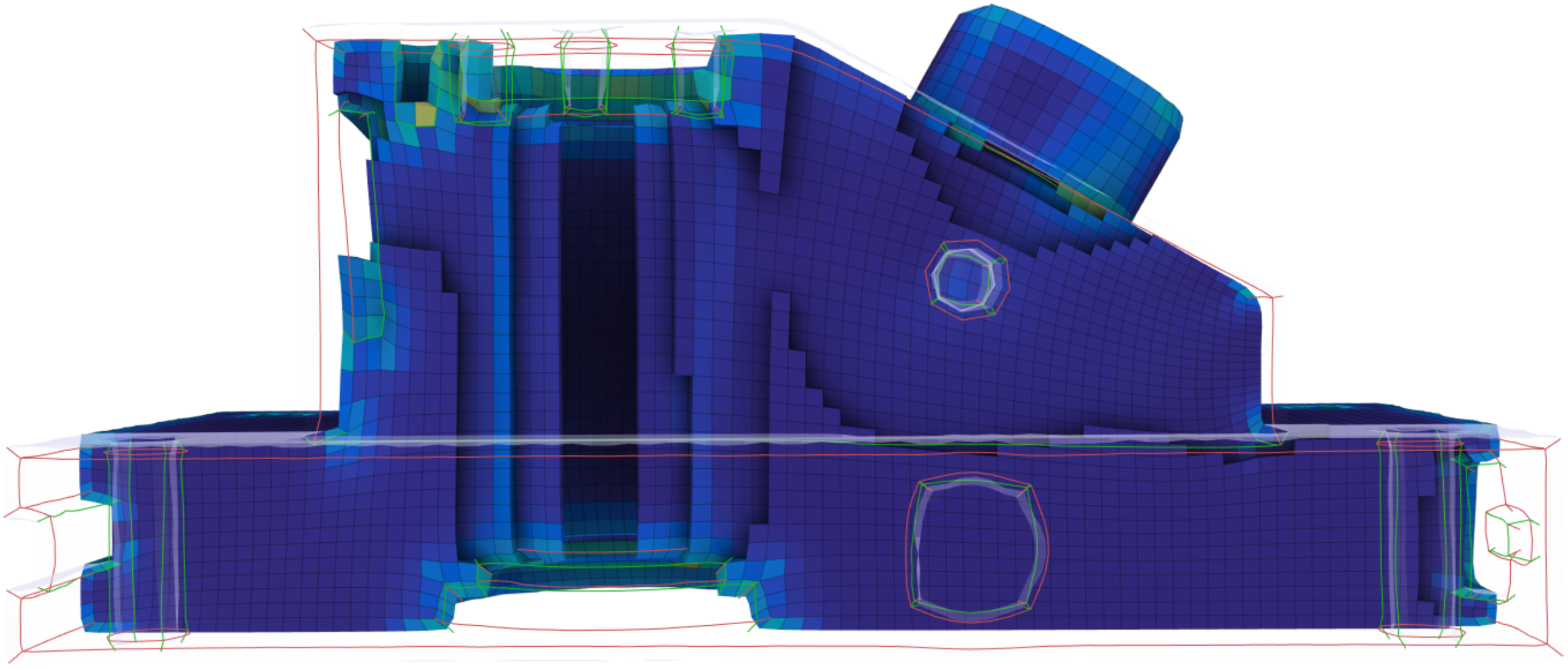}
 \includegraphics[width=0.32\linewidth]{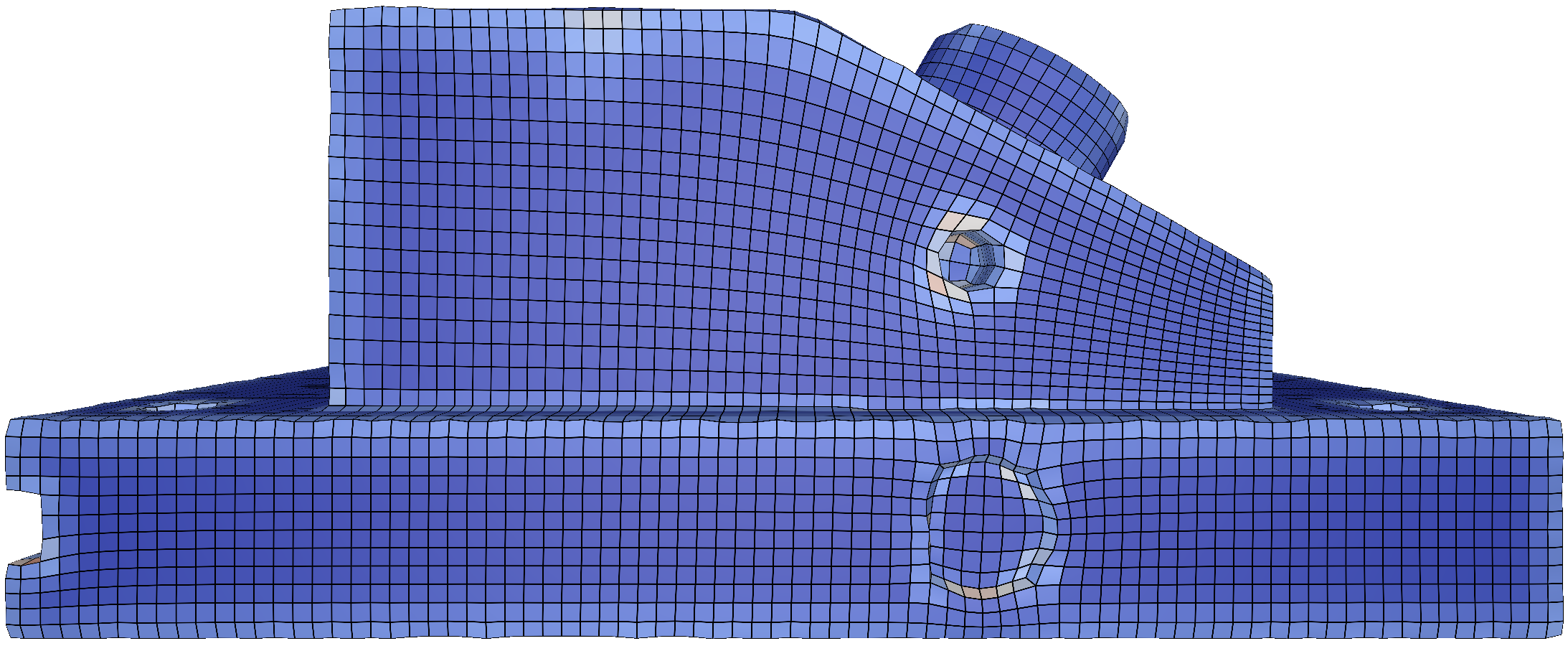}
 \includegraphics[width=0.32\linewidth]{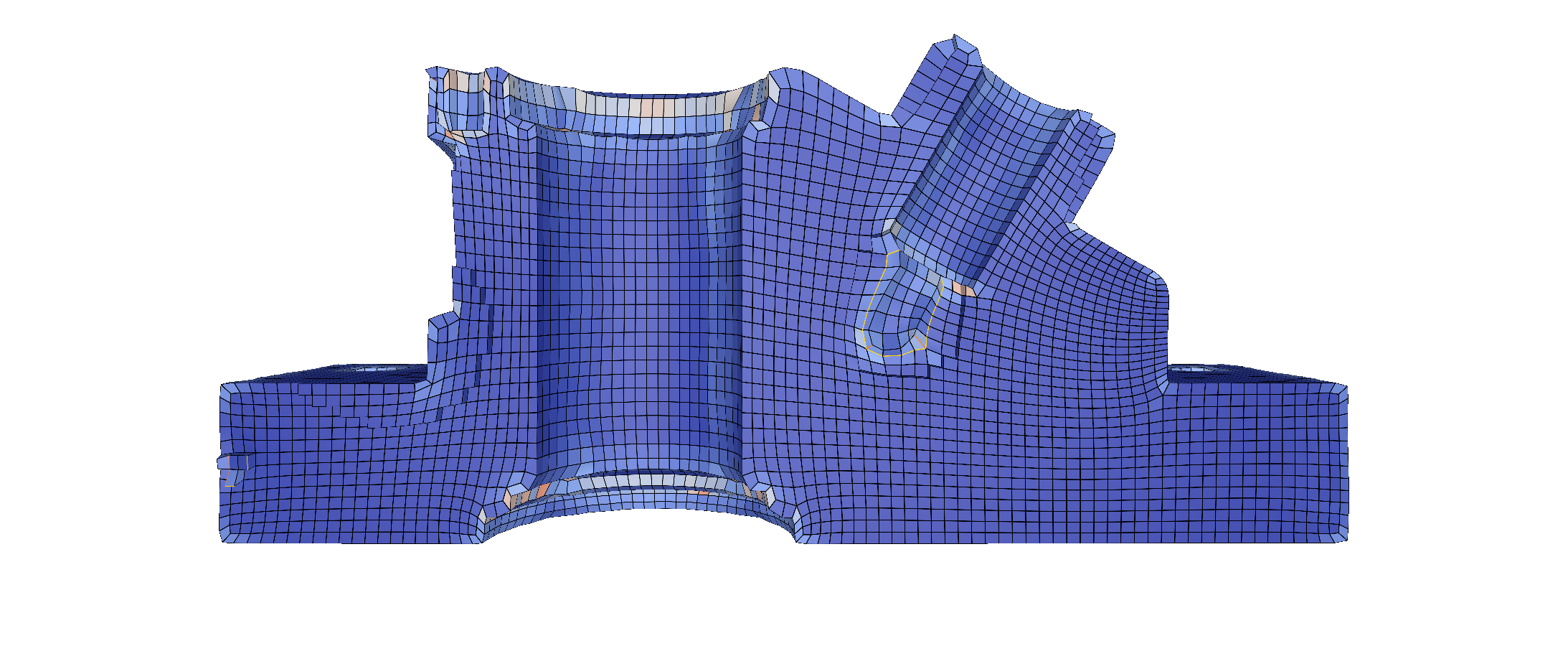}
 \includegraphics[width=0.32\linewidth]{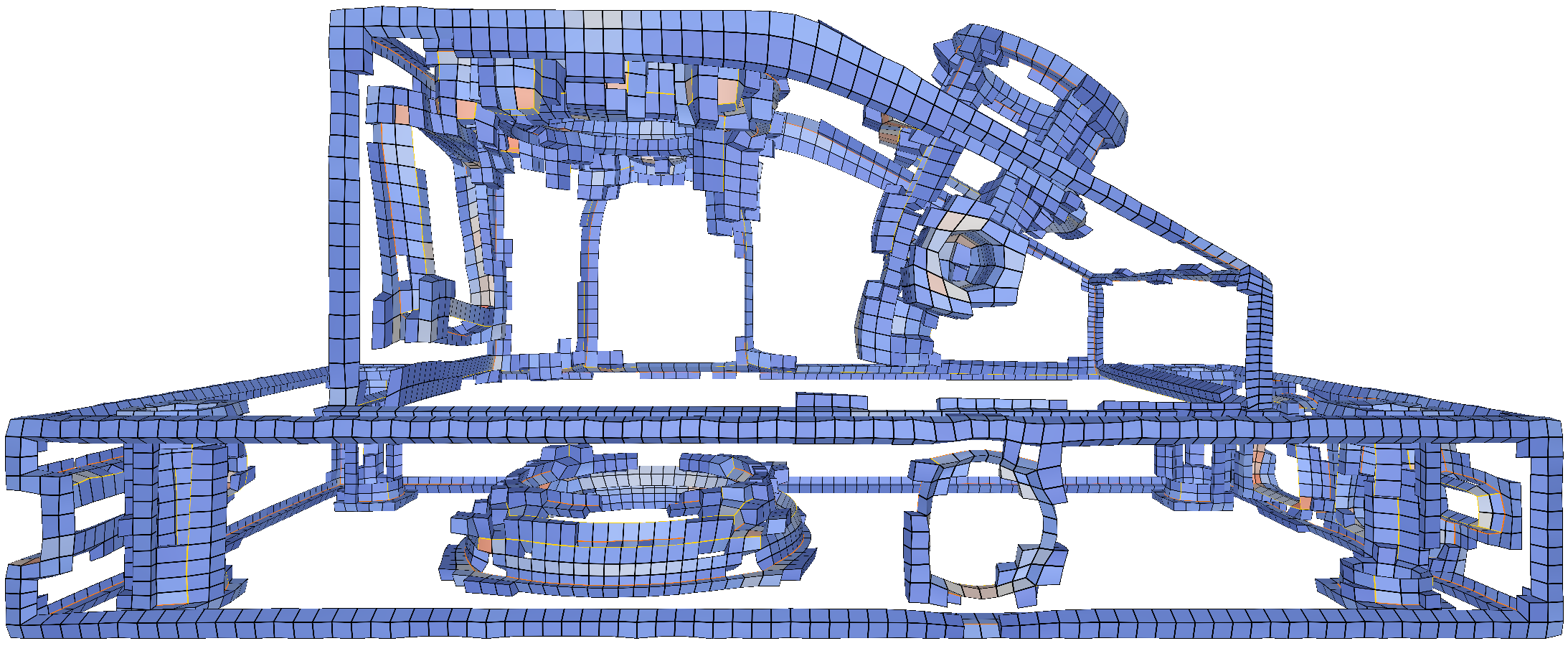}
  \caption{From top left to bottom right: F+C visualization (ours). Slicing in HexaLab \cite{Bracci:2019} with singular edges. Fully opaque surface rendering. Slicing. Quality-based cell filtering.
  Model anc101\_a1 courtesy of \cite{HexMeshSGP2011}.}
 \label{fig:Comp1}
\end{figure*}

To evaluate the potential of the proposed visualization tool for hex-mesh inspection, we performed an informal user study with the goal to assess the strengths and weaknesses of our tool compared to the one by Bracci and co-workers~\cite{Bracci:2019}. With each tool, the users visualized 3 different hex-meshes. The users were asked to comment on how effectively they understood the overall shape of the objects, determined the regions with highly deformed cells, and could assess the spatial relationships between regions with different deformation strengths and the concrete deformation characteristics of cells in regions comprised of highly deformed cells. Visual comparisons to HexaLab \cite{Bracci:2019} and the main sheet extraction method of Xu et al.~\cite{Xu:2018:TVCG} are given in \autoref{fig:Comp4}, \autoref{fig:Comp2}, \autoref{fig:CompAll} and \autoref{fig:Comp1}. In the user study we did not consider the method by Xu et al., since its focus is on a topological mesh analysis and not on the visualization of cell deformations. The major findings from the pursued user study are as follows:

\begin{itemize}
\item \textbf{Global view} Users appreciate that the global context is always visible when using our tool. Due to the use of volume rendering with deformation strength-based classification in the context region, all regions with high deformation cells can be perceived in relation to each other. The visualization hints on all potentially interesting regions. With only volume rendering, however, users sometimes loose the depth perception and feel that the global mesh structure cannot be understood well. This limitation becomes obsolete when coarse-scale edge structures are blended into the context region, which enhances the understanding of the global mesh structure without introducing clutter. HexaLab, in comparison, supports the rendering of singular edges in filtered mesh regions and a transparent mesh outline to maintain some global context. Users perceived as a minor limitation the resulting sparseness of regions in which cells are filtered out. 
\item \textbf{LoD structure} Unlike when using slicing, peeling and quality-based cell filtering, where cells are removed entirely based on a binary decision criterion, users appreciate the smooth LoD-based transition from focus to context and high to low deformations provided by our tool. This effectively reveals how the cell quality changes globally, and whether these changes are rather smooth or occur abruptly. 
\item \textbf{Edge-based rendering} When inspecting regions via the screen space lens and edge rendering, users were able to quickly access both the relation of deformed cells to their surroundings and how the cells are deformed. When rendering opaque surfaces, almost all neighbors of a cell would need to be filtered out---increasingly removing context information---in order to see the cell edges and be able to perform a fine-granular deformation analysis. 
\item \textbf{Scalar field visualization} Two users from computational science found it very appealing that also scalar values given per vertex or cell can be visualized in turn using volume rendering (cf. \autoref{fig:femur}). In particular when hex-meshes are used as simulation grids, this option becomes very effective for visualizing the relationships between simulation results and errors on the one hand, and the underlying cell structures on the other hand. 
\item \textbf{Interactive modification of visual parameters} It was perceived very supportive of a detailed mesh analysis that all rendering parameters could be changed interactively, and, thus, groups of elements could be quickly \mbox{(de-)}emphasized while enabling less and more attention on the global mesh structure.
\end{itemize}

User have also pointed out potential limitations of our approach, some of which could be overcome by only minor adjustments. 

\begin{figure*}[hp]
 \centering
 \includegraphics[width=0.49\linewidth]{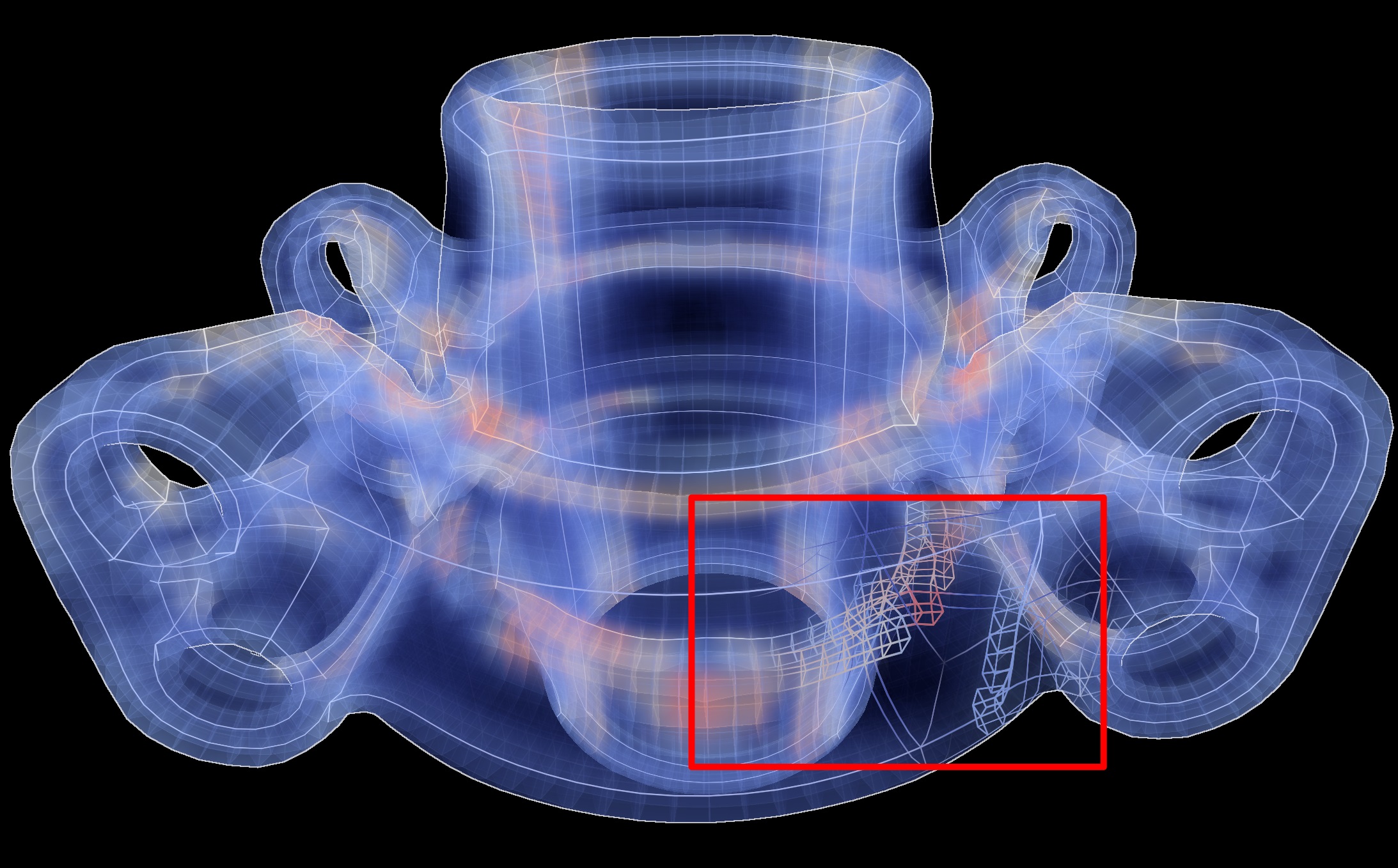}
 \includegraphics[width=0.49\linewidth]{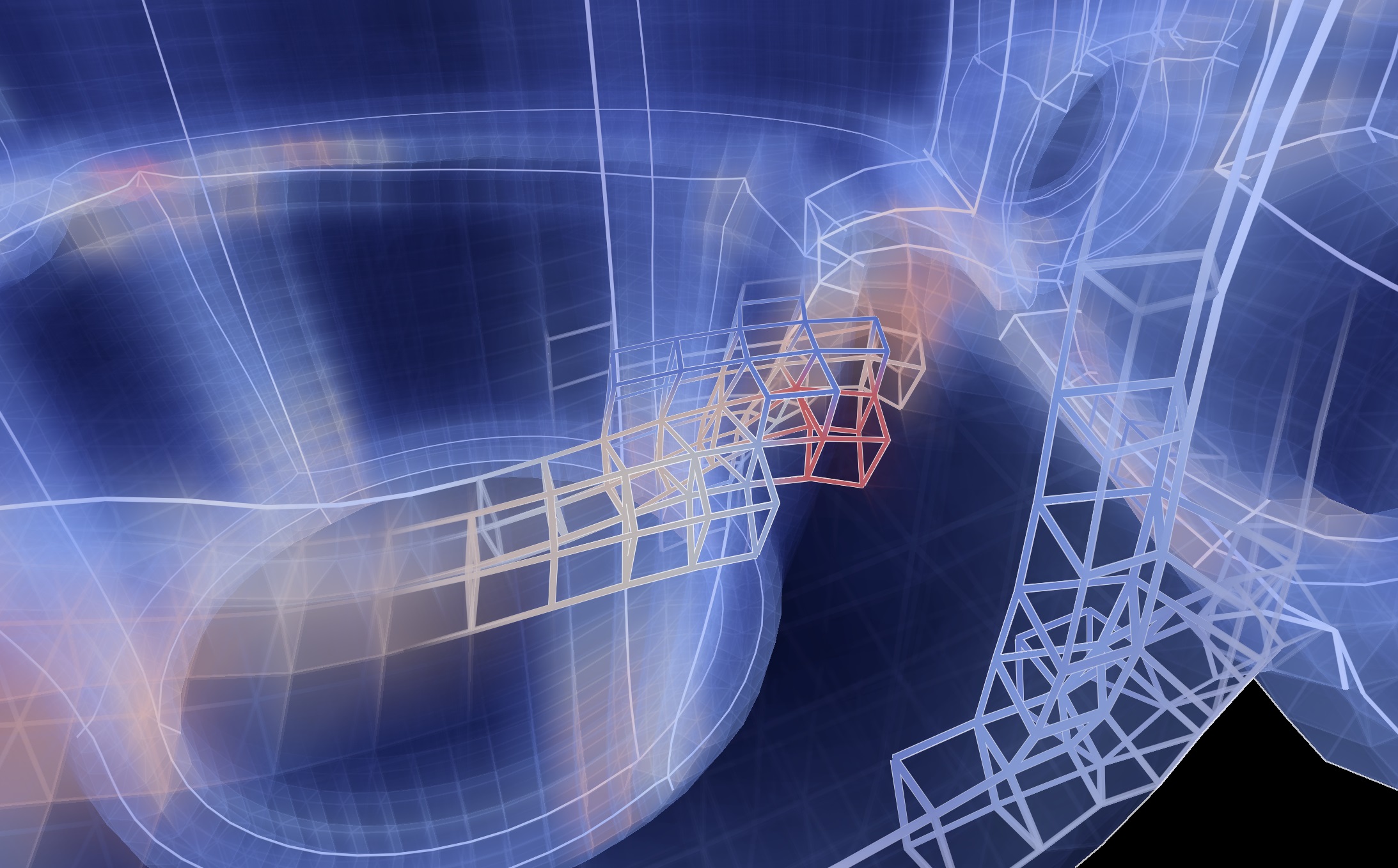}
 \vspace{0.1cm}
  \caption{In the first F+C view, important regions are effectively revealed. Framed region shows sub-structures that have been selected via the focus lens. Zoom-in and focus size adjustment enables a fine granular cell analysis. Surrounding cells with high deformation are still present in the context. 
  Model grayloc courtesy of \cite{AllHex2016}.
  }
 \label{fig:sidebyside-grayloc}
\end{figure*}

\begin{figure*}[hp]
 \centering
 \includegraphics[height=4.0cm]{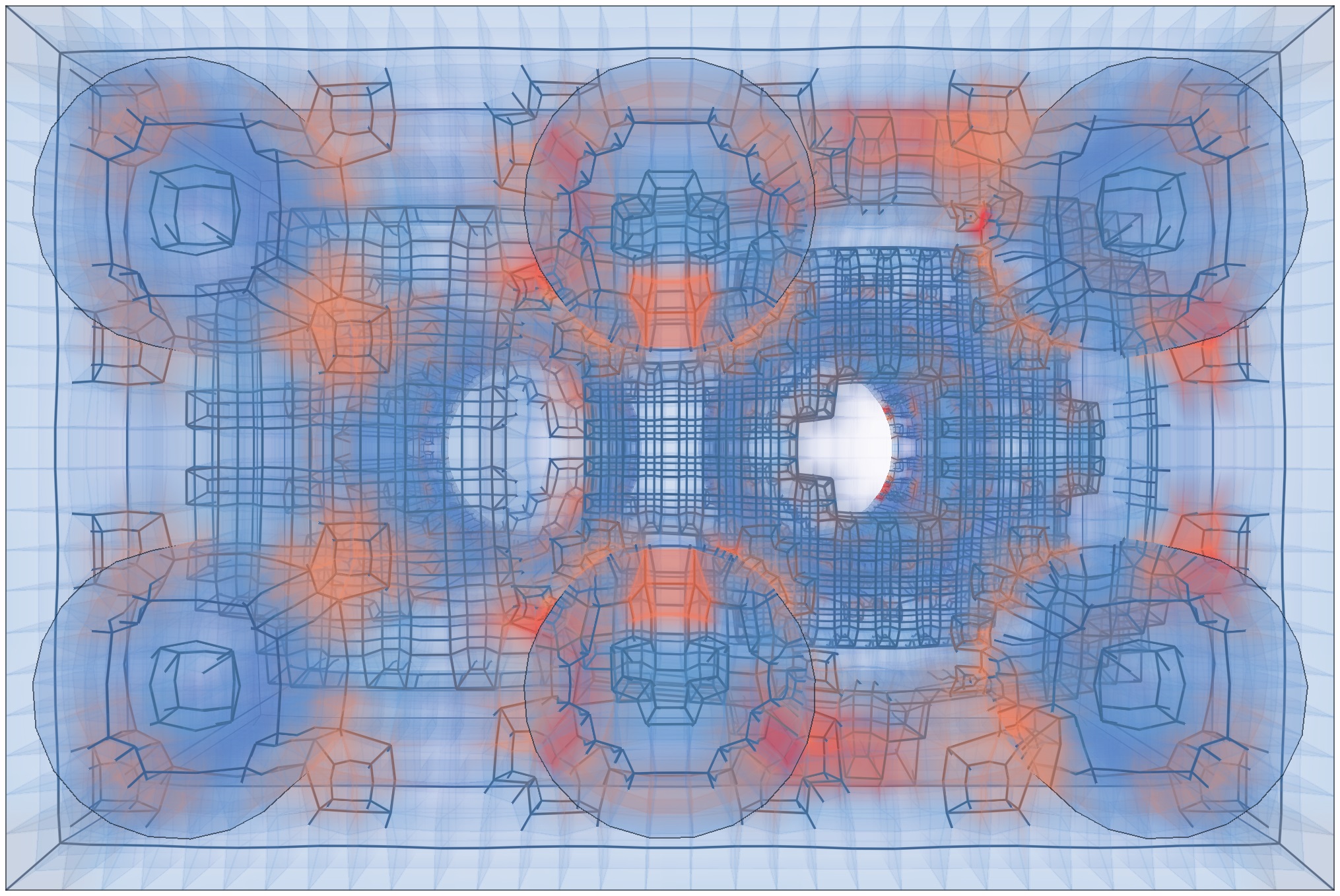}
 \hspace{2.0cm}
 \includegraphics[height=4.0cm]{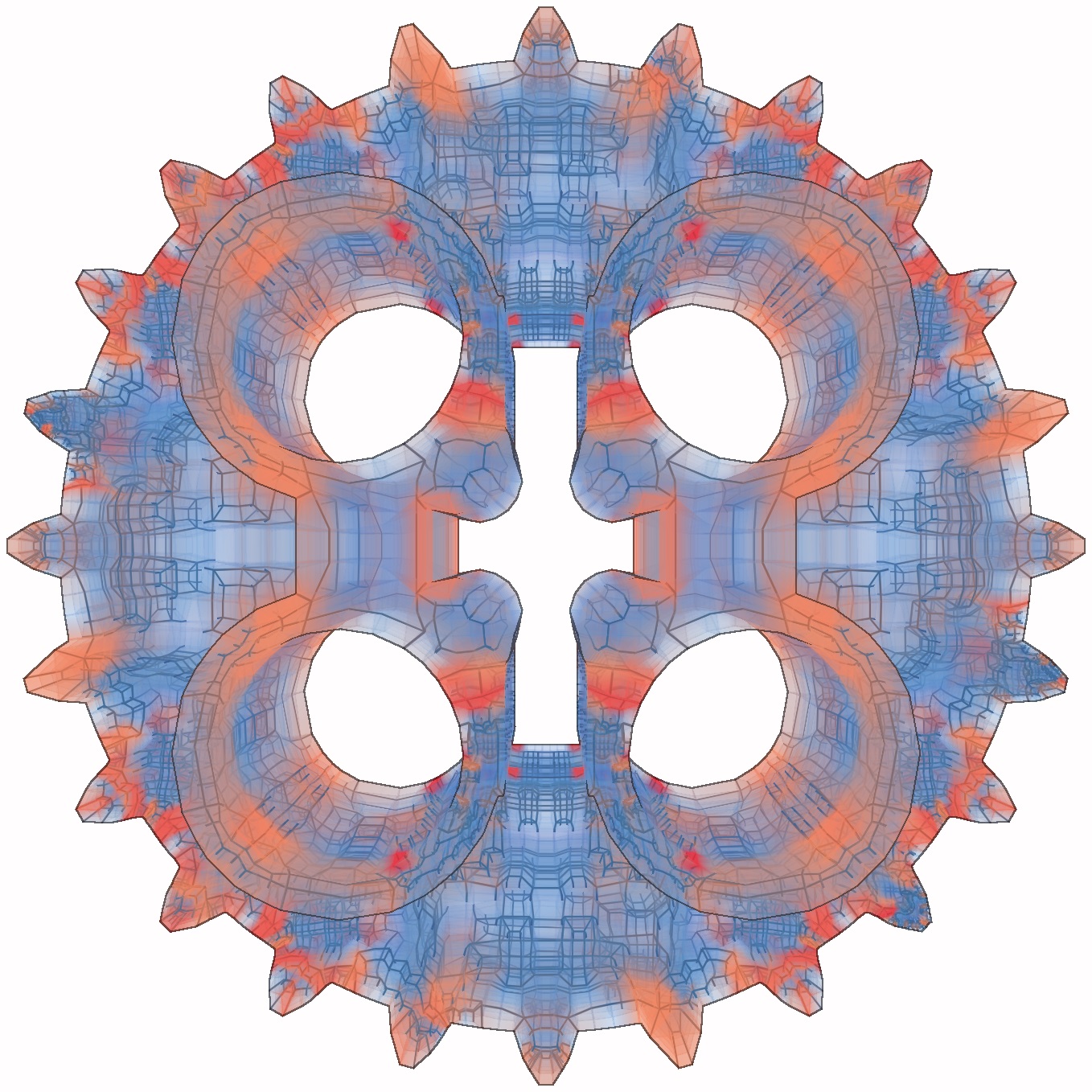}
  \caption{Contextual volume rendering of meshes from an octree-based meshing approach. Due to the highly irregular topological structure, many short line segments are created. Models courtesy of \cite{Octree2019}.}
 \label{fig:Octree}
\end{figure*}

\begin{itemize}
\item It was stated that unshaded lines impact the ability to correctly observe spatial relations when looking at a still image (cf. \autoref{fig:Comp4}). We use depth cues in order to counteract this effect in the focus region by slightly desaturating fragments further away from the camera in the focus region. This is especially useful for lines that are further away from each other.
\item When using a screen space lens, clutter was perceived and some important regions couldn't bee seen, as all elements along the viewing cone are put into focus. Therefore, we also provide a mechanism similar to an object space lens (cf. \autoref{fig:sidebyside-grayloc}). The user can click on the mesh to select the initial object space lens position by picking the closest mesh surface point along the viewing ray. The initial viewing ray is saved, and the user can move the object space lens position along this ray into the object using the mouse wheel. When the user moves the camera, the object space position and the moving ray of the lens stay unchanged.
\item When a meshing technique produces meshes with a very high number of singularities (e.g., octree-based meshing techniques, cf. \autoref{fig:Octree}), the LoD structure is cluttered as well and becomes less useful. This drawback also affects tools like HexaLab \cite{Bracci:2019}, which renders singular edges in regions where cells are filtered out. Xu et al.~\cite{Xu:2018:TVCG} also state regarding their method that ``it is still hard to show the structure of an octree or tet-split hex-mesh due to the overly complex structure and a large number of extracted main sheets''.
\end{itemize}

\section{Conclusion and future work}
In this paper, we have introduced an interactive F+C rendering technique for hex-meshes using fragment-based edge and face rendering. We have demonstrated that even high-resolution meshes can be rendered at high visual quality by using a carefully designed combination of detailed cell information and surrounding contextual information. To achieve this, we have introduced the use of hexahedral sheets for extracting a hierarchical LoD edge structure that provides important shape cues in the context region. This allows to significantly reduce occlusions and visual clutter. By using a purely fragment-based rendering approach, which smoothly transitions between highly detailed edge rendering and volumetric face blending, interactive rendering of data sets comprised of up to a few millions of elements is achieved on current GPU architectures. 
Our results indicate the potential of the proposed rendering technique for an interactive visual inspection of hex-meshes, supported by an automated guidance to important mesh regions. 

In the future, we will shed light on approximate rendering techniques for transparent fragments that can avoid the use of per-pixel fragment lists, such as Multi-Layer Alpha Blending ~\cite{salvi2014multi} or Moment-Based Order Independent Blending ~\cite{munstermann2018moment}. These techniques do not render the fragments in correct visibility order, yet since transparency is mostly used in the context region with more emphasis on closer mesh structures, such techniques might be able to provide a meaningful approximation.    
We further envision an AR-based stereoscopic inspection of hex-meshes to provide an improved spatial understanding of shape variations. Here it will be interesting to analyse whether a purely fragment-based approach is suitable for stereoscopic rendering. We will further shed light on the use of the proposed method for irregular meshes, such as tetrahedral meshes. For such meshes, the construction of an LoD structure is not possible at first hand, and alternative hierarchical representations need to be developed. Furthermore, visualizing highly topologically irregular hex meshes generated by, e.g., octree-based techniques without resulting in excessive clutter is also a problem that still needs more investigation. Finally, we intend to extend the rendering method to perform volume rendering of physical fields given at the hexahedral cells or vertices. This includes in particular the use of extended barycentric interpolation for deformed hex-cells and the rendering of implicit isosurfaces going through the cells.



\ifCLASSOPTIONcaptionsoff
  \newpage
\fi



\bibliographystyle{IEEEtran}
\bibliography{IEEEabrv,template}

%


\section{Acknowledgments}
The authors would like to thank Maximilian Bandle, Technical University of Munich, for his support concerning the efficient implementation of per-pixel fragment sorting on GPUs, and the various authors for providing the hex-meshes we have used.
This work was partially funded by the German Research Foundation (DFG) under grant number WE 2754/10-1 ``Stress Visualization via Force-Induced Material Growth''.

\begin{IEEEbiography}
	[{\includegraphics[width=1in,height=1.25in,clip,keepaspectratio]{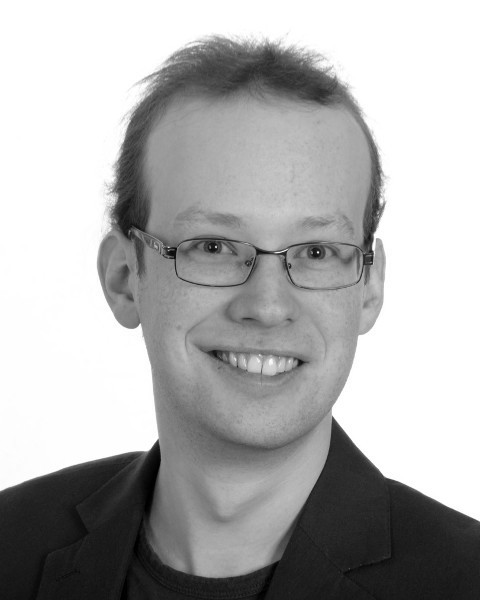}}]{Christoph Neuhauser}
	is a graduate research assistant at the Computer Graphics and Visualization Group at the Technical University of Munich. He received his B.Sc. in computer science from TUM in 2019. Major interests in research comprise scientific visualization and real-time rendering.
\end{IEEEbiography}
\begin{IEEEbiography}
	[{\includegraphics[width=1in,height=1.25in,clip,keepaspectratio]{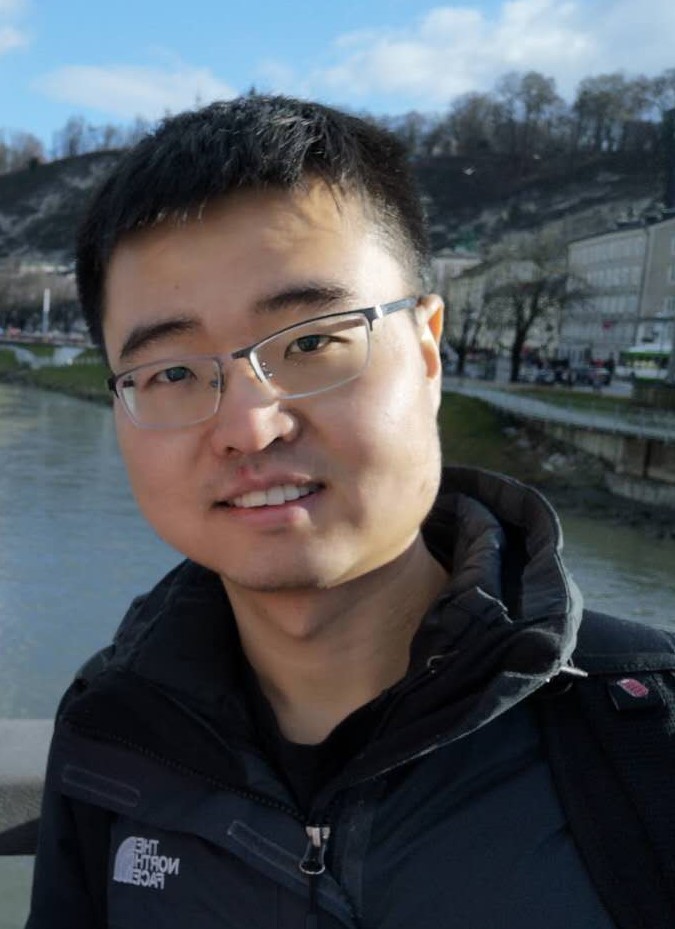}}]{Junpeng Wang}
	is a PhD candidate in the Computer Graphics and Visualization Group at Technical University of Munich, Germany. He received his Bachelor and Master's degrees in Aerospace Science and Technology in 2015 and 2018, respectively, both from Northwestern Polytechnical University, China. Currently, his research is focused on tensor field visualization and numerical simulation for solid mechanics.
\end{IEEEbiography}
\begin{IEEEbiography}
	[{\includegraphics[width=1in,height=1.25in,clip,keepaspectratio]{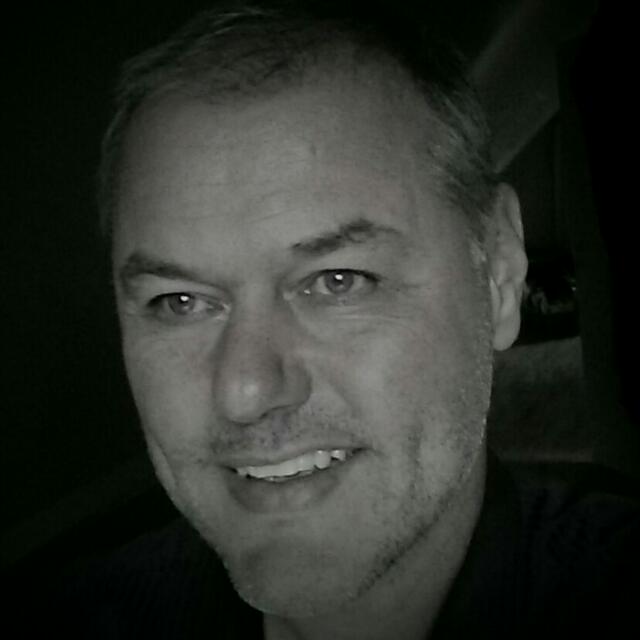}}]{R\"udiger Westermann}
	studied computer science at the Technical University Darmstadt and
	received his Ph.D. in computer science from
	the University of Dortmund, both in Germany. In
	2002, he was appointed the chair of Computer
	Graphics and Visualization at TUM. His research
	interests comprise scalable data visualization and
	simulation algorithms, GPU computing, real-time
	rendering of large data, and uncertainty visualization.
\end{IEEEbiography}
\vfill




\end{document}